\documentclass{CVM_author}

\CVMsetup{
type      = {Research Article},
doi       = {s41095-0xx-xxxx-x},
title     = {Continuous Indexed Points for Multivariate Volume Visualization},
author    = {Liang Zhou$^{1}$\cor{}, Xinyi Gou$^{2}$, and Daniel Weiskopf$^{3}$\\
\note{(Please place \cor{} after the corresponding author; use full names, not initials, for all the authors' names)}},
runauthor = {L. Zhou, X. Gou, D. Weiskopf},
abstract  = {
We introduce continuous indexed points for improved multivariate volume visualization.
Indexed points represent linear structures in parallel coordinates and can be used to encode local correlation of multivariate (including multifield, multifaceted, and multiattribute) volume data.
First, we perform local linear fitting in the spatial neighborhood of each volume sample using principal component analysis, accelerated by hierarchical spatial data structures.
This local linear information is then visualized as continuous indexed points in parallel coordinates: a density representation of indexed points in a continuous domain.
With our new method, multivariate volume data can be analyzed using the eigenvector information from local spatial embeddings.
We utilize both 1-flat and 2-flat indexed points, allowing us to identify correlations between two variables and even three variables, respectively. 
An interactive occlusion shading model facilitates good spatial perception of the volume rendering of volumetric correlation characteristics.
Interactive exploration is supported by specifically designed multivariate transfer function widgets working in the image plane of parallel coordinates.
We show that our generic technique works for multi-attribute datasets.
The effectiveness and usefulness of our new method is demonstrated through a case study, an expert user study, and domain expert feedback.
},
keywords  = {Volume visualization, multivariate volumes, multifield, correlations, indexed points, parallel coordinates},
copyright = {The Author(s)},
}
\newcommand{\myvec}[1]{\mathbf{{#1}}} 

\definecolor{revBlue}{RGB}{77,139,255}
\newcommand{\revcvm}[1]{#1}
\newcommand{\revcvmmj}[1]{#1}
\newcommand{\secRevCvm}[1]{#1}
\newcommand{\rev}[1]{#1}

\begin{document}

\maketitle

    \begin{figure}[b] \vskip -4mm
    \small\renewcommand\arraystretch{1.3}
        \begin{tabular}{p{80.5mm}} \toprule\\ \end{tabular}
        \vskip -4.5mm \noindent \setlength{\tabcolsep}{1pt}
        \begin{tabular}{p{3.5mm}p{80mm}}
    $1\quad $ & Institute of Medical Technology, Peking University Health Science Center, and the
    National Institute of Health Data Science, Peking University, Beijing, 100029, China. E-mail: zhoulng@pku.edu.cn.\\
    $2\quad $ & Peking University People's Hospital, Beijing, 100044, China. E-mail: gouxinyi1998@163.com.\\
    $3\quad $ & Visualization Research Center (VISUS), University of Stuttgart, 70569 Stuttgart, Germany. E-mail: weiskopf@visus.uni-stuttgart.de.\\
&\hspace{-5mm} Manuscript received: 2022-01-01; accepted: 2022-01-01\vspace{-2mm}
    \end{tabular} \vspace {-3mm}
    \end{figure}

\section{Introduction}

\textit{Correlation} is important for multivariate data analysis.
Local correlation analysis using linear data fitting can reveal linear and nonlinear relationships at different parts of the data. 
Well-established multivariate visualization methods such as scatterplots and parallel coordinates have been extended for viewing results of such analysis.
\begin{figure*}[tb]
    \centering
      \includegraphics[width=\linewidth]{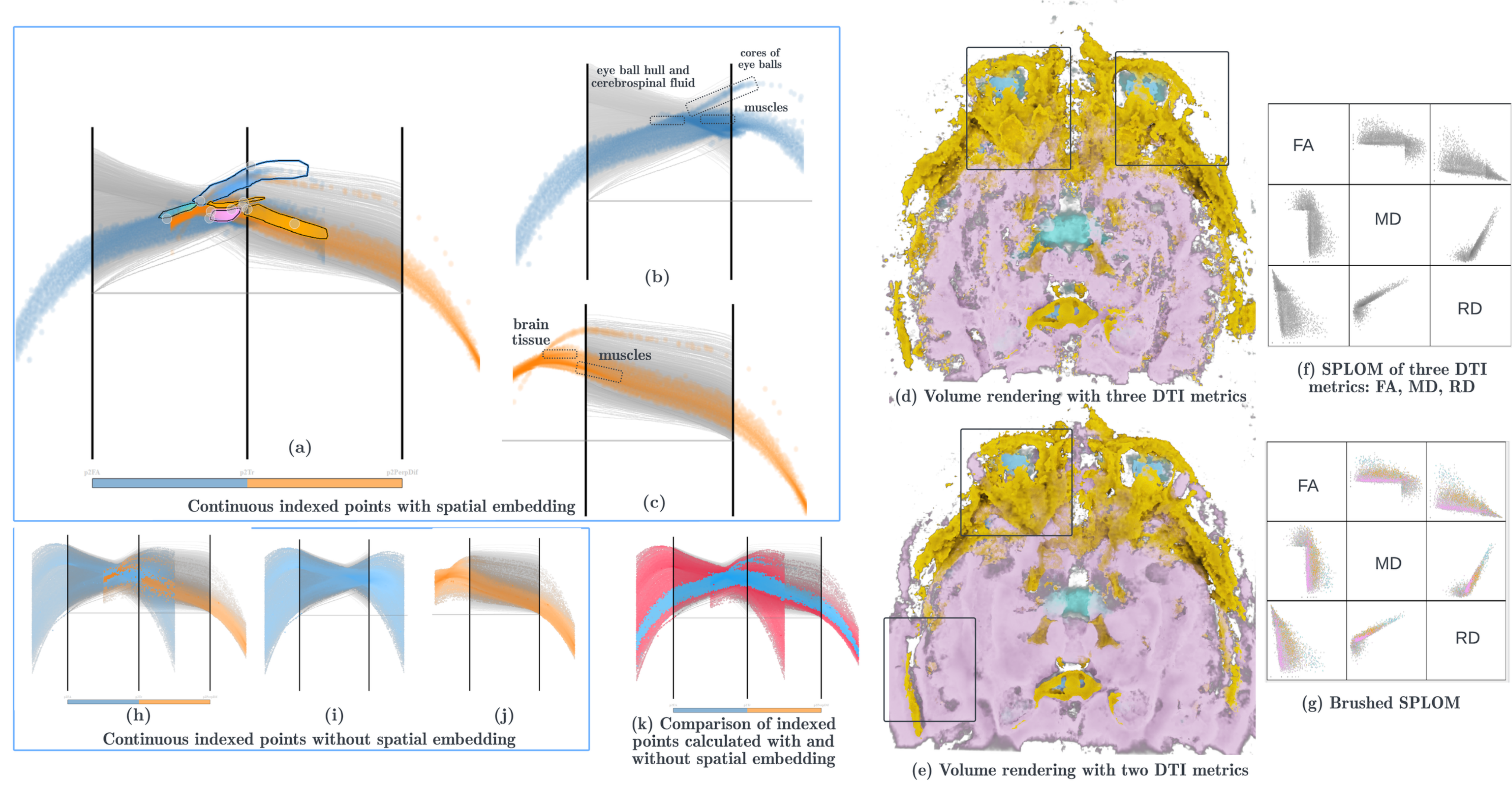}
  \caption{
 Visualization of scalar metrics of a DTI scan for studying the thyroid eye disease. 
 Local linear information of the data is shown as (a--c) a density representation of indexed points with spatial embedding. Patterns in the density-based indexed points guide the brushing of the features to classify (d, e) spatial features that are volume-rendered with an occlusion shading model. These features are not identifiable in (f, g) the scatterplot matrix. Indexed points computed without spatial embedding (h--j) give a large number of false positives (k, red regions) compared to our method (k, blue regions) with spatial embedding. \secRevCvm{Underlying DTI dataset courtesy of Peking University People’s Hospital.}
  }\label{fig:teaser}
\end{figure*}

Local correlation analysis can also be used as a powerful tool for visualizing multivariate data with spatial embedding, e.g., multivariate volumetric data~\cite{Sauber2006}.
In this paper, we propose a local correlation model for multivariate volume data that considers data values in their local spatial neighborhoods.
Volumetric regions can be classified based on their multivariate data distributions---local linear relationships of different angles and various nonlinear structures can be identified using linear fitting. 
Another aspect of our method is the representation of local linear relationships as a density on a continuous domain. Although multivariate volumetric data are defined in a continuous spatial domain, they are typically treated as independent collections of discrete data points defined on grids. In contrast, we propose a continuous-domain extension of discrete indexed point representations in parallel coordinates~\cite{Zhou:2018} to capture the full information contents, consider their spatial configuration, and achieve a high-quality visualization.

Figure~\ref{fig:teaser} shows an example of diffusion tensor imaging (DTI) metrics for studying the thyroid eye disease.
Here, the features are identified as coherent patterns with medium or high density of indexed points in parallel coordinates (Figures~\ref{fig:teaser}~(b,c)), which is the basis for selecting parts of the volume by brushing (Figure~\ref{fig:teaser}~(a)).
The corresponding structures in the spatial domain are volume-rendered using an adapted occlusion shading model (Figure~\ref{fig:teaser}~(d)). For comparison, Figures~\ref{fig:teaser}~(f, g) show the multivariate data in a scatterplot matrix (SPLOM). Here, it is apparent that many of the features would be hard or even impossible to be brushed in scatterplots. 
This demonstrates that relevant features are missed in traditional multivariate plots. With our technique, we aim to provide a new multivariate visualization designed to fit to the interplay between volume rendering and feature identification in parallel coordinates. 

The contributions of this paper are as follows:
\begin{itemize}
    \item An approach to indexed points that represents local correlations in spatial neighborhoods of volumetric data. 
    \item Continuous indexed points as a density model of correlations in parallel coordinates. 
    \item Volume rendering with an advanced shading model using multivariate transfer functions of indexed points. 
\end{itemize}
One benefit of our method is its versatility: it works for multivariate volume data in a generic sense. We demonstrate that the method can be applied to classification and transfer function design for multifield, multiattribute, or multifaceted data \cite{Kehrer2013}. 
With several examples, a case study including a domain expert, and a user study with visualization experts, we show the versatility and illustrate how users can read visual patterns and signatures from the visualizations.

Another benefit is that our technique can be easily integrated in existing brushing-and-linking systems that work with volume visualization and parallel coordinates. The key elements of the implementation are the identification of local correlation, which can even be done in a pre-processing step, and a slight extension of parallel coordinates that takes into account the indexed points computed from local correlation.

[Note: We will provide source code for indexed point computation and visualization along with the publication of the paper, if accepted.]
\section{Related Work}
Our method is related to correlation visualization of multivariate data, continuous and density-based visualization, as well as multivariate volume visualization.

\subsection{Correlation Visualization}
Visualizing correlation is important for understanding multivariate data.
Scatterplots, SPLOMs, or variants are typically used for visualizing correlations in Cartesian coordinates~\cite{Li:IV:2010}.
There, correlations are shown as linear structures. 
Correlations in parallel coordinates are indicated by intersections of polylines or curves based on the point-line duality between Cartesian and parallel coordinates~\cite{InselbergPCPbook}.
Visual mappings in Cartesian coordinates can be combined with parallel coordinates to improve the analysis of correlation,
by placing scatterplots close to parallel coordinates~\cite{Qu:VIS2007,Steed:VAST09,Holten:CGF:CGF1666}.
Scatterplots of multidimensional data after dimensionality reduction can be embedded into parallel coordinates within adjacent axes~\cite{Yuan:VIS:2009}.
Alternatively, correlations are calculated for local neighborhoods of multidimensional data, and are visualized using accurately projected glyphs in multidimensional projection methods~\cite{Bian2021}.

Another strategy unifies various types of parallel coordinates with SPLOMs in a flexible manner.
A user-driven technique supports customized visualizations in parallel coordinates or SPLOM variants~\cite{Claessen:TVCG:2011}.
SPLOMs and 2D and 3D parallel coordinates are unified with smooth transitions in the P-SPLOM method~\cite{Viau:TVCG:2010}.

Direct visualizations of linear relationships in parallel coordinates are available.
Local linear relationships of two data dimensions are displayed using the point-line duality in the DSPCP technique~\cite{Nguyen:TVCG:2017}.
There is a general model that represents multivariate linear structures or flat surfaces as indexed points in parallel coordinates~\cite{Zhou:2018}. We adopt the basic model of indexed points and local linear relationship visualization and extend it to the requirements that come with volume data and their visualization. 

A general issue of parallel coordinates is that the representation of positive and negative correlations is asymmetric and intersections of positive correlations extend to infinity.
This issue has been addressed using an angle-uniform transformation that maps parallel coordinates to a bounded space with symmetric patterns for positive and negative correlations alike~\cite{ZhangZhou2023}. We extend angle-uniform parallel coordinates for visualizing indexed points and, in particular, continuous indexed points.

\subsection{Continuous and Density-Based Multivariate Plots}
Volume datasets are defined on a continuous domain (typically, a cube), but this continuous nature is often lost and ignored when the data is stored as discrete samples on grids.
Faithful visualization requires reconstruction on continuous data spaces.
Such visualizations are based on finding the density of the data domain by applying the preservation of mass from the spatial to the data domain using integration and geometric measure theory. 
An accurate and generic mathematical model is available for continuous scatterplots, along with a 2D density-based visualization~\cite{Bachthaler:2008:VIS}.
The density of continuous scatterplots can be mapped to parallel coordinates based on the point-line duality to arrive at continuous parallel coordinates~\cite{Heinrich:VIS:2009}. We adopt the concept of continuous scatterplots and parallel coordinates to combine and extend them to density representations of indexed points. 

\subsection{Multivariate Volume Visualization}
Much prior work addresses the visualization of multivariate volume data.
A comprehensive survey is available on the broader topic of multifaceted scientific data visualization~\cite{Kehrer2013}.
Parallel coordinates are typically used for exploring the volumes via multidimensional transfer functions using brushing-and-linking~\cite{Akiba:2007:Eurovis,Zhao2010,Guo:2011:PacificVis,Zhou:PVis:2013}.
Dimensionality reduction methods and SPLOMs are also used to design multidimensional transfer functions~\cite{Guo:2011:PacificVis,Zhou:PVis:2013}.
Principal component analysis (PCA) and correlation analysis are performed globally on volume attributes, and the exploration of volumes is achieved with scatterplots~\cite{Oeltze2007}. 
Local correlation coefficients are calculated to form correlation fields and are represented in multifield-graphs for multivariate volume visualization~\cite{Sauber2006}.
\revcvmmj{
A graph representation of hierarchical state transition information measured by the distance of probability distributions can be constructed to explore and visualize time-varying volumes through brushing-and-linking~\cite{Gu2011}.
A RadViz-based method maps multidimensional volume samples to a 2D radial space for transfer function design to visualize multivariate or multi-band volumes~\cite{Kumar2024}. 
}
\revcvm{An adaptive sampling technique is available for computing spatial correlations of ensembles of 3D volumes~\cite{Neuhauser2024}.}

\begin{figure*}[tb]
     \centering
    \includegraphics[trim={0 0 0 0},clip,width=\linewidth]{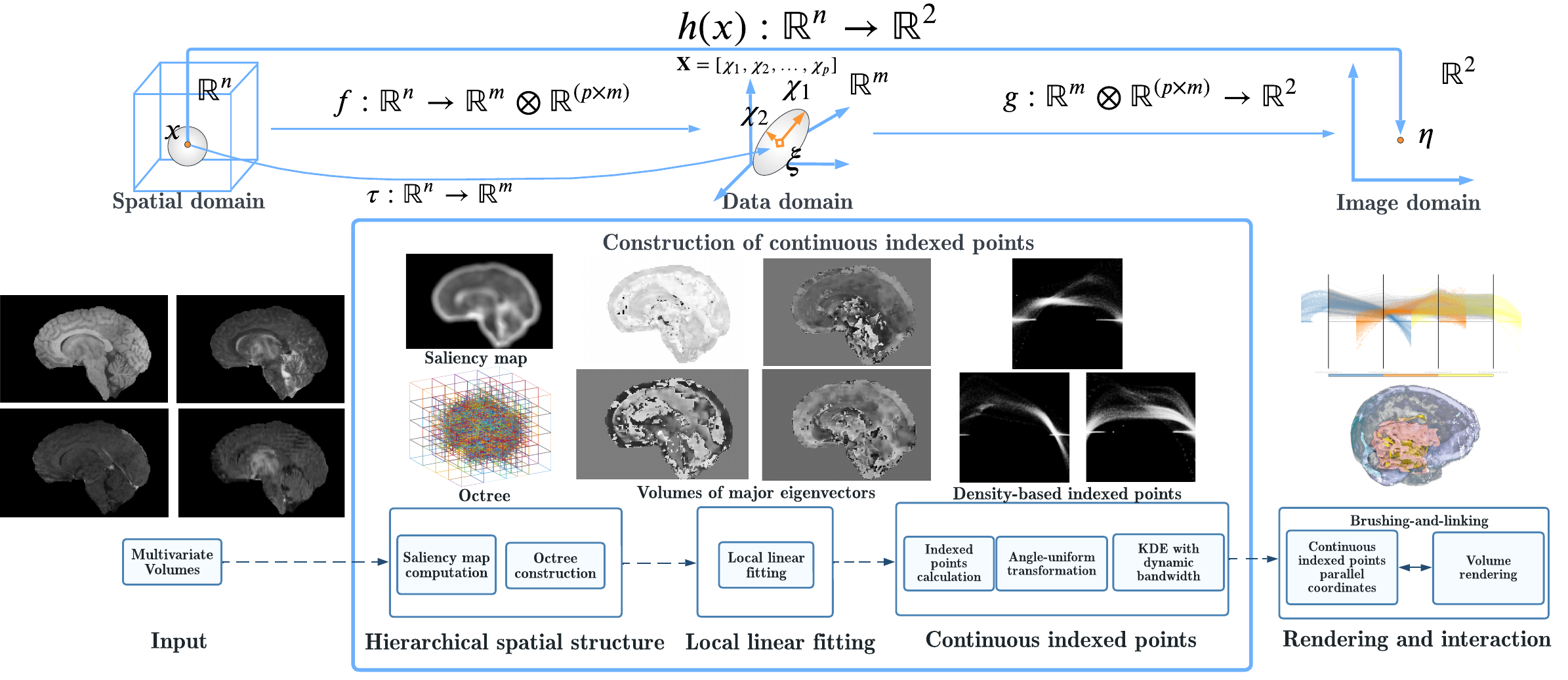}
    
    \caption{Overview and workflow of our continuous indexed points method for local correlation visualization for multivariate volumes.
    \revcvmmj{For each point $\bm{x} \in \mathbb{R}^n$ in the spatial domain, local linear information of its neighborhood is calculated. The local linear information $(\bm{\xi}, \mathbf{X})$ includes the data value $\bm{\xi} = \tau(\bm{x}) \in \mathbb{R}^m$ and a matrix $\mathbf{X} \in \mathbb{R}^{(p\times m)}$ containing a number of $p$ (the dimensionality of indexed points, namely, $p$-flat indexed points) eigenvectors fitted to the data in the neighborhood. Continuous indexed points $\bm{\eta} \in \mathbb{R}^2$ are then computed by transforming $(\bm{\xi}, \mathbf{X})$ to the 2D image domain for visualization and interaction. \secRevCvm{Underlying MRI dataset by Menze et al.~\cite{Menze2015}.}}
    }
    \label{fig:pipeline}
\end{figure*}

\rev{
Alternatively, multivariate volumes can be visualized with indirect methods.
For example, fiber surfaces~\cite{Carr2015} extend isosurfaces to bivariate volumes.
Positional probabilities of fiber surfaces are quantified and visualized for uncertain bivariate volumes~\cite{Athawale2023}.
A visual analysis method is available for bivariate fields of electronic transition data using fiber surfaces and continuous scatterplot~\cite{Sharma2021}.
Feature level sets---generalized level sets for multivariate volumes---unify isosurfaces and fiber surfaces~\cite{Jankowai2020}. 
Linearization is another strategy for multivariate volume visualization.
Space-filling curve methods are used to linearize volume datasets with the preservation of spatial coherency and support the comparison of different variables~\cite{Weissenboeck2019,Zhou2021}.
}
\revcvm{
Trait-based merge trees are introduced to visualize multi-field volume datasets~\cite{Jankowai2023}.
Visual analysis of multi-channel volumetric tissue data is achieved with visualization embedded on edges between cells~\cite{Moerth2024}.
}

In contrast to all techniques described above, our method visualizes the data domain of multivariate data simultaneously with local correlation information in a single parallel coordinates view without context switching.
The new visual representation can be used to explore multivariate volumetric data using its correlation-based transfer function space for data classification.

\section{Overview and Mathematical Model}
In this section, we first provide an overview of the steps taken to compute the continuous indexed points before we present a mathematical description of our approach. 

\subsection{Method Overview}
\revcvm{We aim to show correlation in multivariate volume field data. 
Therefore, we target continuous density modeling of the index points and modeling rooted in the spatial relationships (coherence) of the data. 
Our method involves two stages---a precomputation stage and an interactive stage---to achieve this overall goal as shown in Figure~\ref{fig:pipeline}, which illustrates the visualization process along with the involved transformations and spaces.}

We start with a multivariate volume dataset as input.
\revcvmmj{A precomputation stage (the blue box) calculates spatial correlations of multivariate volumes (feature extraction) and transforms this information to continuous indexed points (in a parallel coordinates plot) as explained in Section~\ref{sec:idxPtCompute}. }
There, the computational parts are important for scalability of the large datasets that come with volume visualization.
The angle-uniform transformation is to show the distribution of indexed points in a finite area for good interaction, i.e., brushing-and-linking.

\revcvm{A subsequent interactive stage realizes the actual visualization (Section~\ref{sec:idxVis}) and interaction (Section~\ref{sec:interactions}) techniques.
The rendering of continuous indexed points aids the identification of correlated feature patterns, while the occlusion shading provides important depth cues in the volume rendering.
Interactions are designed to facilitate the exploration of correlations in the volume rendering and indexed points in parallel coordinates through brushing-and-linking.
}

\subsection{Mathematical Model}
\label{sec:mathModel}

\begin{figure}[tb]
\centering
    \includegraphics[width=0.95\linewidth]{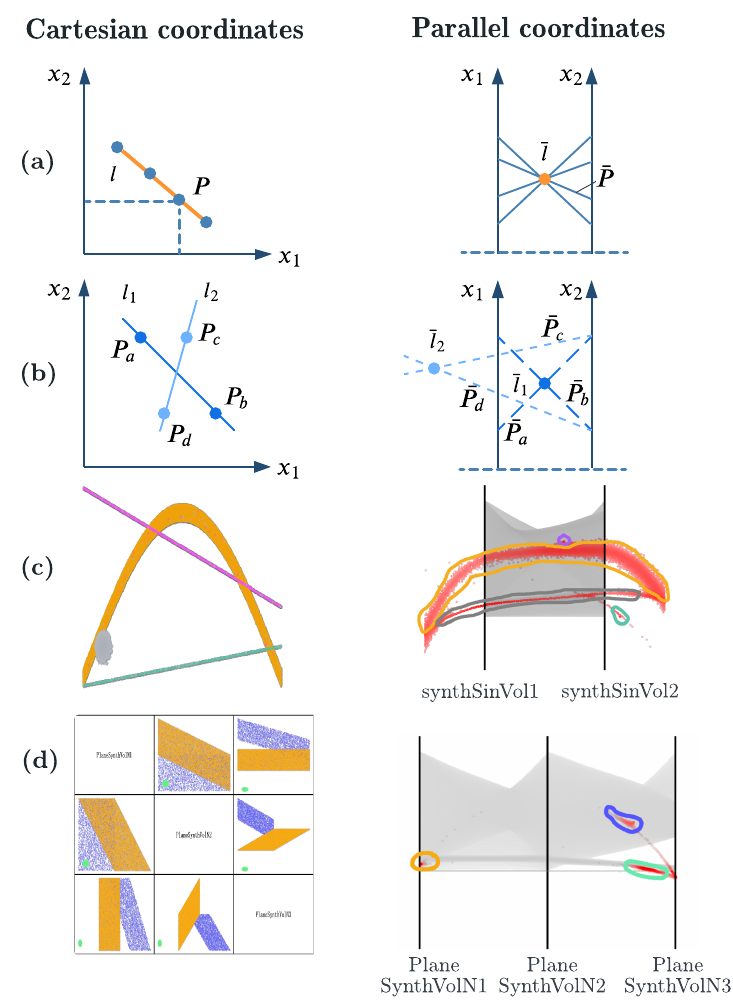}
    \caption{
    \secRevCvm{The relationship between Cartesian coordinates and parallel coordinates.
    The point-line duality is shown for the example of a single point $P$ (blue) and negatively correlated points on a line segment $l$ (orange) in Cartesian coordinates~(a), and two points $\bar{l}_1$ and $\bar{l}_2$ in parallel coordinates~(b).
    }
    More complex density patterns~(c) are brushed to highlight a part with negative linear correlation (violet), positive linear correlation (green), nonlinear correlation (orange/yellow), and Gaussian blob (gray). Cartesian planes (d--left, orange and blue) and non-flat volumes (d--left, green) are mapped to patterns of 2-flat indexed points (red points in d--right) with the corresponding brushes.
    }
    \label{fig:pointLine}
\end{figure}

Let us start by briefly summarizing the concept of indexed points, which serve as a main mathematical building block of our technique.
\secRevCvm{
The concept relies on the point-line duality that specifies the transformations between Cartesian coordinates and parallel coordinates: a Cartesian point maps to a polygonal line (polyline) in parallel coordinates and a Cartesian line maps to a point in parallel coordinates (Figure~\ref{fig:pointLine}).
}

An indexed point is a point representation of a Cartesian line (1-flat) or plane (2-flat) in parallel coordinates.
The Cartesian coordinates are associated with the space in which data points are represented.
Visually,  can be identified with the canvas of a scatterplot in the 2D case.
A data point $P$ in Cartesian coordinates is mapped to a line $\bar{P}$ in parallel coordinates (Figure~\ref{fig:pointLine}~(a)).
In fact, that is how a traditional parallel coordinates visualization is constructed from input data. 
\secRevCvm{Conversely, a line in Cartesian coordinates is mapped to a point in parallel coordinates, i.e., 1-flat indexed point, which is the intersection of a group of lines in parallel coordinates. 
See the orange line $l$ and point $\bar{l}$ in Figure~\ref{fig:pointLine}~(a).   Figure~\ref{fig:pointLine}~(b) shows an example with two Cartesian lines and Figure~\ref{fig:pointLine}~(c) more complex configurations with dense patterns.} 
The idea indexed points can be extended recursively to represent a plane in 3D Cartesian coordinates using a 2-flat point in parallel coordinates (Figure~\ref{fig:pointLine}~(d)).
\secRevCvm{For more details, we refer to the book by Inselberg~\cite{InselbergPCPbook} or our previous work~\cite{Zhou:2018}.}


For the general case, higher-dimensional linear structures in $(p+1)$-dimensional Cartesian coordinates can be represented with indexed points, i.e., $p$-flat indexed points~\cite{InselbergPCPbook,Zhou:2018}.
Indexed points of $p$-flats can be used to visualize local multivariate correlations by local linear fitting in neighborhoods in Cartesian coordinates~\cite{Zhou:2018}.

In this paper, we consider the case of multivariate data with spatial embedding in 3D and focus on 1-flat and 2-flat indexed points for local linear relationships of two and three variables, respectively (Figure~\ref{fig:pointLine}). 
1-flat indexed points are the first natural step to including information about local correlation, which is traditionally ignored in multivariate volume visualization. 2-flat indexed points go even further by enabling the comparative analysis of three variables, which was previously not possible. An extension to higher-dimensional indexed points is mathematically straightforward, but we leave the investigation of its use for visual data analysis to future research.

For volumetric data, the calculation of indexed points involves three different domains (\revcvmmj{Figure~\ref{fig:pipeline}-top}): the $n$-dimensional spatial domain, the $m$-dimensional data domain of the multivariate data values living in the spatial domain, and the 2D image domain in which the indexed points are drawn.
The conversions from the spatial domain to the data domain, and, eventually, to the image domain can be described by two maps.
\revcvmmj{The map $\tau:  \mathbb{R}^n \rightarrow \mathbb{R}^m$ represents the mapping from the spatial domain to the data domain, i.e., the mathematical representation of the dataset (along with interpolation).} Since we focus on multivariate volumes, we have $n=3$ for volumetric data. For full generalizability, we will keep an arbitrary $n$ in the following mathematical description.   
The number of dimensions of the attached multivariate data is given by $m$. Therefore, $\tau$ is a mathematical description of the dataset to be visualized.

In the data domain, the representation of an indexed point is anchored at its data point and comprises the direction information from local linear fitting. We model such information for each data point by
\begin{equation}
    f: \mathbb{R}^n \rightarrow \mathbb{R}^m \otimes \mathbb{R}^{(p\times m)} \,, \ \bm{x} \mapsto (\bm{\xi}, \mathbf{X}) = f(\bm{x}) \;,
\end{equation}
with the data value $\bm{\xi} = \tau(\bm{x})$ (\revcvmmj{Figure~\ref{fig:pipeline}-top left}). The interesting part is the fitted directions, written as a matrix $\mathbf{X}$ of size $p\times m$ whose $p$ rows store the $m$-dimensional vectors $[\bm{\chi}_1, \bm{\chi}_2, \ldots, \bm{\chi}_p]$ living in the vector space over the data domain. For our case of 1-flat and 2-flat indexed points, $p=1$ or $p = 2$. Since the data domain is flat, we can use the same notation for points and vectors.
Prior work~\cite{Zhou:2018} computed the vectors $\mathbf{X}$ based only on data values in the neighborhood of $\bm{\xi}$. A critical difference is that we take into account the location in the spatial domain, $\bm{x}$, as well. 

The final step takes the linear fit from the data domain to the image domain (\revcvmmj{Figure~\ref{fig:pipeline}-top right}):
\begin{equation}
g: \mathbb{R}^m \otimes \mathbb{R}^{(p\times m)} \rightarrow \mathbb{R}^2 \,, \  (\bm{\xi}, \mathbf{X}) \mapsto \bm{\eta} = g(\bm{\xi}, \mathbf{X}) \;,
\end{equation}
where a reduction to the respective two parallel coordinates axes and the point-line duality are employed for 1-flat indexed points.
For 2-flat indexed points, the conversion is from flat planes to the associated three axes.

Finally, the two mappings can be combined (\revcvmmj{Figure~\ref{fig:pipeline}-top}),
\begin{equation}
h: \mathbb{R}^n \rightarrow \mathbb{R}^2 \,, \  \bm{x} \mapsto \bm{\eta} = h(\bm{x}) = (g \circ f)(\bm{x}) \;,    
\end{equation}
to arrive at a transformation from spatial to image domain. Here, it is important to note that $h$ is a function that takes points to points. This is different from the traditional use of parallel coordinates, where points are mapped to polylines. 
Therefore, our model of continuous indexed points draws from both continuous scatterplots \cite{Bachthaler:2008:VIS} (with their point-to-point mapping) and continuous parallel coordinates \cite{Heinrich:VIS:2009} (with their mapping to the parallel coordinates plane). With the assumption of a continuous spatial domain, $\mathbb{R}^n$, we formulate the continuous indexed points problem as finding the density $\rho$ on the image domain:
\begin{equation}
    \rho : \mathbb{R}^2 \rightarrow \mathbb{R}, \; \bm{\eta} \mapsto \rho(\bm{\eta})\;.
\end{equation}
This density depends on $h$ and the source density in the spatial domain, $s$ (which is often assumed to be constant with a value of 1).
Virtual mass $M$ is conserved throughout the transformation and can be written using integrals as follows:
\begin{equation}
    M = \int_V s(\bm{x})\,d^n \bm{x} = 
    \int_{\Phi=h(V)} \rho(\bm{\eta})\,d^2\bm{\eta}
    \;.
\end{equation}
This equation has to hold for any volume $V$ and mass $M$. 

In this paper, we use a notation that specifically denotes the dimensionality of the integration domain in the superscript of the differential, as in earlier work on continuous plots~\cite{Bachthaler:2008:VIS,Heinrich:VIS:2009} and common in physics literature (see, e.g., Jackson~\cite{Jackson:1962}). For example, $d^n \bm{x}$ indicates integration over an $n$-D domain. 
Using geometric measure theory similar to the case of continuous scatterplots~\cite{Bachthaler:2008:VIS}, we can compute
\begin{equation}
\label{eq:densitymeasuretheory}
   \rho(\bm{\eta}) = \int_{h^{-1}(\bm{\eta})} 
   \frac{s(\bm{x})}{|\text{Vol}(D h)(\bm{x})|} d^{(n-2)}\bm{x} \;,
\end{equation}
with the volume measure $|\text{Vol}(D h)|$ defined by the partial derivatives of $h$ over the space that is orthogonal to $h^{-1}(\bm{\eta})$. For the case of volumetric data with $n=3$, the integration is in 1D space. 

The workflow of computing $\rho$ for continuous indexed points from the input data is illustrated in Figure~\ref{fig:pipeline}. 
We detail the numerical computation of 1-flat and 2-flat continuous indexed points (i.e., the map $h$) and a forward (scattering) algorithm for Equation~\ref{eq:densitymeasuretheory} in Section~\ref{sec:idxPtCompute}.
The visualization of indexed points is explained in Section~\ref{sec:idxVis}.

\section{Numerical Computation of Continuous Indexed Points}
\label{sec:idxPtCompute}

The core of our method is the transformation of the density of a region in the spatial domain all the way to the image plane as continuous indexed points: see the blue box in Figure~\ref{fig:pipeline}. 
Adopting sampling on the spatial domain (similar to Heinrich et al.~\cite{Heinrich:2011:splatting}), we aim to transform points from the spatial to the image domain, finally reconstructing a density image from splatted point samples.
\revcvm{Note that these steps are a precomputation stage of our pipeline that is independent of the interactive visualization.}

For each point $\bm{x}$ in the spatial domain, a neighborhood $D \subset \mathbb{R}^n$ is found. Please note that we use $\bm{x}$ to denote a point and its location alike.
The data value $\bm{\xi}$ of $\bm{x}$ in the data domain is given by $\tau$, and data values of points in the neighborhood are denoted as $\tau(D) \subset \mathbb{R}^m $ in the data domain.
We perform local fitting, using PCA for the footprint of the spatial neighborhood in the data domain. 
The resulting local linear information from $f$ is then converted to continuous indexed points in the 2D image domain by $g$.
Our technique first converts local linear information to indexed points with an angle-uniform transformation \cite{ZhangZhou2023} to the parallel coordinates plane.
Subsequently, kernel density estimation with dynamic bandwidth is used to generate a density representation for continuous indexed points in the image domain.

\subsection{Hierarchical Spatial Structure Construction}
For simplicity, we assume $\bm{x}$ to be nodes on a regular grid. However, in our general model, $\bm{x}$ is not necessarily restricted to be grid points or sampled with equal distance. 
But even with this simplification, computing local PCA for the neighborhood of each volume sample in a multivariate volume is costly. 
Therefore, a hierarchical structure---an octree in our case---is used to accelerate the process by skipping neighboring samples that have the same local neighborhood \revcvmmj{(Figure~\ref{fig:pipeline}-Hierarchical spatial structure)}.

Unlike typical octrees for univariate volumes, our octree has to be constructed by considering every volume attribute because the computation of local PCA involves all $m$ dimensions of the multivariate volume.
\revcvm{The octree should have a granularity that equals to the finest detected feature contained in each attribute at every location.
We use a saliency map $S_f(\bm{x})$ to describe the detected features.
}
For a given location, we have to decide whether it has exactly the same multivariate data values for all volume nodes in the local neighborhood as its adjacent locations.

\revcvm{The steps of building the saliency map $S_f(\bm{x})$ is as follows.}
First, the gradient magnitude is normalized for each volume attribute and summed up, where $\tau_i(\bm{x})$ is the $i$-th attribute of the volume data:
\begin{equation}
    S(\bm{x}) = \sum_{i=1}^{m}\left||\nabla \tau_i(\bm{x}) \right||\;.
\end{equation}
Afterward, the impact of the local neighborhood size is modeled by convolving $S(\bm{x})$ with a low-pass filter $\zeta$:
\begin{equation}
    S_f(\bm{x}) = \zeta(\bm{x})*S(\bm{x})\;.
\end{equation}
The filtered saliency map $S_f$ is then used to guide the computation of the octree.
A node is subdivided if the difference of the minimum and maximum values, i.e., the value range of the corresponding region of $S_f$, is greater than the subdivision threshold $t_s$.

\begin{figure}[!htb]
     \centering
    \subfloat[Spatial domain computation (without transformation) ]{\includegraphics[width=0.8\linewidth]{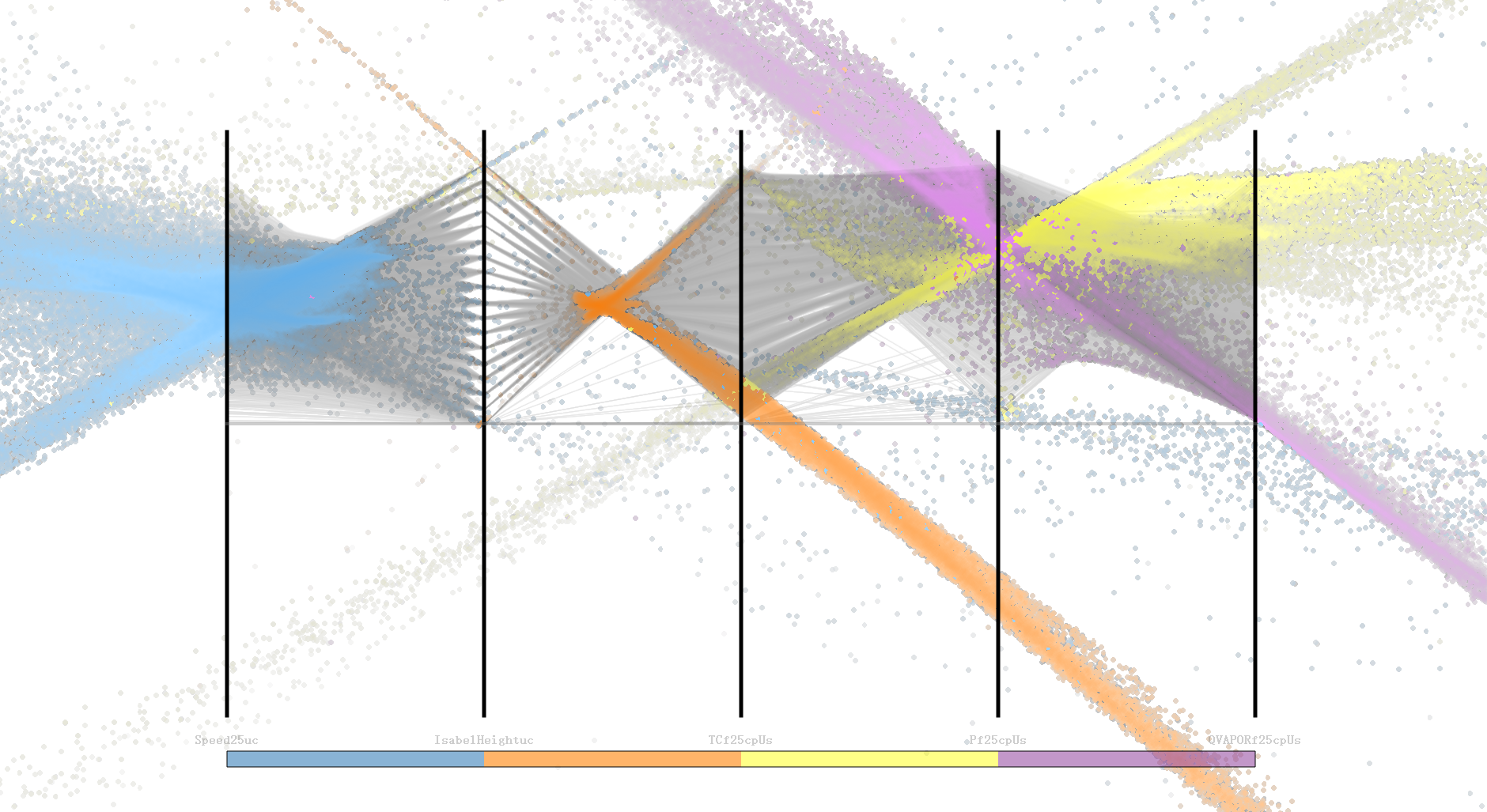}}\\
    \vspace{-1em}
     \subfloat[Spatial domain computation (with transformation)]{\includegraphics[width = 0.8\linewidth]{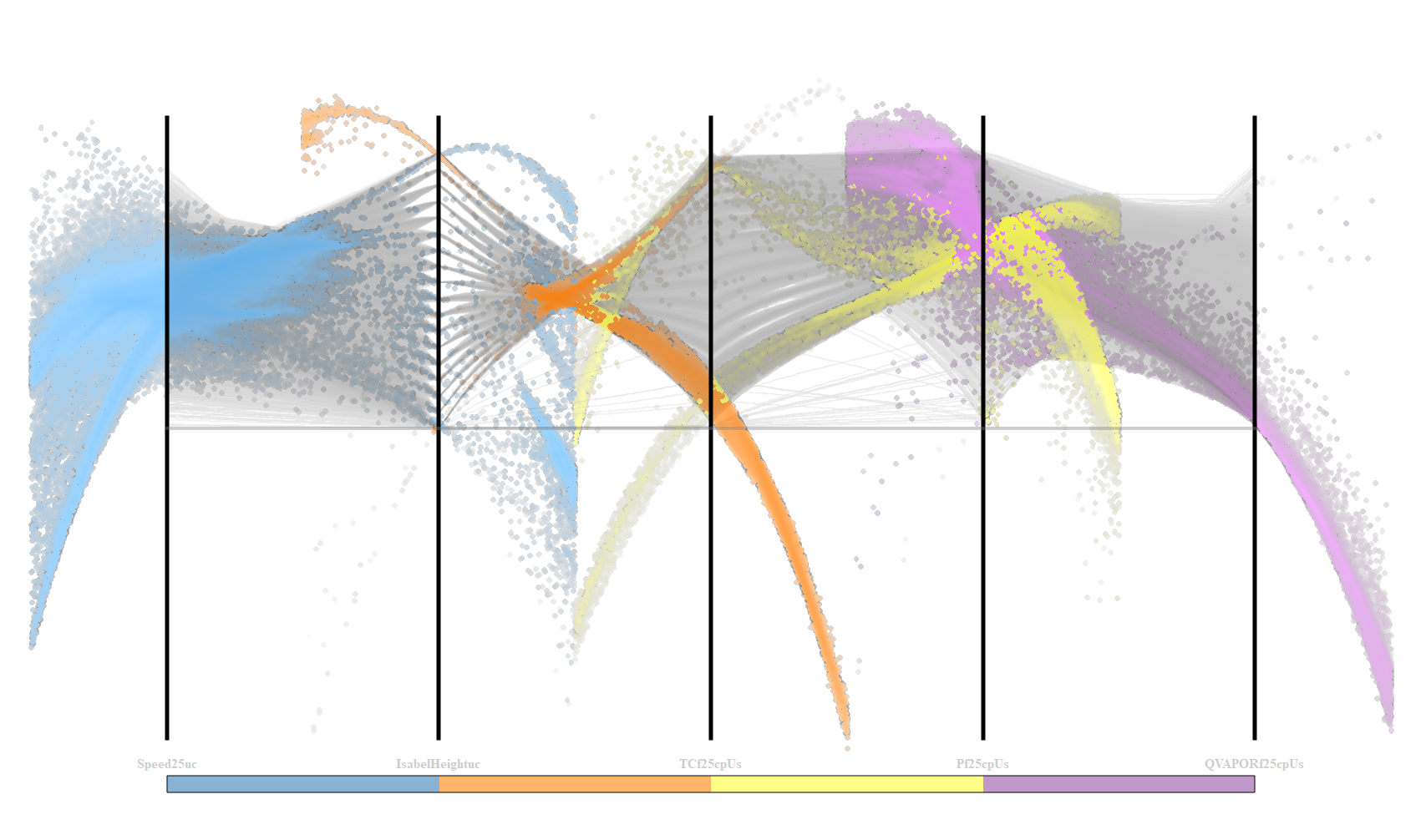}}\\\vspace{-1em}
    \subfloat[Data domain computation (with transformation) ]{\includegraphics[width=0.8\linewidth]{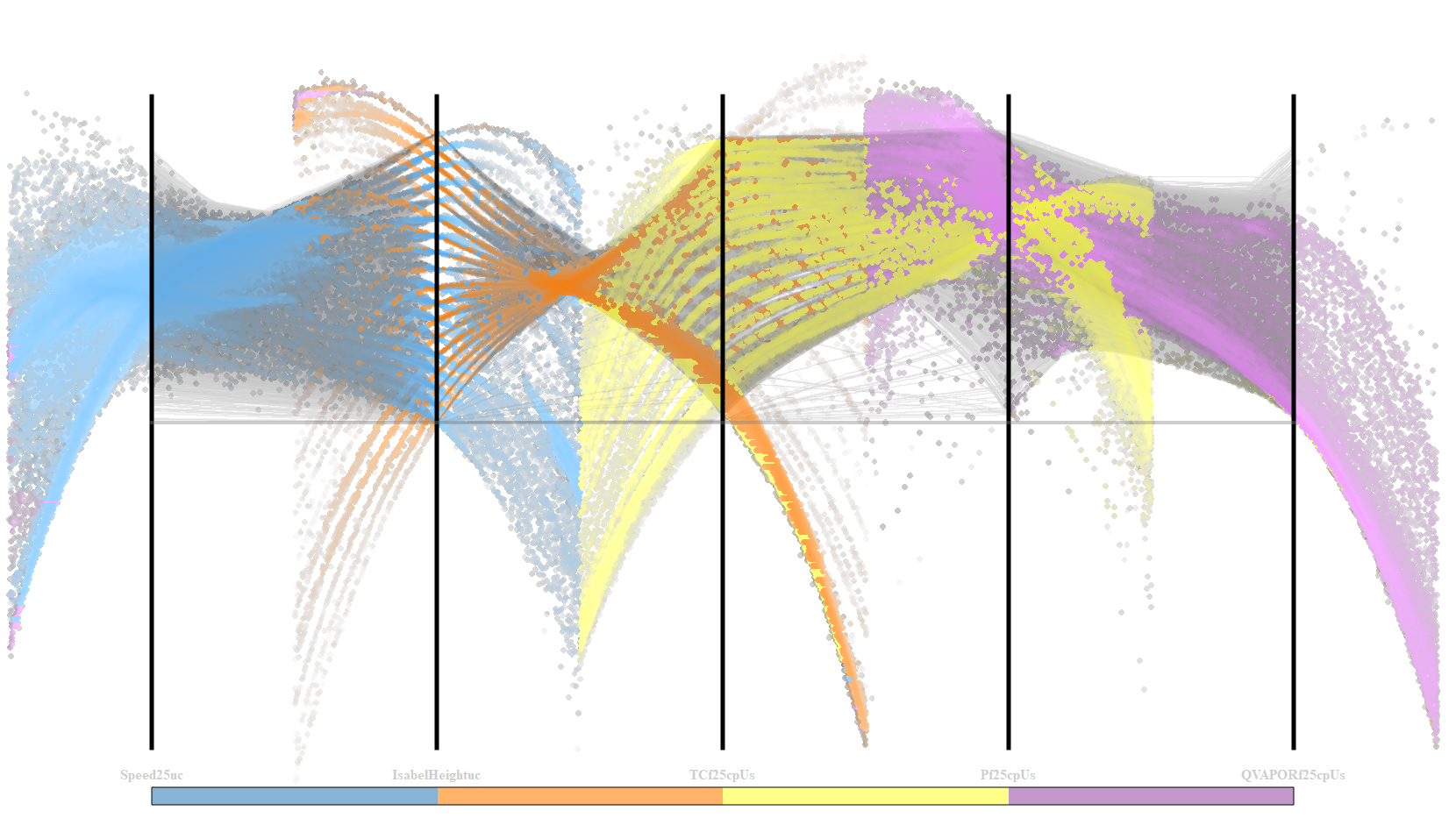}}
    \caption{1-flat indexed points computed from (a, b) the spatial domain and (c) the data domain. The angle-uniform transformation makes the originally unbounded indexed points (a) bounded in the image plane (b, c). \secRevCvm{Underlying Hurricane Isabel dataset by U.S. National Center for Atmospheric Research (NCAR)~\cite{HurricaneIsabel:2004:VisContest}.}}
    \label{fig:idxDomains}
\end{figure}

\newcommand{\contIdxHeight}{4.4cm}
\begin{figure*}[!tb]
     \centering
    \subfloat[Discrete]{\includegraphics[height=\contIdxHeight]{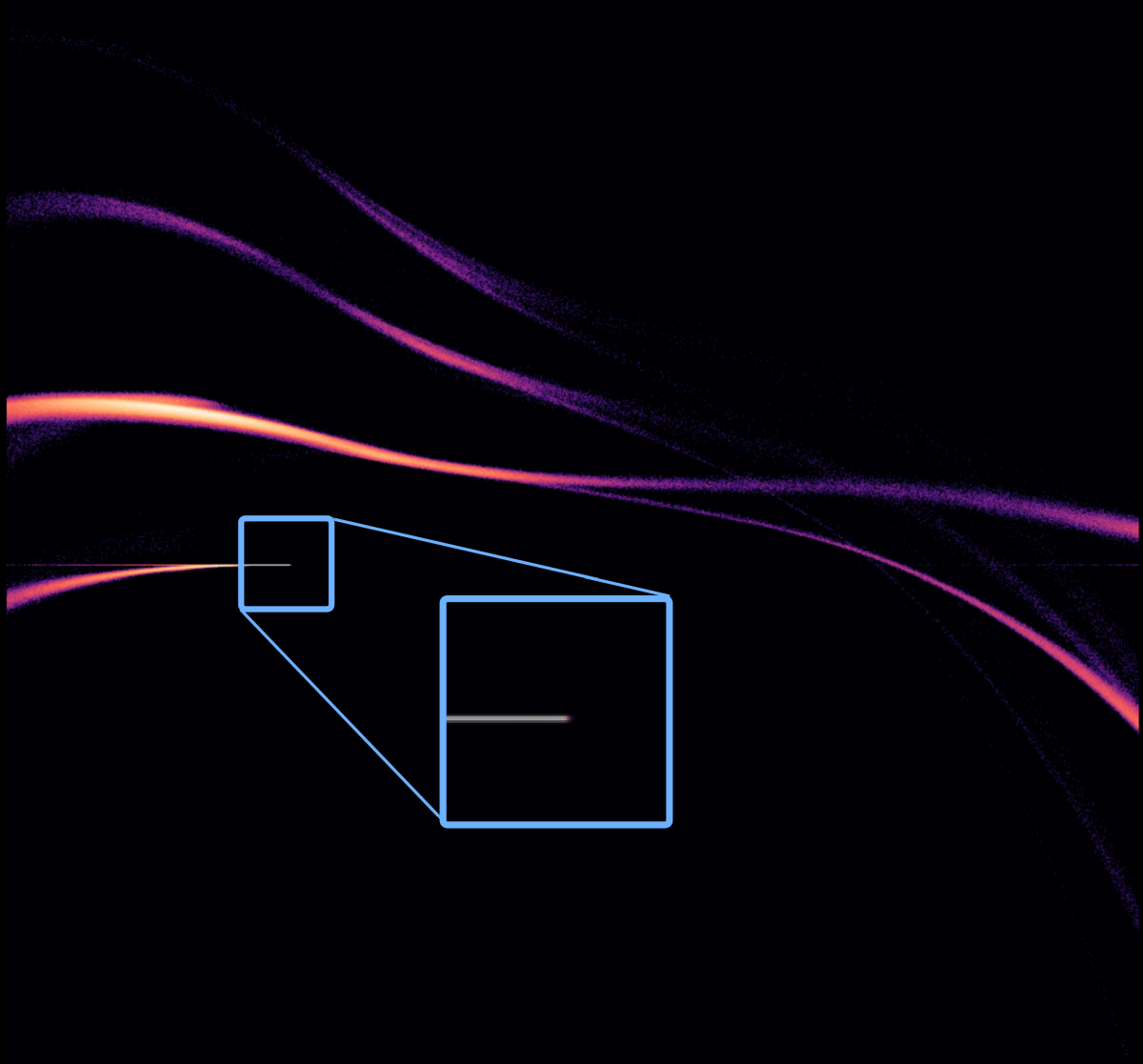}}\hfill
    \subfloat[Fixed bandwidth]{\includegraphics[height=\contIdxHeight]{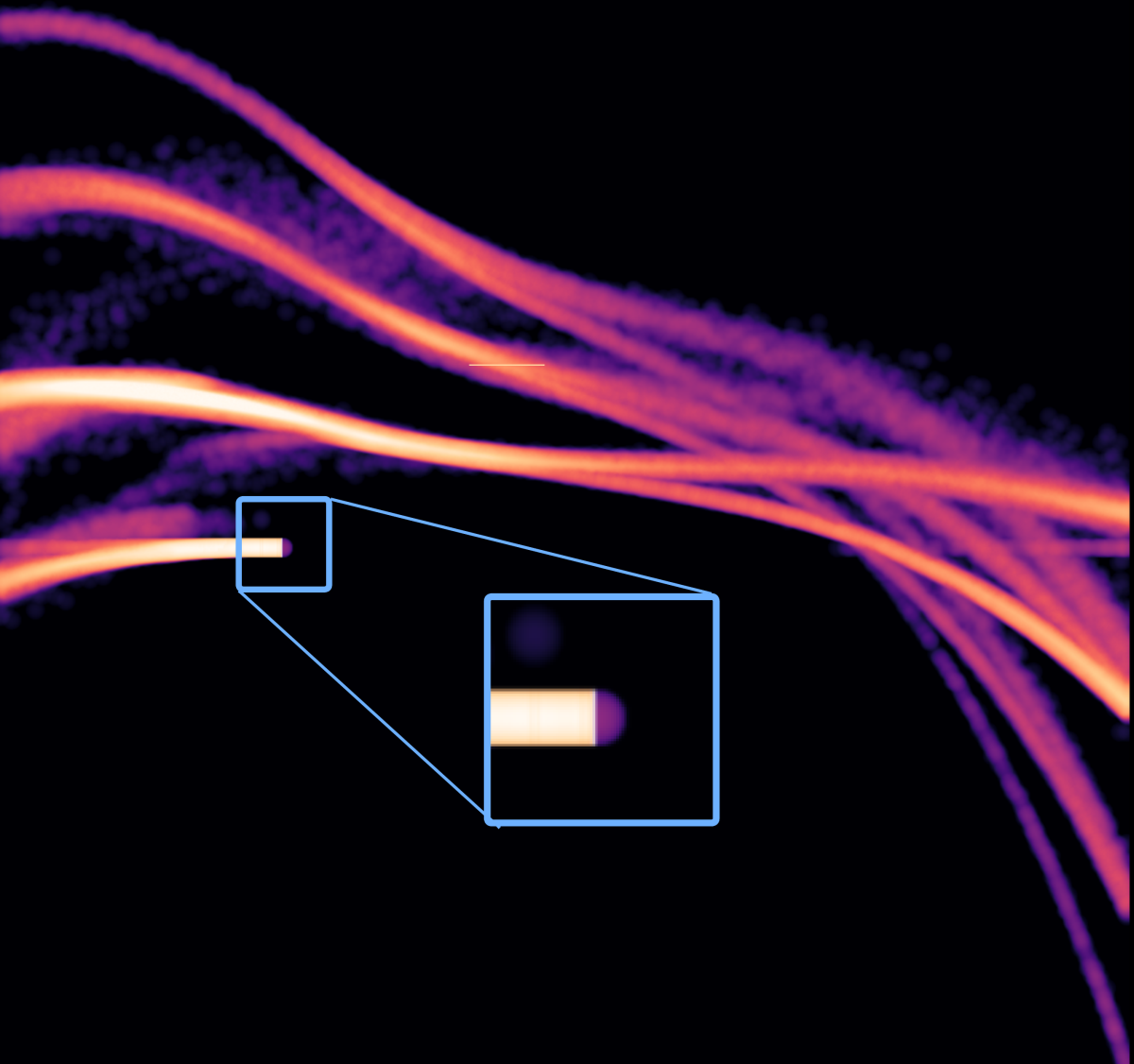}}\hfill
    \subfloat[Dynamic bandwidth]{\includegraphics[height=\contIdxHeight]{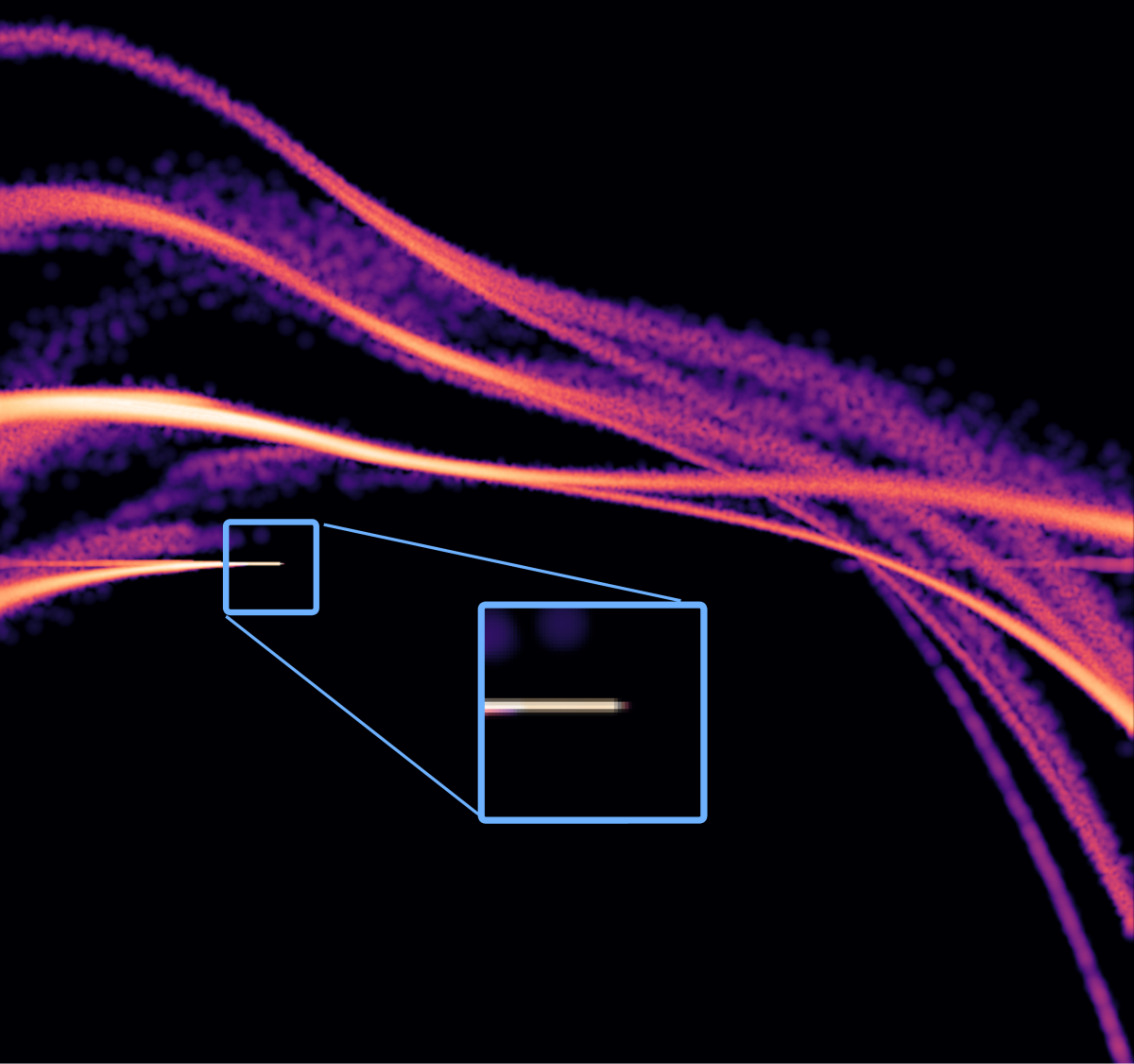}}
    \caption{Different density reconstructions of indexed points for a multivariate data of CT scan with scalar value and derived gradient magnitude (black-purple-white colormap). An ablated region of high-density is zoomed-in (blue box) in each figure. \secRevCvm{Underlying CT Tooth dataset by GE Aircraft Engines~\cite{Pfister2001}.}}
    \label{fig:kderesults}
\end{figure*}
\subsection{Local Linear Fitting}
For each leaf node in the octree, local fitting by PCA is performed for the spatial neighborhood $D$ of the central sample $\bm{x}$ of the node \revcvmmj{(Figure~\ref{fig:pipeline}-Local linear fitting)}.
\revcvmmj{
Local fitting extracts local linear features that can form global nonlinear patterns---enabling the selection of a wide range of features.
Any linear fitting technique could be adopted in our general model.
Specifically, we choose PCA as it minimizes the L2-norm of the fitted data and offers good explainability.
The spatial neighborhood constraint is a spatial filter that removes spurious correlations of data points far apart spatially, which is essential for analyzing spatial data~\cite{Ester2001} (see Appendix A.3 for an example).
}

The computation starts by Monte Carlo sampling the neighborhood of $\bm{x}$ in the continuous spatial domain to approximate the volumetric integration needed for local PCA.   
Then, we evaluate the function $f$ by performing PCA on the multivariate data values $\tau(D_c)$ of the samples in the neighborhood.
The PCA eigenvectors are sorted with descending order of component variance and denoted as $E: \left\{\myvec{e}^1, \myvec{e}^2, \ldots, \myvec{e}^m \right\}$.

For our case of 1- and 2-flat indexed points, only the information of the major and second major eigenvectors are required~\cite{Zhou:2018}. 
The data point $\bm{\xi}$ and the major eigenvector $\myvec{e}^1$ and the second major eigenvector $\myvec{e}^2$ are then the local fitting information:
\begin{equation}
f(\bm{x}) =
   \begin{cases}
  (\bm{\xi}, \myvec{e}^1)\;\;\quad\quad\text{for 1-flat}\\
    (\bm{\xi}, \myvec{e}^1,\myvec{e}^2)\;\quad\text{for 2-flat}
   \end{cases}.
\end{equation}

A comparison of visualizations of 1-flat indexed points created from the data domain and the spatial domain can be found in Figure~\ref{fig:idxDomains}.
With the spatial constraint (Figure~\ref{fig:idxDomains}~(b)), the visualization is less cluttered and indexed points appear much clearer and in fewer locations than the data domain computation  (Figure~\ref{fig:idxDomains}~(c)) as data correlations of locations that are far apart in the spatial domain are filtered out. Therefore, our new approach is the key element in facilitating visual analysis of data within its spatial context.

\subsection{Indexed Points Computation}
\label{sec:eigenToIdx}

In the next step, we reconstruct $(\bm{\xi}, \myvec{e}^1, \myvec{e}^2)$ for every sample and store this information in a new multivariate volume.
\revcvmmj{Our tests showed that the direction information $(\myvec{e}^1, \myvec{e}^2)$ is more sensitive than the point information $\bm{\xi}$ (each dimension is bounded by [0,$t_s$]) for subsequent indexed point computation within leaf nodes, and, therefore, using only the data of the central volume sample is sometimes not sufficient for accurate direction reconstruction.
Therefore, we introduce another threshold $t_e$ for nodes that are not strictly homogeneous but have value ranges smaller than $t_s$.}
If the leaf node has a variance smaller than $t_e$, the linear fit computed for the node is simply used for every corresponding sample.

For nodes with variance between $t_e$ and $t_s$, we interpolate local fitting information, i.e., eigenvector interpolation, from a coarser level to approximate that of each volume sample of the node.
Component-wise interpolation of eigenvectors~\cite{Kindlmann:2000:TVCG} is simple, yet error-prone. In contrast, logarithmic Euclidean~\cite{Arsigny2006}, quaternion-based~\cite{Collard2014}, and spectrum-sine~\cite{Wang2022} strategies create better results with fewer errors than component-wise interpolation.
We extend the spectrum-sine interpolation for its good performance and simplicity.
\revcvm{
Two parameters $d_i$ and $\theta$ are required for weight computation, where $d_i$ is associated with the distance of the point to be interpolated to its neighbors, and $\theta$ is the largest angle enclosed by any two eigenvectors:
\begin{equation}
    \theta = \max{\arccos(\myvec{e}^k_i, \myvec{e}^k_j)}, i,j = 1, 2, \ldots, 8, i\neq j, k = 1, 2\;.
\end{equation}
}
The linear fit has to be computed only once for a multivariate volume dataset, and the size of the resulting volume is not necessarily the same as the input volume although full-sized volumes are used in the paper for high-quality visualization.

Subsequently, the local fitting information $(\bm{\xi}, \myvec{e}^1)$ or $(\bm{\xi}, \myvec{e}^1, \myvec{e}^2)$ is used to compute the associated 1-flat indexed points $\bm{\eta}_{1,2}, \bm{\eta}_{2,3}, \ldots, \bm{\eta}_{m-1,m}$ and 2-flat indexed points $\bm{\eta}_{1,2,3}, \bm{\eta}_{2,3,4}, \ldots, \bm{\eta}_{m-2,m-1,m}$ as described in the indexed points parallel coordinates method~\cite{Zhou:2018} (\revcvmmj{Figure~\ref{fig:pipeline}-Continuous indexed points/Indexed points calculation}).
\revcvm{Let us consider the point $\bm{\xi}$ and the end point of the major eigenvector $\bm{\xi} + \myvec{e}^1$ and the second major eigenvector $\bm{\xi} + \myvec{e}^2$ in the data domain. 
To compute 1-flat indexed points, we project the eigenvector $\myvec{e}^1$ and point $\bm{\xi}$ to all ($m-1$) 2D subspaces spanned by adjacent attribute pairs in parallel coordinates.
For 2-flat indexed points, the normal vector $\myvec{n}$ of the 2-flat is computed as the cross product: $\myvec{n}=\myvec{e}^1 \times \myvec{e}^2$. 
Then, $\bm{\xi}$ and the end point of $\bm{\xi}+\myvec{n}$ are projected to all ($m-2$) 3D subspaces that are visualized as points on 2D image planes over parallel coordinates.}

\subsection{Angle-Uniform Transformation of Indexed Points}
Standard indexed points computed in the previous section are not bounded and are asymmetric for positive and negative correlations.
These issues forbid the visualization of all indexed points in the limited area of the image plane, and further limits the subsequent exploration in volume visualization.
Therefore, we apply the angle-uniform transformation for a bounded and symmetric representation of indexed points~\cite{ZhangZhou2023} (\revcvmmj{Figure~\ref{fig:pipeline}-Continuous indexed points/Angle-uniform transformation}).

\revcvm{We generalize the method from 1-flats to 2-flats by computing indexed points in their original infinite image planes and transform them to the bounded image plane with the same point transformation~\cite{ZhangZhou2023} and then translate them to the corresponding axes.
For an indexed point $\bm{\eta}$, we rewrite the 2D coordinates of its original image plane in homogeneous coordinates, $[c_1, c_2, c_3]$, and compute the orientation $\phi$ of the associated 2D Cartesian line using the point-line duality.}
Then, the angle-uniform transformation maps the tuple $(c_1, c_2, c_3, \phi)$ to new, bounded 2D coordinates $(u,v)$ in the image plane.
As a result, the indexed points become bounded in the parallel coordinates in an area with the width of three adjacent vertical axes, and have symmetric visual pattern for positive and negative correlations  (Figures~\ref{fig:idxDomains}~(b) and~(c)).
In this way, we have derived the mapping from a location $\bm{x}$ in the spatial domain to the indexed points $\bm{\eta}$ in the image plane of parallel coordinates. 
In the following, we describe the method for creating the density representation of continuous indexed points.

\subsection{Density Estimation of Continuous Indexed Points}

Our approach to computing Equation~\ref{eq:densitymeasuretheory} uses a forward-mapping of point samples to the image plane. 
To arrive at the density field of continuous indexed points, we rely on adapted kernel density estimation for good visualization quality (\revcvmmj{Figure~\ref{fig:pipeline}-Continuous indexed points/KDE with dynamic bandwidth}).
We adopt kernel density estimation~\cite{Shirley1995,Jensen1996,Walter1997,Hey2002} and use it to compute the density of continuous indexed points for every pixel in the image domain:
\begin{equation}
        \rho(\bm{\eta}) = \frac{1}{\pi r^2} \int_{\Xi} H(\rho(\hat{\bm{\eta}}))\rho(\hat{\bm{\eta}})d^2\hat{\bm{\eta}}\;,
\label{eqn:KDEcont}
\end{equation}
where $\rho$ is the density, $H$ is the kernel function, and $\Xi$ is a disk neighborhood around $\bm{\eta}$ with radius $r$.
Equation~\ref{eqn:KDEcont} is approximated using Monte Carlo integration:
\begin{equation}
    \rho(\bm{\eta})
    \approx \frac{1}{N\pi r^2 }\sum_{i = 0, \hat{\bm{\eta}}\in\Xi}^{N} H(\rho(\hat{\bm{\eta}}_i))\rho(\hat{\bm{\eta}}_i)\;,
\label{eqn:KDE}
\end{equation}
where $N$ is the number of samples in the neighborhood $\Xi$, and $\hat{\bm{\eta}}_i$ is the $i$-th sample.
We choose the Epanechnikov (parabolic) kernel as $H$ because of its optimality with respect to mean square error~\cite{Epanechnikov1969,Wand1994}, making it superior to other commonly used choices of $H$ including the Gaussian, exponential, triangle, cosine, and box kernels.
Visually, the Epanechnikov has a smooth fall-off with a bounded support---striking a good balance between the smoothness and sharpness.

We accelerate the computation of Equation~\ref{eqn:KDE} by separating the location information of indexed points and the sample density to simplify the computation as many indexed points fall into same locations. 
An image of sample density is calculated by accumulating indexed point samples in the image domain.
A kd-tree ($k=2$) is built to accelerate the query of samples in the nearest neighborhood $\Xi$.
Only samples with unique locations in the image domain are considered in the kd-tree computation, which significantly reduces the amount of samples.

Using a fixed bandwidth for $H$, kernel density estimation tends to ablate regions of high densities and overly smooth results.
Figure~\ref{fig:kderesults} illustrates this problem. The fixed bandwidth approach (Figure~\ref{fig:kderesults}~(b)) thickens the high density feature (in the blue box) compared to the original discrete indexed points shown in Figure~\ref{fig:kderesults}~(a).   
To address this issue, we propose a dynamic adaptation of bandwidth.
The bandwidth is inversely related to the sample density $\rho(\bm{\eta})$, i.e., a large neighborhood is searched for low density regions and vice-versa. 
A maximum radius parameter and a maximum accumulated density of the neighborhood are used in the kernel density estimation to control the neighborhood size.
The result can be seen in Figure~\ref{fig:kderesults}~(c). 
Note how our method preserves the thin high density region on the bottom left ``branch'' (highlighted with the blue box) of the image while retaining the smooth look of low density regions as with the fixed bandwidth approach.

The combination of the forward mapping of point samples from the spatial domain to the image space (similar to Heinrich et al.~\cite{Heinrich:2011:splatting}) and the subsequent kernel density estimation take care of all density effects that come with the non-linear transformations. 
We perform scattering of correctly sampled points from the spatial domain to the angle-uniform plane of deformed parallel coordinates. 
Since points stay points in this process, we can guarantee conservation of virtual ``mass,'' and the final density estimation with the Epanechnikov kernel (or similar alternatives) does not affect the integrated densities in local neighborhoods.

\begin{figure*}[tb]
    \centering
    \subfloat[Phong shading]{\includegraphics[width=0.28\linewidth]{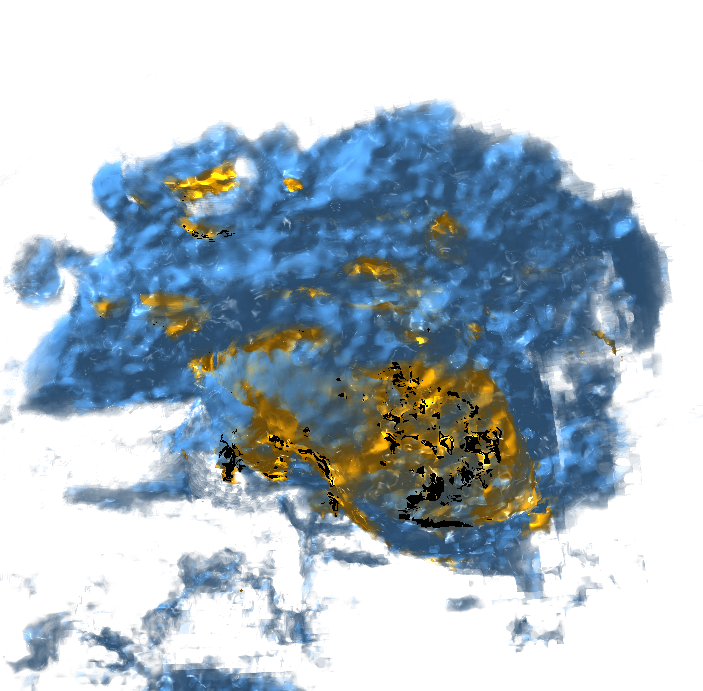}}\hfill
    \subfloat[Directional occlusion shading]{\includegraphics[width=0.28\linewidth, trim={0.2cm 0 0 0},clip]{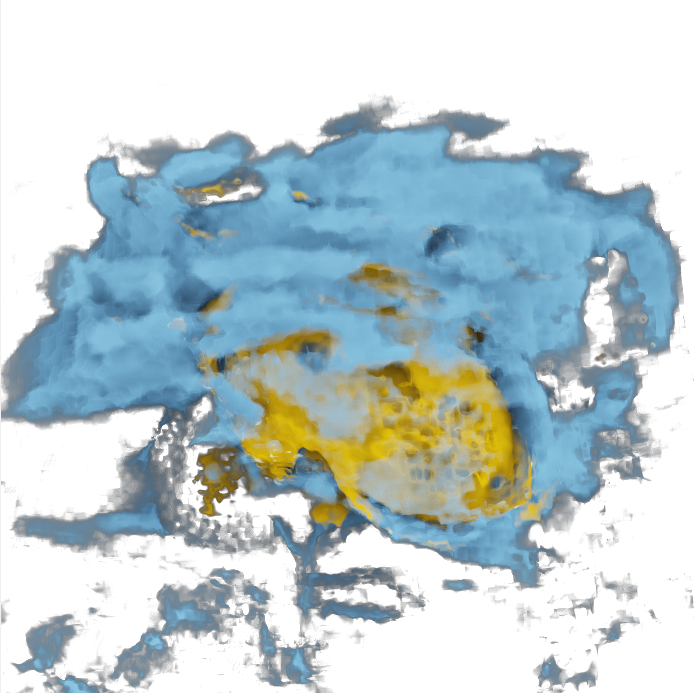}}\hfill
         \subfloat[Extinction optimized]{\includegraphics[width=0.28\linewidth, trim={0 0 0.2cm 0}, clip]{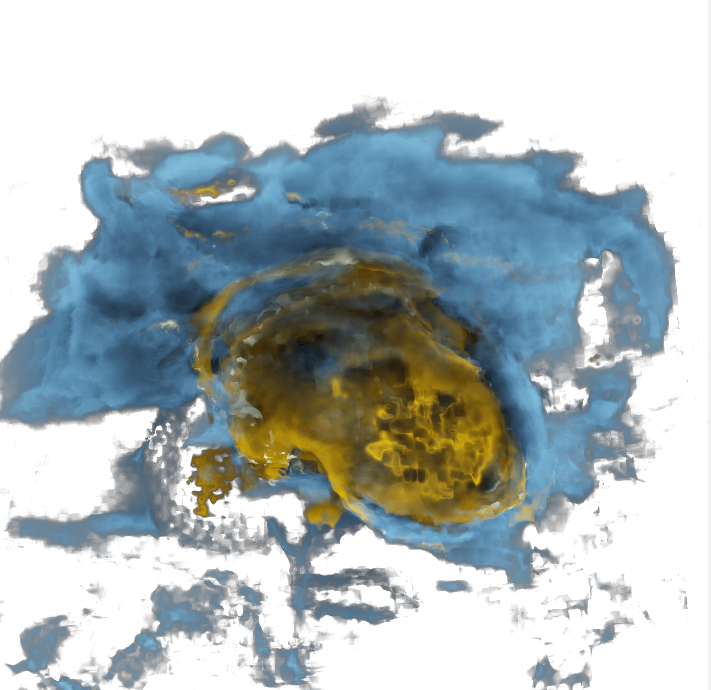}}
    
    \caption{A comparison of different shading models for volume rendering of a tumor (yellow) and the edema (blue) in a multimodal magnetic resonance imaging (MRI) dataset. \secRevCvm{Underlying MRI dataset by Menze et al.~\cite{Menze2015}.}  }

    \label{fig:renderingCompare}
\end{figure*}

Another aspect concerns the densities of the curved lines in angle-uniform parallel coordinates. The typical density model for parallel-coordinate lines is based on ``counting'' lines along the $y$-direction \cite{Heinrich:VIS:2009}. The main deformations in angle-uniform parallel coordinates occur in the $x$-direction, thus not affecting density. The minor deformations in the $y$-directions are considered in our implementation by  using a forward mapping from the spatial domain to image space.

\section{Rendering} 
\label{sec:idxVis}

In this section, we detail the visualization of continuous indexed points and the rendering of multivariate volumes (\revcvmmj{Figure~\ref{fig:pipeline}-Rendering and Interaction}).
Continuous indexed points are drawn in the same image plane of parallel coordinates so that no context switching is needed.
Multivariate volumes are visualized using multivariate transfer functions for local linear information with directional occlusion shading for good spatial perception.

\subsection{Rendering Continuous Indexed Points}
For each subspace, continuous indexed points are stored in an image buffer of high dynamic range, and we convert them to low dynamic range for visualization in parallel coordinates.
For each pixel in the buffer, the logarithm of its value, i.e., the density $\rho(\bm{\eta})$, is taken and normalized using the minimum and maximum value of this buffer.
The normalized value is then color-mapped using a discrete categorical colormap~\cite{Brewer:AMS:1999} based on the subspace it belongs to.
The colormap is shown below the label of parallel axes.
The user can change the contrast of the visualization with a slider controlling the gamma mapping.

Then, the indexed point layer is alpha-blended on top of the original parallel coordinates plot.
The blending weight can be easily and flexibly changed by the user. 
Examples of visualizations of continuous indexed points can be seen in Figures~\ref{fig:teaser},~\ref{fig:ui} to~\ref{fig:dti2flats}, and~\ref{fig:synth1flatsCompare} to~\ref{fig:toothCompare} (in the Appendix). 

\subsection{Multivariate Indexed Points Transfer Function}
The association between continuous indexed points and the spatial domain is established in volume rendering using multivariate transfer functions.
During volume rendering, the value is sampled at spatial location $\bm{x}$, and then transformed to indexed point locations in the image space to lookup the transfer function.
Ideally, this transformation takes the local fitting information $(\bm{\xi}, \mathbf{X})$ at $\bm{x}$ and maps it to the indexed point location $\bm{\eta}$ for flexible parallel coordinates manipulations, for example, changing the axis order.
However, this is costly as it has to be computed for every fragment of the render buffer during volume rendering.
In practice, we precompute indexed points for a given order of parallel coordinates axes and store them in volumes such that the retrieved value at $\bm{x}$ readily gives $\bm{\eta}$.

\subsection{Directional Occlusion Shading}
The widely-used Phong illumination model relies on the gradient of a volume. However, the gradient is not defined for multivariate volumes with the exception that attributes are derived from a primary scalar field whose gradient is used for shading. Therefore, the gradient of a random volume attribute is erroneous in the general case and, instead, gradient-free shading methods should be used.

\begin{figure*}[htb]
    \centering
    \includegraphics[width=0.9\linewidth]{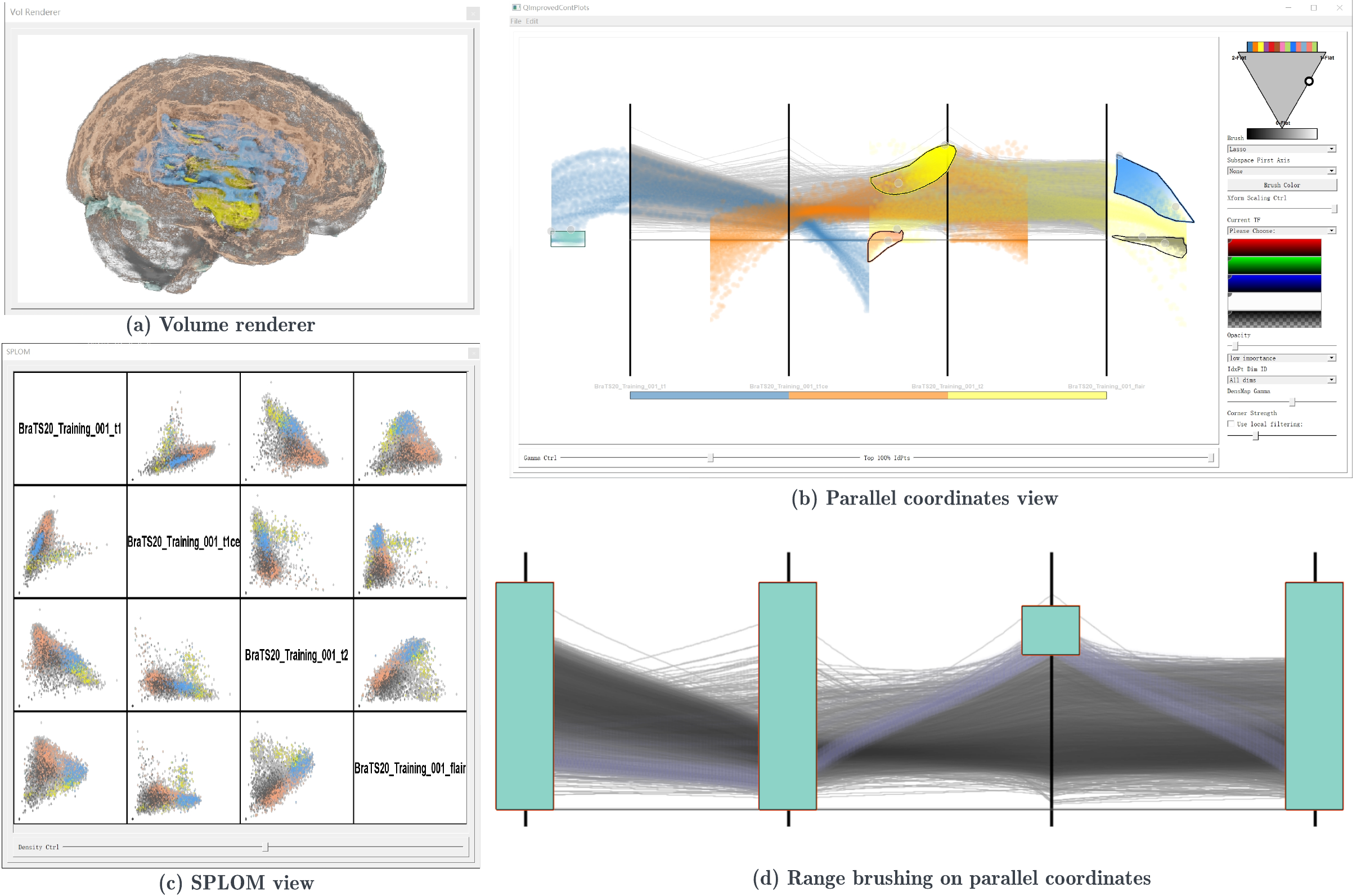}
    \caption{The user interface of our visualization tool with an example of a brain MRI dataset~\cite{Menze2015,Bakas2017}. The volume rendering view (a) shows the classification by transfer functions for continuous indexed points in (b) the parallel coordinates view. The SPLOM view (c) provides a reference of data features with colored data points through brushing-and-linking.  
    Range brushing on parallel axes is also supported (d).\secRevCvm{Underlying MRI dataset by Menze et al.~\cite{Menze2015}.}
}
    \label{fig:ui}
\end{figure*}
Shading models that create occlusion effects or shadows~\cite{Schott:2009:Eurovis,Campagnolo2019,Ament:VIS:2013} are a feasible choice.
Such techniques can generate plausible depth cues that are important for spatial comprehension of the data.
We use directional occlusion shading~\cite{Schott:2009:Eurovis,Campagnolo2019} as it can provide high quality occlusion effects and can be efficiently accelerated with the GPU.
To highlight interior structures that may be occluded by others, we optionally calculate the optimized extinction using a closed-form solution~\cite{Ament2017} with the importance function of transfer function, where larger opacity indicates higher importance.

Figure~\ref{fig:renderingCompare} compares volume renderings of the tumor region of a multimodal brain MRI scan from the BraTs database~\cite{Menze2015,Bakas2017} classified using the multivariate indexed points transfer functions.
Phong illumination (Figure~\ref{fig:renderingCompare}~(a)) gives erroneously shaded regions as the gradient is chosen from one attribute and does not match the surface of the classified structure, and it also provides little depth cues.
In contrast, the directional occlusion shading model (Figure~\ref{fig:renderingCompare}~(b)) yields a more appropriate visualization than the other models.
The rendering shows occlusions revealing fine details of the structures and providing vital depth cues for correct spatial perception. 
\rev{Figure~\ref{fig:renderingCompare}~(c) shows that the extinction optimized mode reveals internal structures better than the standard mode (Figure~\ref{fig:renderingCompare}~(b)). Conceptually, the extinction optimized mode simplifies the task of designing the opacity of transfer functions to setting low or high importance of transfer functions. 
}

\section{User Interaction and Implementation}
\label{sec:interactions}
User interactions and the implementation of our method \revcvmmj{(Figure~\ref{fig:pipeline}-Rendering and Interaction)} are discussed in this section.
The user interface of our tool is shown in Figure~\ref{fig:ui}.

\subsection{Transfer Function Widgets}

Transfer functions of indexed points are set using rectangular or free-hand lasso brushes in the 2D image plane of parallel coordinates, as shown in Figure~\ref{fig:ui}~(b). 
Colors and opacity can be manipulated using a color gradient with a circular fall-off within the brushes.
\rev{The setting of the opacity function is done with a slider or a drop-down box.}

Because of the occlusion of indexed points from neighboring subspaces, the user may want to focus on a specific subspace.
The subspace ID of a transfer function widget can be set by the user with a drop-down option.
Axis brushes that allow the user to select combinations of the value range of each original volume variable are also supported (Figure~\ref{fig:ui}~(d)).

Details of the underlying mechanism of the transfer function lookup can be found in Appendix~A.1.

\subsection{Brushing-and-Linking}
In addition, the analysis of indexed points is supported by data brushing and linking between the original parallel coordinates and continuous indexed points in the parallel coordinates view (Figure~\ref{fig:ui}~(b)), and the SPLOM view (Figure~\ref{fig:ui}~(c)).
These linked views complement each other and help the user understand the underlying data.

Balanced kd-trees are built to accelerate queries of samples. 
A balanced kd-tree is set up for each layer of parallel coordinates: one for parallel coordinates lines for the original data values, and one for continuous indexed points.
The dimensionality of the data domain, i.e., $k=m$, is used for the computation of kd-tree for the original data.
For the indexed points, we use downsampled indexed points data from the volume and construct 2D trees as they live in the 2D image plane of parallel coordinates.
The data value, 2D position, and subspace ID of the indexed point and its percentile and strength are stored in a kd-tree node. 
Thanks to kd-trees, brushing-and-linking between indexed points, parallel coordinates, and SPLOM is typically interactive or takes the most a few seconds for a large number of samples.

\subsection{Implementation}
Our method was implemented using C++ and Matlab, and tested on a machine with 3.5\,GHz Intel i7 CPU, 32\,GB main memory, and an NVIDIA Quadro P6000 graphics card. 
\revcvm{Most of the data processing stages---the construction of the octree, local linear fitting, and density estimation of indexed points---were implemented in Matlab as precomputation.
For local linear fitting, a neighborhood size of $7\times7\times7$ voxels and 100 Monte Carlo samples are used as this configuration balances the quality of results and computational time after we tested various settings.
The acceleration of octrees is data-dependent, and a reduction of 38\% to 73\% of the preprocessing time was achieved in our experiments with simulation data and MRI scans. However, for noisy CT data, octrees are not advantageous and per-voxel computation is recommended.} 
Preprocessed data items computed in Matlab are loaded into our visualization tool for interactive visualization and exploration. 

The visualization tool is based on C++, Qt, and OpenGL. 
Directional occlusion rendering of multivariate volumes extends the cone sampling technique~\cite{Campagnolo2019}. 
In a typical ray casting framework, the method separates rendering from the creation of extinction coefficients.
The volume of extinction coefficients is updated only when transfer functions are changed. 
During rendering, interactivity is achieved with efficient GPU acceleration using compute shaders that realize multivariate transfer functions and cone sampling. 
We use compute shaders as they are the fastest implementation for GPU-based ray casting according to a performance study~\cite{Sans2017} and our own experiments comparing against an implementation using vertex and fragment shaders.

For the Brain MRI dataset with the size of $4\times240\times240\times155$, an average of 14 frames per second (fps) was achieved with a viewport of $600\times600$ for volume rendering with the directional occlusion shading at a sampling rate of 0.3, which provides sufficient quality for exploration.
For better rendering quality, 11 and 4 fps are achieved for sampling rates of 0.5 and 1.0, respectively.
The average numbers for the rendering with extinction optimization are 8, 4, and 1 fps, respectively, for the same sampling rates. 
Updates of brushing-and-linking took 4 seconds on average.
A similar performance was achieved for the Hurricane Isabel dataset~\cite{HurricaneIsabel:2004:VisContest} (size $5\times500\times500\times100$).
More details on the computational performance can be found in Appendix~A.2.

\section{Example Results and Case Study}
Several multivariate volume datasets are used to demonstrate the effectiveness of our method. First, we use synthetic datasets to illustrate the interpretation of patterns in continuous indexed points, 
which is an important step for subsequent visualization and analysis of volumetric data. 
Afterward, we show the usefulness of our method using real-world multivariate volumetric datasets.

\subsection{Interpretation of Indexed Points Patterns}
\label{sec:patternInterp}

With synthetic datasets (Figure~\ref{fig:pointLine}~(c, d)),
we show continuous indexed point patterns of data features with various correlations.
Please refer to Appendix~A.3 for another more complicated example.
\revcvmmj{
Thanks to the local linear fitting,  both local linear and global nonlinear features can be extracted with our method.
Therefore, a variety of features can be visualized with patterns of indexed points.
Spatial filtering removes false positives of correlations that are not within spatial neighborhoods.
}
\secRevCvm{
The horizontal coordinate of 1-flat indexed points encodes the angular information of local linear features in the corresponding 2D subspace~\cite{ZhangZhou2023}, while the vertical coordinate maps the intercept of the fitted features which indicates the value offsets in Cartesian coordinates (the supplemental material,~\cite{Zhou:2018}).
Therefore, nonlinear shapes of 1-flat indexed point patterns can be associated with features in Cartesian coordinates: horizontal dispersion of the 1-flat indexed point patterns indicates correlation changes but data values remain similar; vertical dispersion indicates fixed correlations with value changes; elongated 1-flat indexed points patterns map to smoothly changing arches in Cartesian coordinates; smeared round patterns correspond to isotropic Cartesian features. 
}

\begin{figure*}[htb]
    \centering
    \subfloat[Volume rendering (1-flat)]{\includegraphics[height = 3.3cm]{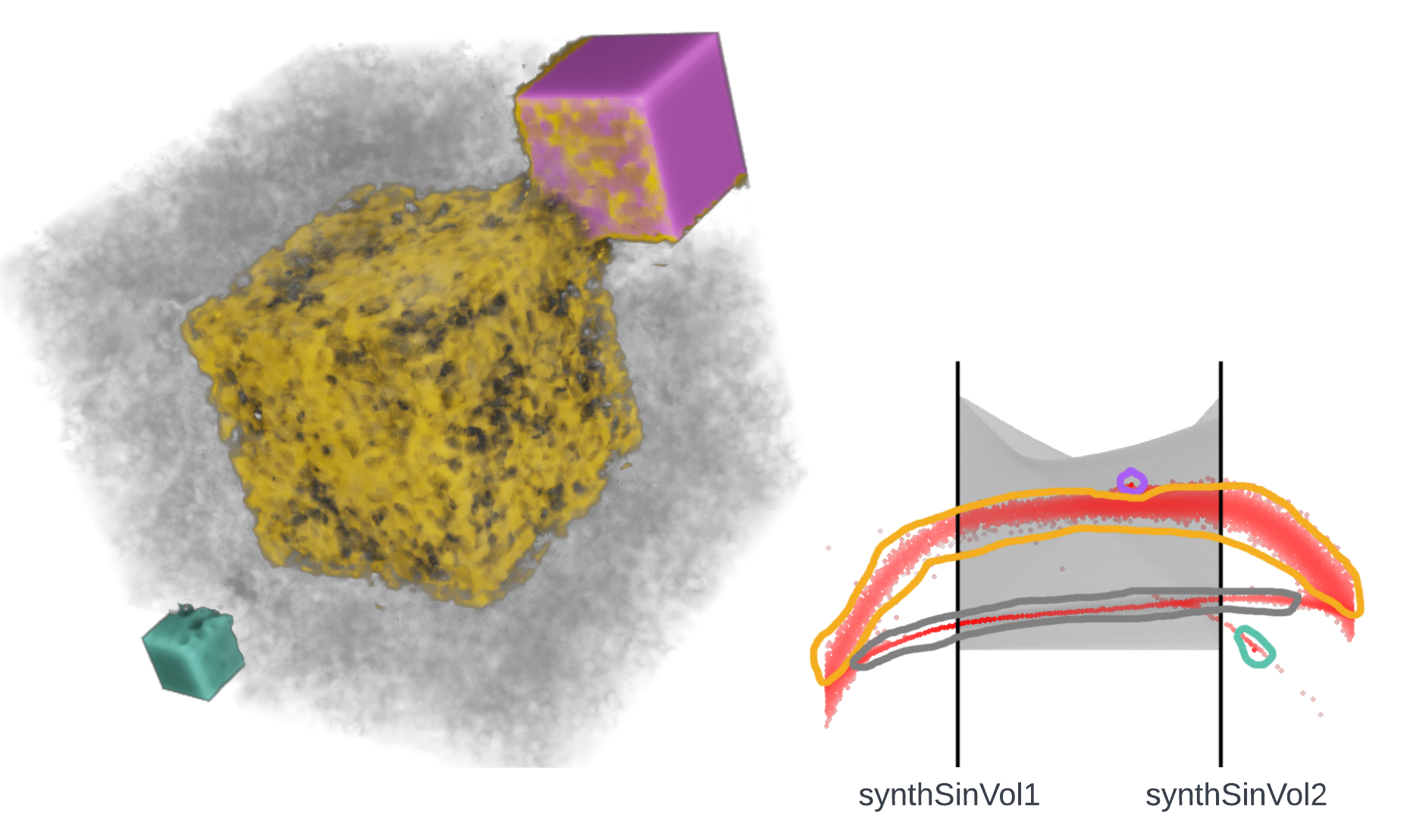}} 
    \subfloat[Volume rendering (2-flat)]{\includegraphics[height = 3.3cm]{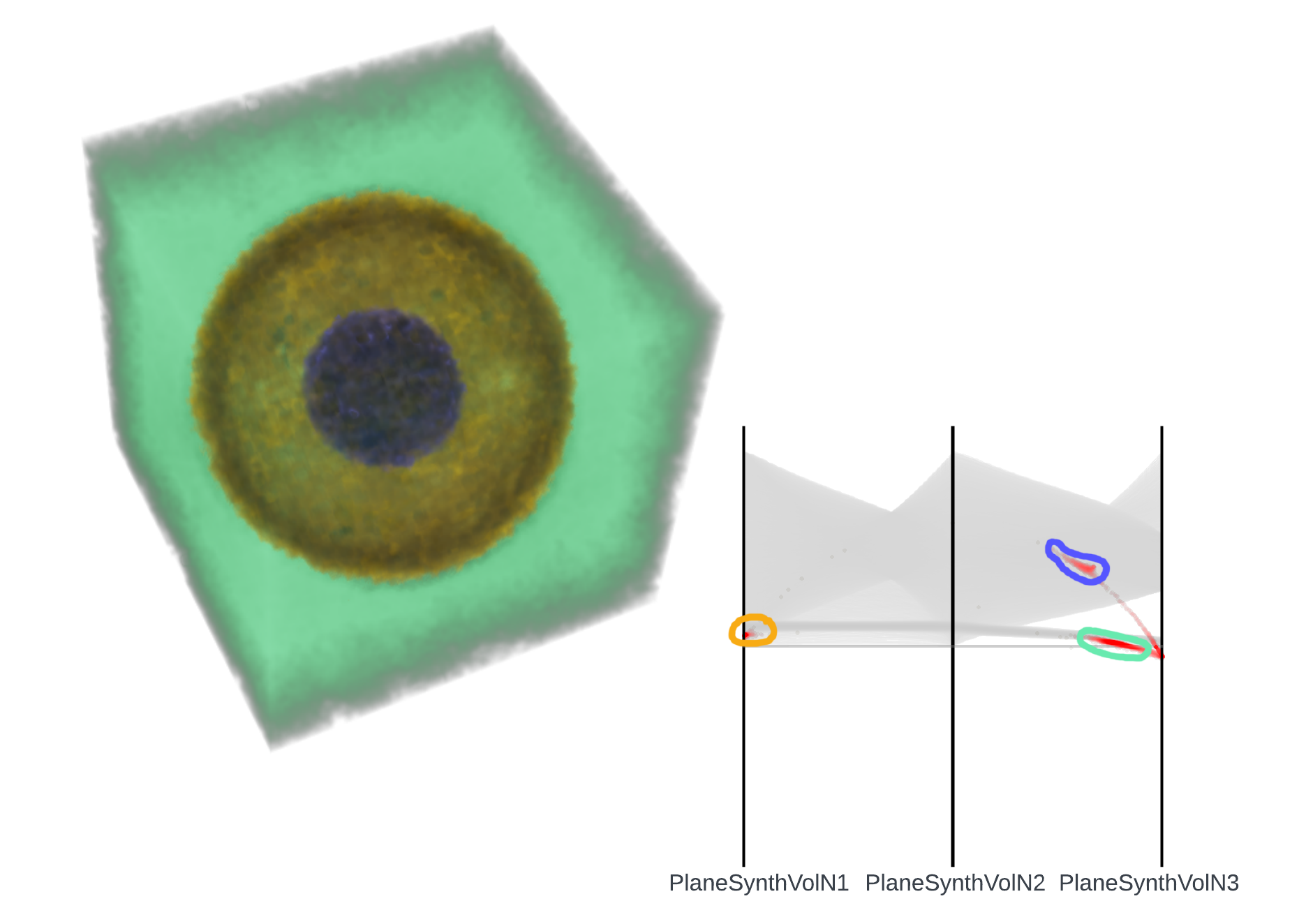}} \hfill
    \subfloat[Domain comparison (1-flat)]{\includegraphics[height = 3.3cm]{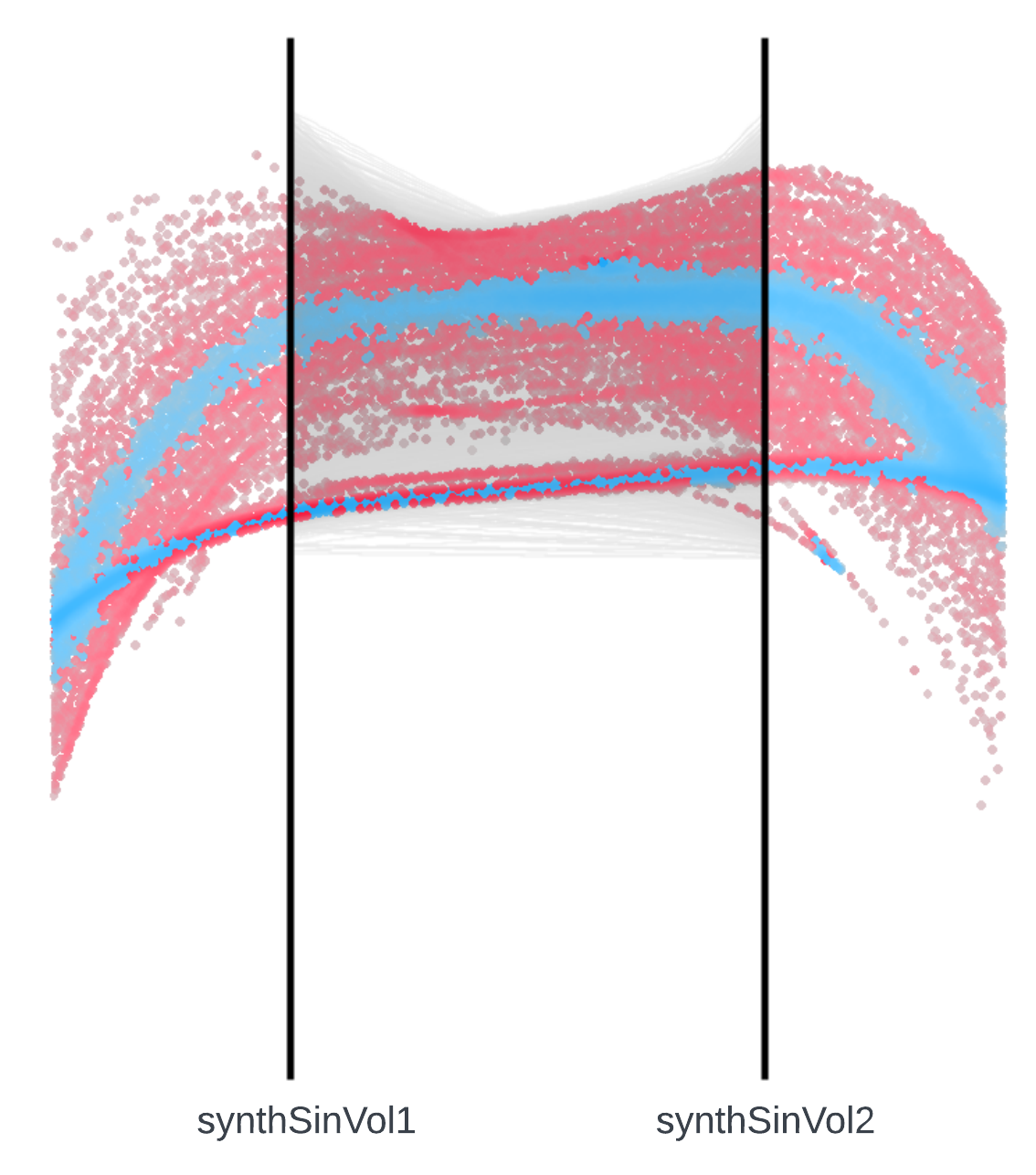}}   \hfill
    \subfloat[Domain comparison (2-flat)]{\includegraphics[height = 3.3cm]{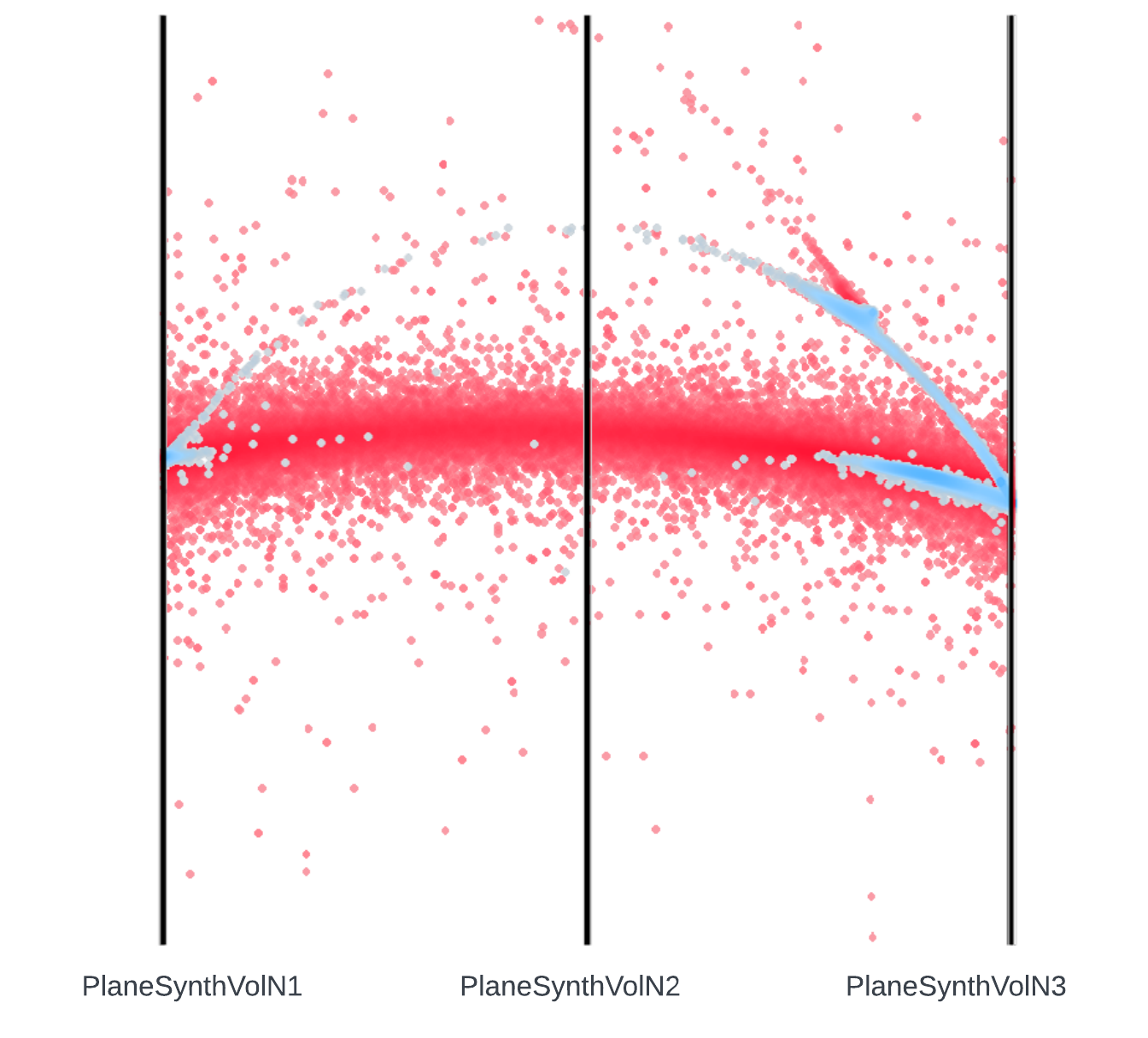}}  
    \caption{Interpretation of indexed points patterns: synthetic examples with (a) 1-flat and (b) 2-flat indexed points computed with spatial and data domains (c, d). 
    Shared regions by both methods are in blue and unique regions by the data domain method are in red.
    }
    \label{fig:synDomCompare}
\end{figure*}

Patterns of medium and high densities are shown by indexed points, as in Figures~\ref{fig:pointLine}~(c, d) and~\ref{fig:synDomCompare}~(a).
For 1-flat indexed points, a high-density arch that spans the entire horizontal axis of the two attribute axes is shown in orange in Figure~\ref{fig:pointLine}~(c).
Strong negatively correlated patterns are shown between the axes (Figures~\ref{fig:pointLine}~(c, purple)).
High-density positively correlated patterns can be seen outside of the axes (Figures~\ref{fig:pointLine}~(c, green)).

For 2-flat indexed points (Figures~\ref{fig:pointLine}~(d),~\ref{fig:synDomCompare}~(b)), two high-density smeared point patterns are associated with different planes (Figure~\ref{fig:pointLine}~(d, orange and blue), and high-density arches correspond to nonlinear structures (Figure~\ref{fig:pointLine}~(d, green)). 
We are able to classify each feature in the volume, as seen in the volume renderings in Figure~\ref{fig:synDomCompare}~(a, b).
The classification result is also seen in the scatterplots in Figures~\ref{fig:pointLine}~(c, d) through brushing-and-linking. 
Here, the overlapping colored regions indicate that our method successfully classifies features of similar data values using directional information (normal vectors of planes for 2-flat) of local fittings.

Another aspect of indexed point patterns is the domain of computation of indexed points.  
The comparison of the spatial domain indexed points computation with the data domain computation is already shown in Figure~\ref{fig:idxDomains}.
We further demonstrate the difference of the two versions using explicit encoding as shown in Figure~\ref{fig:teaser}~(k) and Figure~\ref{fig:synDomCompare}~(c, d).
\revcvmmj{The visual signatures of indexed points for the two variants of the synthetic data are compared in Figure~\ref{fig:synDomCompare}~(c, d), corresponding to the examples shown in Figure~\ref{fig:synDomCompare}~(a, b), respectively.}
Figure~\ref{fig:synDomCompare}~(c, d) show the visualizations where blue regions are shared by both methods and red regions are from the data domain computation only. 
Further examples comparing volume samples from the data domain computation against the spatial computation can be found in Appendix~A.3.
These indicate that the spatial correlation computation is more reliable than the data domain computation.

\subsection{Multivariate Volume Examples}
We demonstrate the usefulness of our method using representative real-world datasets. These include multivariate data from simulation and multimodal medical imaging data.

\begin{figure*}[tb]
    \centering
    \subfloat[Volume rendering (1-flat)]{\includegraphics[height = 3.6cm]{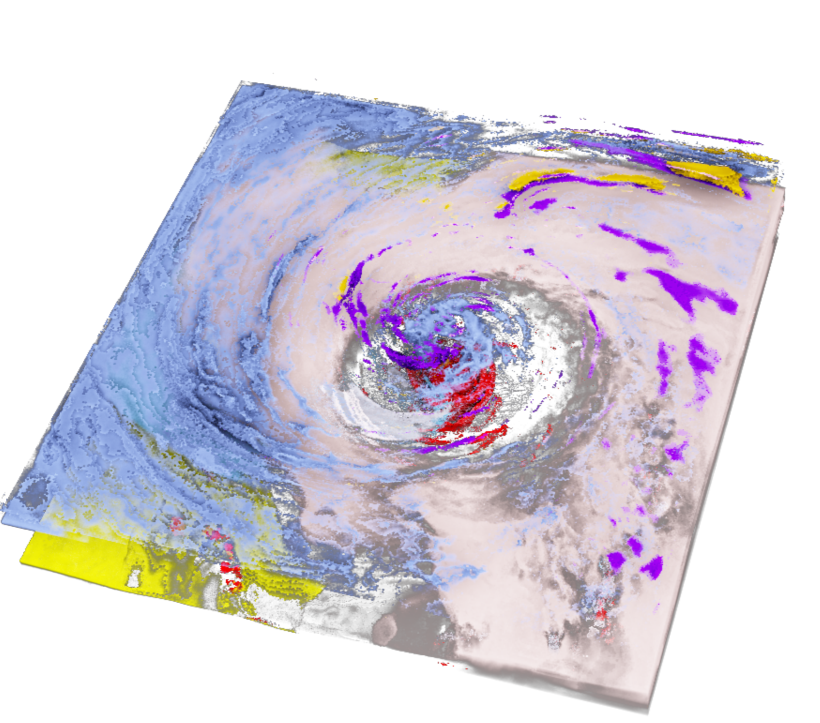}} 
    \subfloat[1-flat continuous indexed points]{\includegraphics[height = 3.8cm]{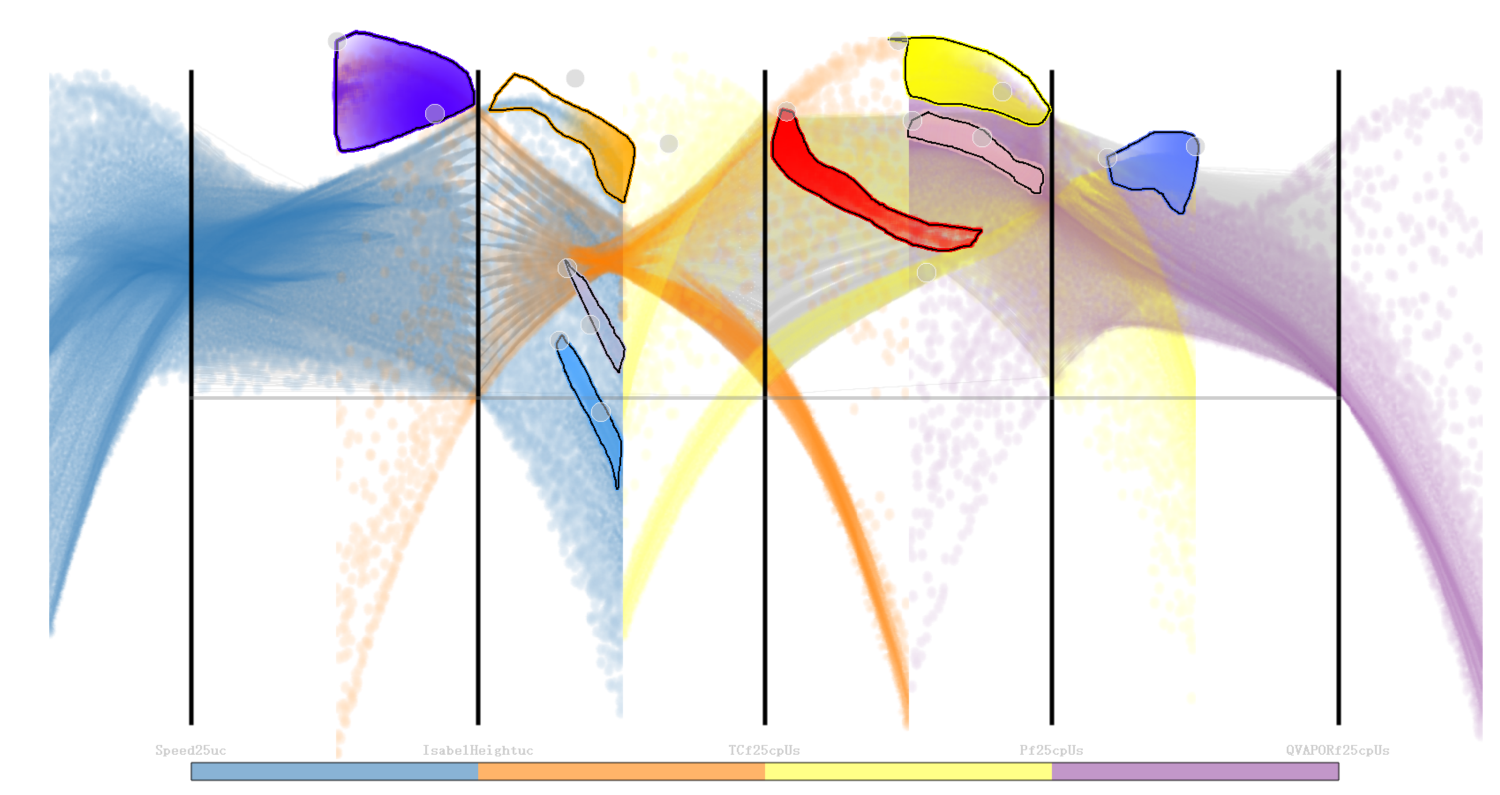}} 
    \subfloat[Brushed SPLOM]{\includegraphics[height = 3.3cm]{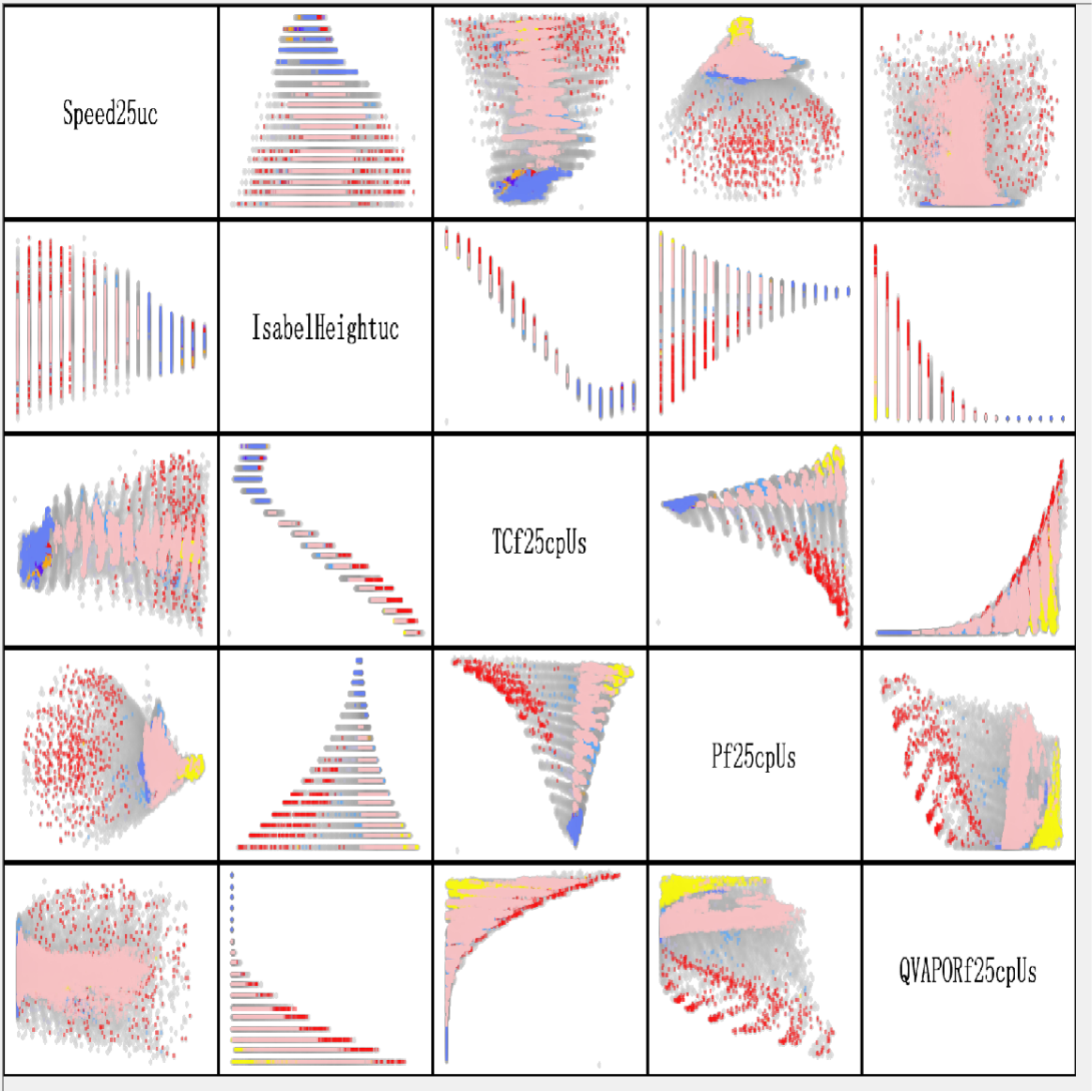}}  \\ 
    \subfloat[Volume rendering (2-flat)]{\includegraphics[height = 3.6cm]{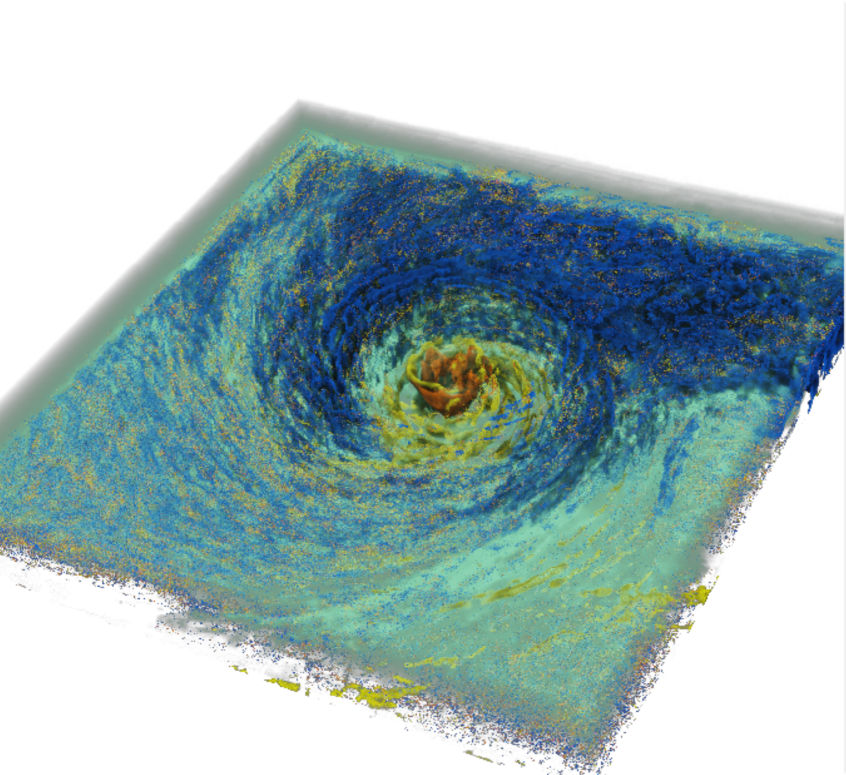}} 
    \subfloat[2-flat continuous indexed points]{\includegraphics[height = 3.9cm]{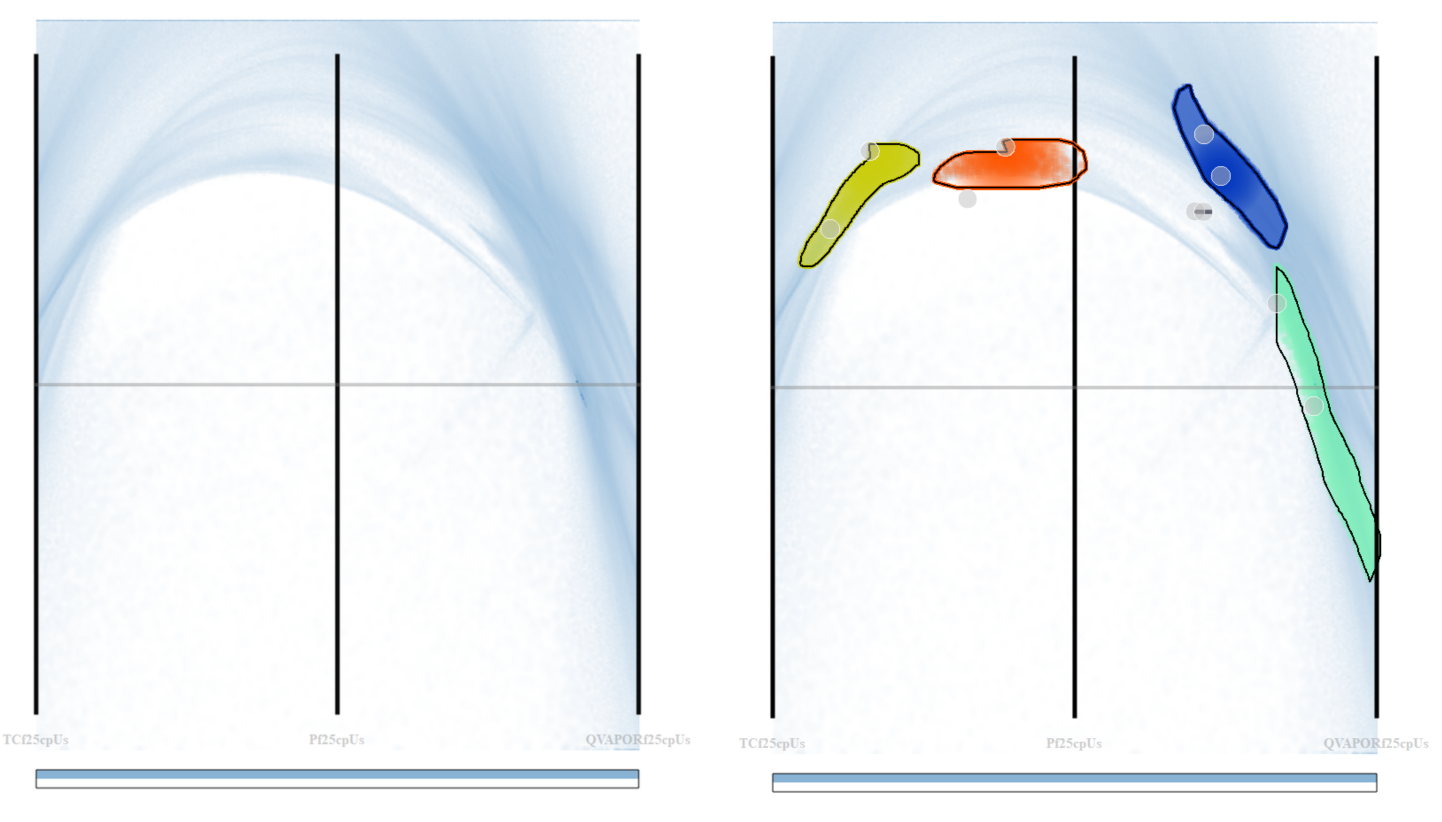}} 
    \subfloat[Brushed SPLOM]{\includegraphics[height = 3.3cm]{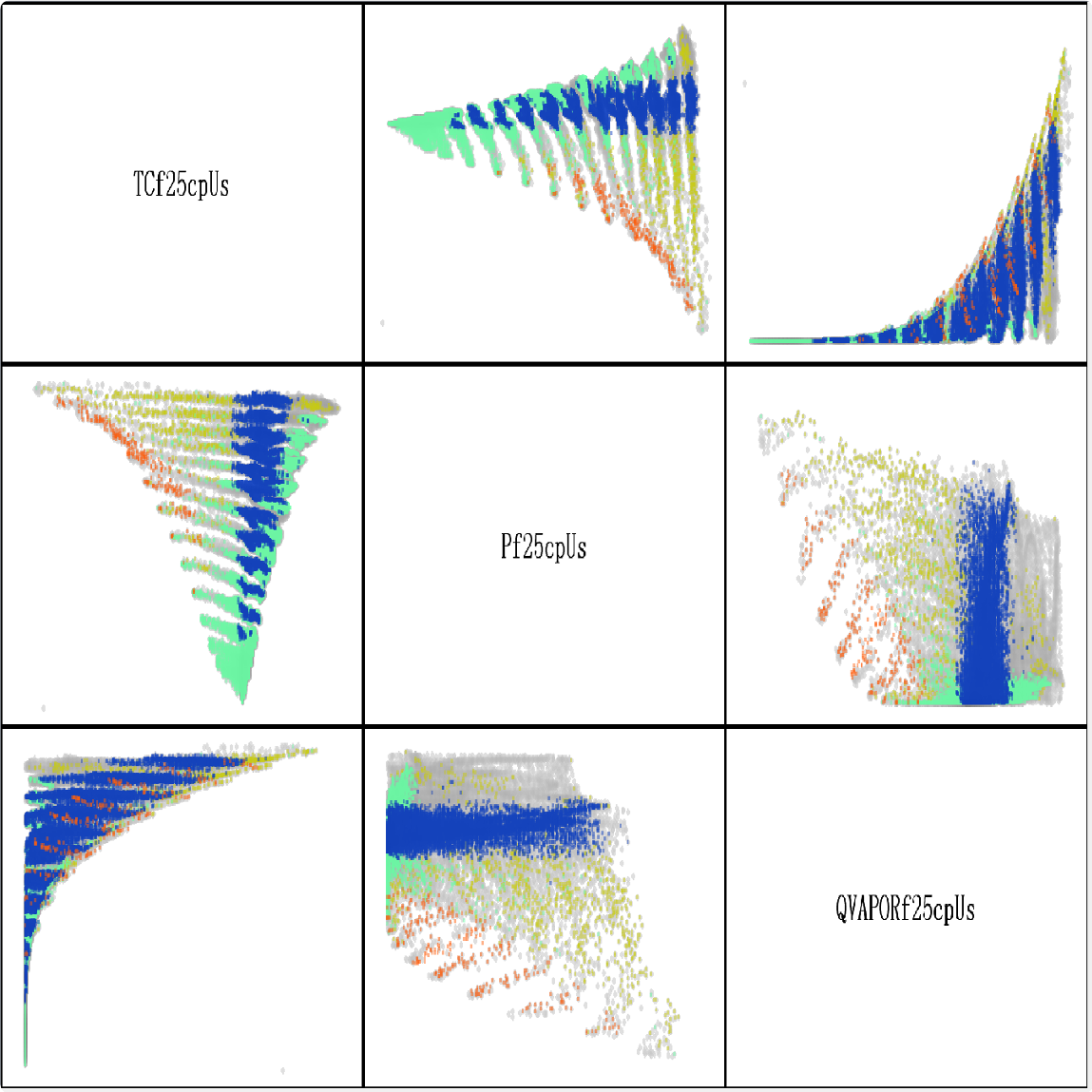}}      %
    \caption{Visualizations of the Isabel data with 1-flat (a--c) and 2-flat (d--f) indexed points. \secRevCvm{Underlying Hurricane Isabel dataset by NCAR~\cite{HurricaneIsabel:2004:VisContest}.}}
    \label{fig:isabel}
\end{figure*}

The Hurricane Isabel dataset~\cite{HurricaneIsabel:2004:VisContest} is a widely-used multivariate simulation dataset.
The visualization of 1-flat indexed points of five attributes (speed, height, temperature, pressure, and vapor) of one timestep in the simulation is shown in Figure~\ref{fig:isabel}~(a, b).
It can be seen that the hurricane eye and whirling structures at different heights are classified using our method.
These features are associated with patterns of medium to high densities of continuous indexed points, but are not visible in the SPLOM. 
\rev{We further explore the data with 2-flat indexed points for three variables---here, temperature, pressure, and vapor.
The 2-flat continuous indexed points contain several high density crossings for strong linear correlations of three attributes and arches for smoothly changing linear relationships (Figure~\ref{fig:isabel}~(e)).
Different structures, e.g., hurricane eye (orange), swirling arms (yellow), and atmosphere at low (blue) and high altitude (cyan) are classified (Figure~\ref{fig:isabel}~(d)) with brushing-and-linking.
These patterns are not recognizable with SPLOM as shown in Figures~\ref{fig:isabel}~(c, f). 
}

A multimodal brain MRI scan of a patient with brain tumor from the BraTs database~\cite{Menze2015,Bakas2017} is shown in Figure~\ref{fig:ui}.
Here, four modalities, namely, T1, T1ce (contrast enhanced), T2, and FLAIR (Fluid-attenuated Inversion Recovery) are visualized with our method.
By identifying features that are off the main ``branch'' in continuous indexed points (Figure~\ref{fig:ui}~(b)), we are able to classify the brain tumor (yellow), the edema (blue), and blood vessels (beige and cyan), as shown in Figure~\ref{fig:ui}~(a). 
Again, these features are not identifiable in the SPLOM (Figure~\ref{fig:ui}~(c)).

\revcvmmj{
Moreover, continuous indexed points can classify volumes using data value and gradient magnitude similar to the classic multidimensional transfer functions.  
An example using the standard benchmark CT Tooth scan is shown in Figure~\ref{fig:toothCompare} in Appendix A.4. }

\subsection{DTI Case Study}
\label{sec:caseStudy}
In this case study, we worked with a collaborating radiologist to analyze scalar metrics of DTI to study thyroid-associated ophthalmopathy (TAO)/Grave's orbitopathy~\cite{Bartalena2021,Genere2019}.
TAO is associated with autoimmune thyroid disease. 
Eye muscles of TAO patients are swollen and stiff and compress the optic nerves, potentially causing severe complications and blindness.
TAO is related to extraocular myocular inflammatory edema, fat infiltration and fibrosis.
Therefore, the diagnosis and the efficacy of the treatment of TAO relies on the evaluation of pathology of eye muscles, which can be aided using DTI.

Scalar metrics derived from DTI, such as fractional anisotropy (FA), radial diffusivity (RD), and mean diffusivity (MD) are sensitive to changes in microstructures~\cite{Chen2023} that could potentially be used for classifying different materials.
Studies have also shown that FA and MD metrics are indicators for identifying pathological and stress changes in muscles~\cite{Klupp2019}, and significant associations of maximal muscle power of the soleus muscle with FA and RD are found~\cite{Scheel2013}.
Therefore, the radiologist suggested studying the scan of a TAO patient with the FA, MD, and RD metrics of the DTI. 
The goal was to classify various materials of the scan, including eye muscles, the eyes, and the brain for further analysis. 

With 1-flat indexed points, we are able to classify muscles (yellow), eye balls and cerebrospinal fluid (blue), and the brain tissue (pink) as shown in Figure~\ref{fig:teaser}~(d).
Brushes used for the classification are shown in Figure~\ref{fig:teaser}~(a). The brushed areas are associated with patterns of strong linearly correlated regions in the continuous indexed points as labeled in the indexed points of sub-dimensions as in Figure~\ref{fig:teaser}~(b, c).
The blue brushes are within the FA-MD subspace, the pink brush is within the MD-RD subspace, and the yellow brushes are located in one each subspace, respectively. 
Using all three DTI metrics improves the classification over two metrics, e.g., FA and MD, as shown in Figure~\ref{fig:teaser}~(e), where the brain tissue cannot be separated from tissues on the skull, and the structure of eye muscles are not as complete (in the boxes). 
Note that the brushed regions are not visible in the SPLOM as shown in Figures~\ref{fig:teaser}~(f, g). 
\revcvmmj{
For comparison, the classification results using traditional multivariate transfer functions on FA-MD, MD-RD, and FA-MD-RD spaces are shown in Figure~\ref{fig:DTImdTF} in Appendix A.4.
There, structures extracted in Figure~\ref{fig:teaser}, especially, the muscles, cannot be clearly revealed with traditional multivariate transfer functions.
}
\begin{figure}[tb]
    \centering
    \subfloat[2-flat continuous\\\quad\;  indexed points]{\includegraphics[height = 3.3cm]{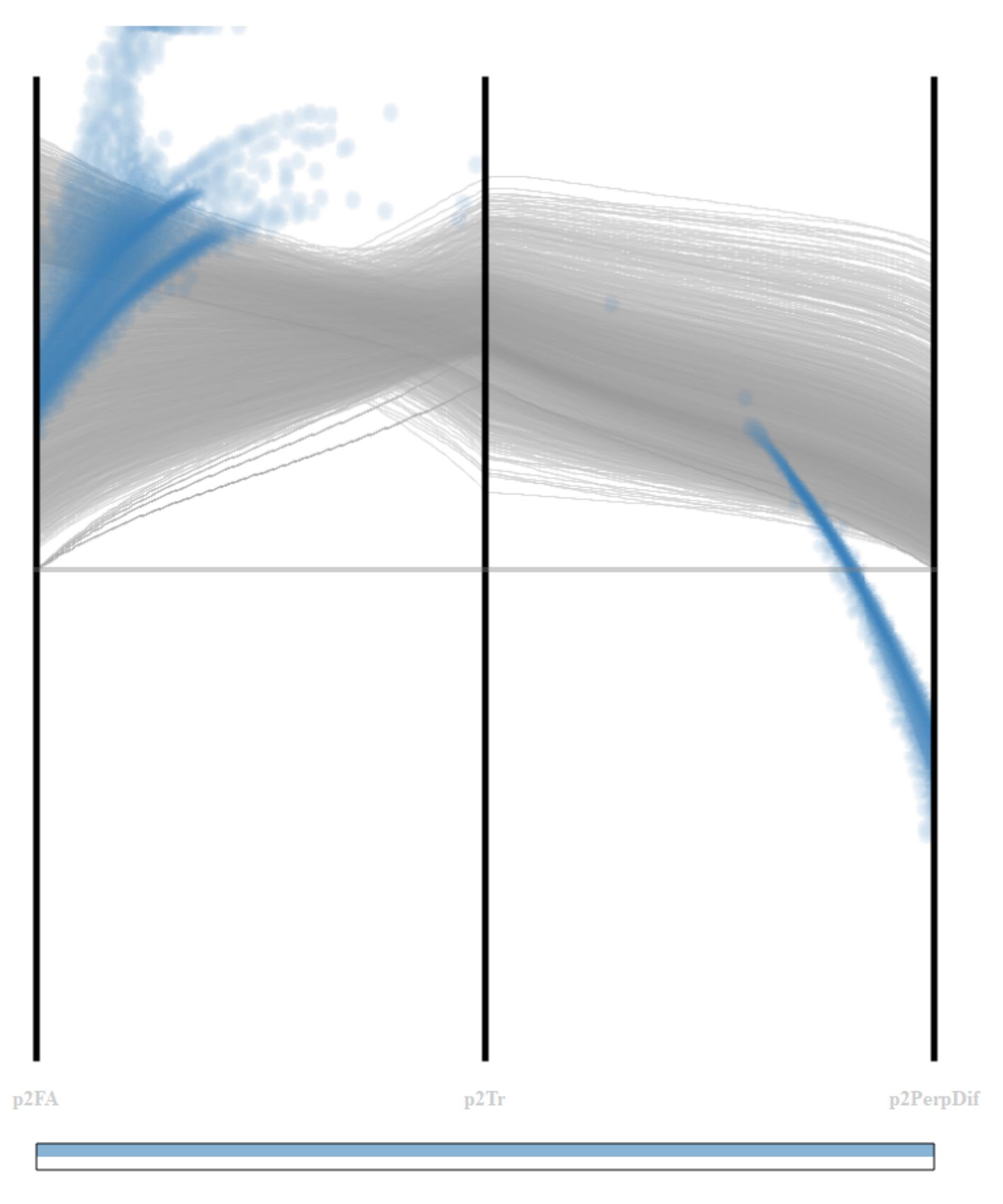}} 
    \subfloat[2-flat indexed points transfer functions]{\includegraphics[height = 3.3cm]{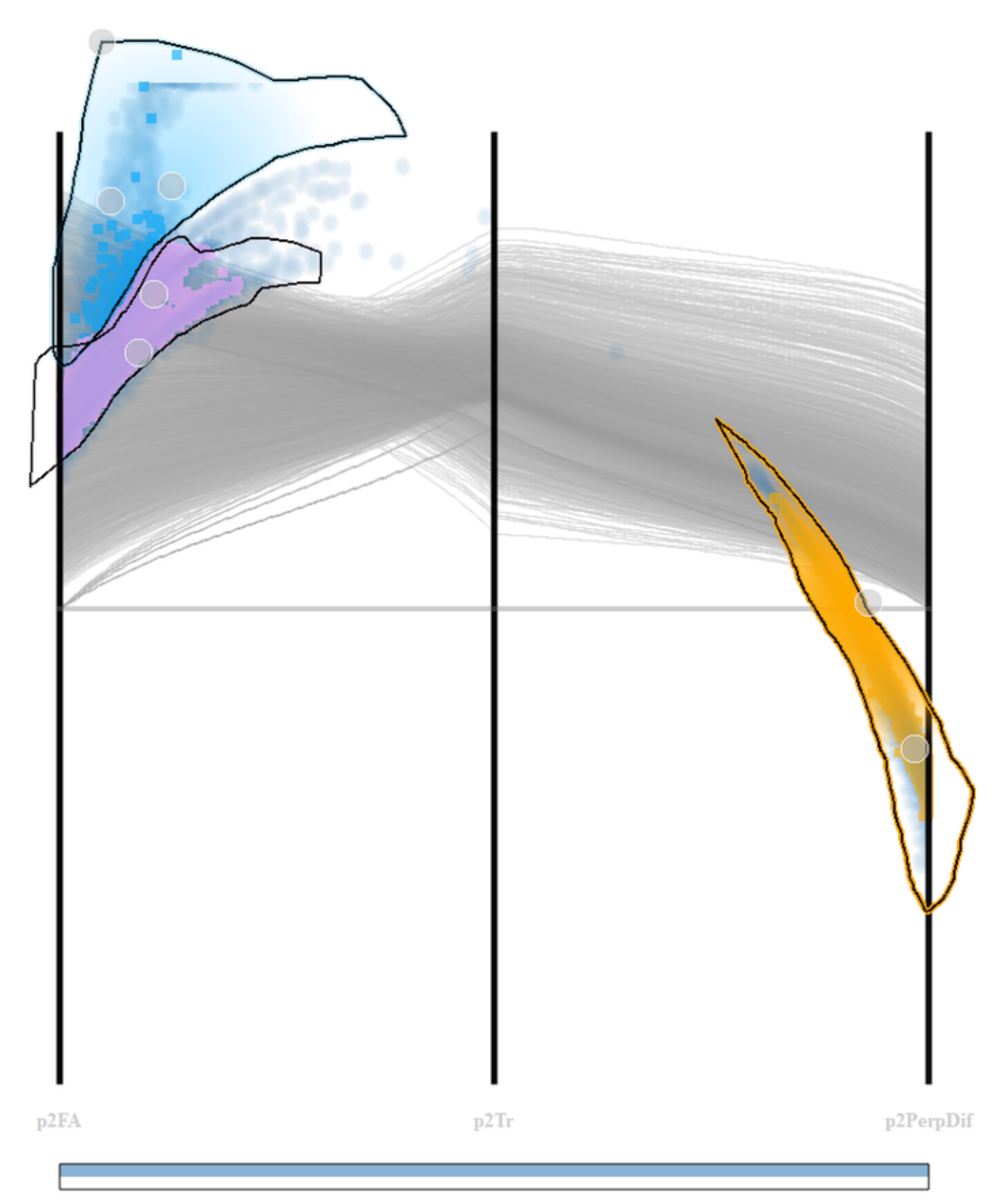}} 
    \subfloat[Volume rendering]{\includegraphics[height = 3.2cm]{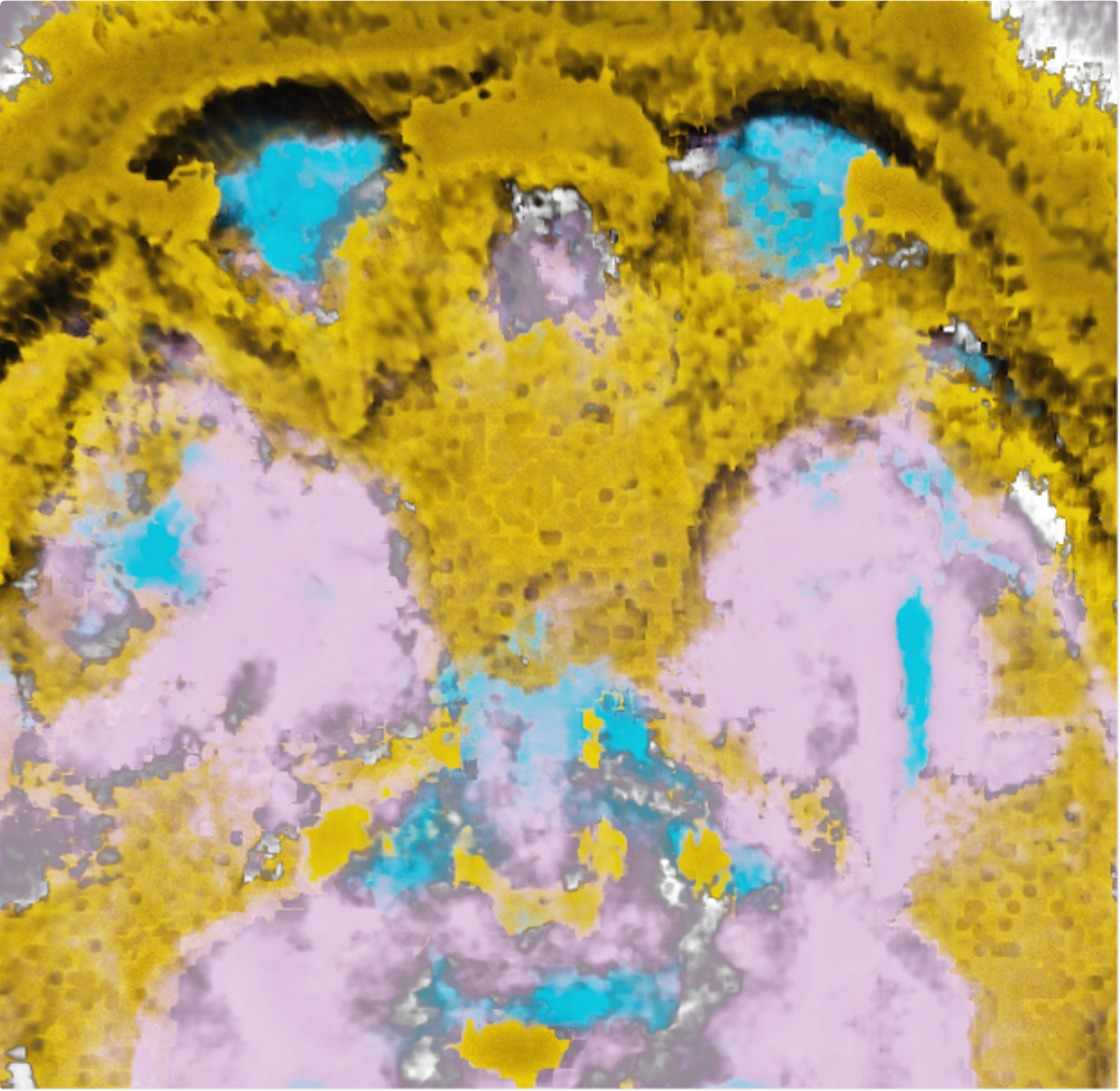}}   
    
    \caption{Visualizations of (a) 2-flat indexed points, and (c) volume rendering classified with (b) transfer functions of the DTI example. \secRevCvm{Underlying DTI dataset courtesy of Peking University People’s Hospital.}
    }
    \label{fig:dti2flats}
\end{figure}

We also calculated 2-flats for the data.
In contrast to the ability of classifying micro structures with 1-flats, 2-flats show the similarity between the three metrics and provide qualitative information.
\revcvmmj{Three branches are visible in the 2-flat visualization (Figure~\ref{fig:dti2flats}~(a)), and the data is classified into three regions (Figure~\ref{fig:dti2flats}~(b, c)) with varying degree of fluid and fat: material contains mainly of fluid (blue), mainly of fat (pink), and in between (yellow).}

\subsection{Process Model}
\secRevCvm{
The user might be roughly inspired by the following  process model to explore multivariate volume data with our visualization tool. 
First, the user might look for high-density regions and clusters by observing the 1-flat indexed points.
By using the shape association rules discussed in Section~\ref{sec:patternInterp}, the user can build a mental map of how the features look in Cartesian coordinates.
Subsequently, with a rectangle or lasso brush drawn over a feature of interest, the user interactively selects the 1-flat indexed points pattern and the volume rendering is updated accordingly. 
In the next step, similar to traditional transfer functions, the feature is fine-tuned through trial-and-error until the volume rendering of the feature is satisfactory.
Specifically, the user fine-tunes the selection by slightly moving the brush around, changing the gradient widget, and modifying the opacity; the brushes can also be removed if not satisfying. 
Adding brushes for all features and repeating the fine tuning steps, the user finishes the exploration and a desired visualization is achieved. In summary, effective visualization typically relies on an iterative and interactive process of pattern recognition in parallel coordinates and brushing-and-linking with the spatial volume rendering view.
In an optional second round, the user might explore the 2-flat indexed points for three attributes using the aforementioned process. 
This could reveal further relationships between attributes. 
}
 
\section{User-Oriented Evaluations}
\revcvmmj{Our method was evaluated with a user study with four visualization experts and by obtaining feedback from one domain expert.}

\subsection{User Study with Visualization Experts}
We used a usability study design with a think-aloud approach and questionnaires in pair analysis sessions to get feedback from four scientific visualization experts. 
They are all experienced experts with experiences in the area from 4 to 10 years. 
After an introduction to indexed points, participants were shown a simple synthetic example to get familiar with our proposed method, and then they were asked to watch the exploration of a synthetic data with 2-flat indexed points and a brain MRI data with 1-flat indexed points with our visualization tool operated by the researcher.
During the session, participants asked questions and instructed the researcher for specific exploration operations of their interests while the researcher operated, listened and talked to them.
The participants were asked to fill out a questionnaire after the task.
\begin{table}[htb]
  \caption{%
  	Mean responses of the questionnare on a Likert scale.%
  }
  \label{tab:studyRes}
  \vspace{1ex}
  \scriptsize%
  \centering%
  \begin{tabular}{p{6.3cm}c}
  	\toprule
  	Statement &  Response   \\
  	\midrule
    Our method extracts features that are not possible with data value attributes. & 4.75 \\
    Patterns of indexes points can be recognized for feature extraction. & 4.25 \\
    The connection between indexes points and correlation can be understood with some explanation. & 4.25\\
    The rendering improves the depth perception. & 4\\
    The tool can help users to gain more insights than multivariate volume data alone. & 5\\
    The visual mapping of indexed points complements existing multivariate visualizations. & 4.25\\
  	\bottomrule
    \end{tabular}
\end{table}

We summarize the mean responses of the questionnaire using a Likert scale (1 for strong disagreement to 5 for strong agreement) on the analysis session as shown in Table~\ref{tab:studyRes}.
For the usability, an average SUS score of 75.6 was achieved, which can be rated as good.

Comments were collected from the free text questions. 
Participants used parallel coordinates, SPLOM, dimensional reduction techniques, and clustering to analyze the data domain of multivariate volumes while transfer functions, slicing, and multivariate colormaps are used for visualizing the spatial domain. 
For comparing volume data, they used correlations and other statistical measures, multivariate transfer functions, and parallel coordinates. 
Regarding limitations, they suggested to find automated technique and/or guidelines for indexed points selection. Understanding and visualizing the sensitivity of indexed points would be helpful, and some improvements in the user interface were mentioned.

Overall, participants agreed that the proposed method was able to provide new insights into multivariate volume datasets, and can effectively classify volumetric features using the local linear information. They agreed that the implemented interactive tool has good usability.

\subsection{Domain Expert Feedback}
The radiologist provided feedback on our method for the DTI use case.
\revcvmmj{
She liked the method for the following reasons.}
The method allows users to visualize spatially embedded local correlations with continuous indexed points to find regions of interest in the spatial domain as medical researchers are interested in the associations between structural and pathological changes and DTI scalar metrics.
The extracted structures are visible as density patterns in continuous indexed points but not easily identifiable with scatterplots.
The method uses the conjunction of multiple DTI metrics to give good classification results which were previously not possible without segmentation tools, e.g., segmentation functions in 3D Slicer. 
She also appreciated the rendering as the occlusion effect allows better judgement on the correctness of classified structures compared to the traditional Phong shading.

Regarding limitations, she thought that 1) understanding the meaning of visual signatures of continuous indexed points was rather difficult, and 2) highly varying sensitivity over different positions of continuous indexed points made the user interaction difficult and a lot of trial-and-errors were required to design a good transfer function.
We agree that the interpretation of indexed points is more difficult than scatterplots of data variables as they are similar to derivatives of data values that require a bit more mental efforts, but the same pattern finding logic holds and users just need to find medium-high density regions and branches.
For the second limitation, it is data-dependent and typical for volume rendering with transfer functions in general and not only with our method.

\section{Conclusion and Future Work}
We have presented continuous indexed points for visualizing local linear information in multivariate volume datasets.
Indexed points encode local linear information in parallel coordinates.
Our new method introduces a mathematical model for transforming the density of multivariate volumes defined on continuous spatial domains all the way to the image plane of parallel coordinates.
We calculate linear fitting for random samples in spatial neighborhoods for multivariate data using PCA.
An octree is built to accelerate the computation by skipping homogeneous regions. 
The local linear information is converted to indexed points with angle-uniform transformation and finally to the density representation using density estimation with a dynamic bandwidth.
Features are interactively classified using multivariate transfer functions, and the associated spatial structures are volume-rendered with an interactive occlusion shading model.
We have demonstrated the usefulness of our method for synthetic and real-world datasets. Our evaluation includes a case study of DTI metrics with a radiologist as a domain expert, an expert user study, and additional feedback from the domain expert.

The main advantage of continuous indexed points is that they support feature classification for multivariate volumes using local correlation information. 
This local correlation information can indicate local linear  features directly, but also global nonlinear features. 
We have illustrated the interpretation of visual patterns in continuous indexed points, and their advantage over scatterplots for feature selection.
Patterns of continuous indexed points retain the same visual signature as point-based density representations that are familiar to scientific visualization users.  
\secRevCvm{
Traditional scatterplots in Cartesian coordinates allow us to identify clusters and correlations but cannot handle correlations of different directions with overlapping samples. 
With continuous indexed points, we are able to perform such correlation related tasks and some of the clustering tasks, and further cluster information can be seen with the original polyline-based parallel coordinates. 
The visual signatures of indexed points are different from scatterplots, and it is up to future work to examine their performances in a user study.
}

Another advantage of our method is that it generalizes the local correlation computation for data defined on continuous spatial domains.
\revcvmmj{With the computation in a spatial embedding, our method filters out false positives of correlations that are outside of spatial neighborhoods.}
The mathematical model is rigorous and generic, and, therefore, supports the design of reliable visualizations of continuous data and is a step toward trustworthy visualization.

For future work, we would like to use parallel computing on the GPU to accelerate the computation of local linear information for the preprocessing step.
We plan to use sampling methods, for example, blue noise sampling, to obtain locations in the spatial domain for local fitting computation instead of using grid points.
It is also of our interest to automatically cluster continuous indexed points for easier feature classifications.
\secRevCvm{
We also plan to unify other analysis techniques for local neighborhoods, for example, fitting or dimensionality reduction methods, with a generic visual representation for more flexible analysis of multivariate volumes.
Comparisons of our method against dimensionality reduced volume attributes can be valuable.
User studies on the perception of patterns of indexed points and scatterplots are also of our interest.} 
Finally, we could identify and visualize uncertainties in the data and processing steps.

\appendix

\section*{Appendix}
We include contents that are left out for conciseness from the paper in this appendix. \revcvmmj{Specifically, we provide more details on transfer functions, the computational performance, the interpretation of 1- and 2-flat indexed points using a complex synthetic data, and comparisons to traditional multidimensional and multivariate transfer functions.}

\subsection*{A.1 Transfer Function Lookup}
\begin{figure}[h]
    \centering
    \includegraphics[width = 0.95\linewidth]{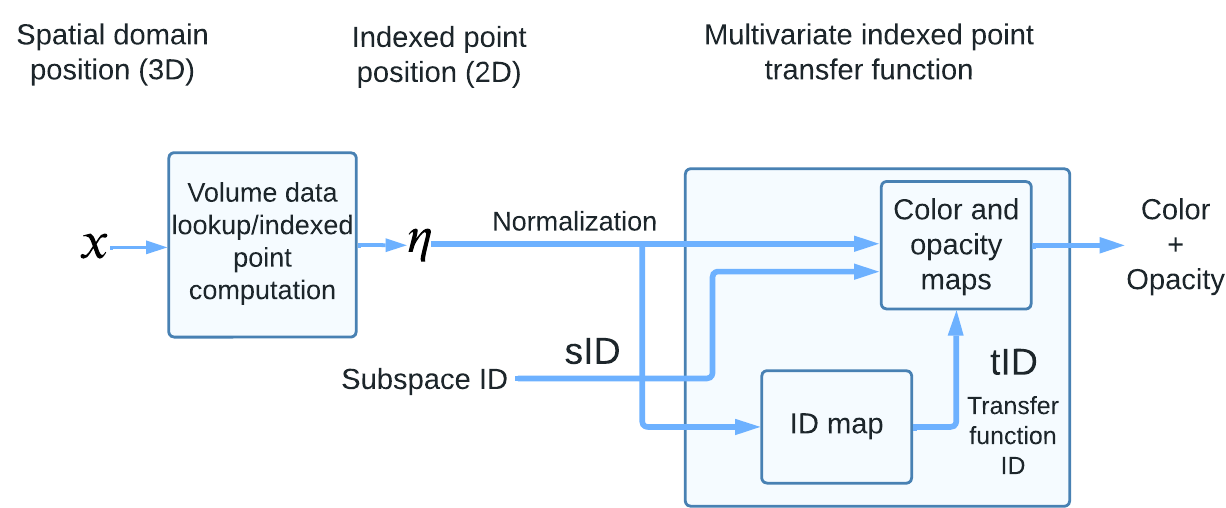}
    \caption{The process of multivariate indexed point transfer function lookup.}
    \label{fig:multiTF}
\end{figure}

A lookup table is used to realize the multivariate transfer functions as shown in Figure~\ref{fig:multiTF}.
Since indexed points of positive correlations overlap with those of neighboring subspaces, a set of color and opacity maps, and an ID map are required for the transfer function. The ID map is used to check the corresponding subspace ID (sID) and the transfer function ID (tID) of the indexed point $\bm{\eta}$. 

The subspace ID sID is used to distinguish subspaces in overlapping regions of indexed points.
An sID of $-1$ is assigned by default, indicating that the widget lives in the image plane across all subspaces, and it can be conveniently changed by the user for the desired subspace.
By default, transfer functions are constructed with the OR-operation of brushes.
Alternatively, the tID is employed to identify and link transfer functions: transfer functions with the same tID are joined together using the AND operation.
Then, $\bm{\eta}$ along with tID and sID are used to look up the color and opacity maps for the volumetric sample.

\subsection*{A.2 Computational Performance}
The computational performance was measured on a machine with 3.5\,GHz Intel i7 CPU, 32\,GB main memory, and an NVIDIA Quadro P6000 graphics card. 
\rev{
The computational time of indexed points volumes in Matlab are reported in Table~\ref{tab:preprocessTime}.
All data are processed with a neighborhood size of $7\times7\times7$ voxels after the normalization of data values.
\revcvmmj{The time varies with the subdivision threshold ($t_s$) and the eigenvector interpolation threshold ($t_e$) of octree nodes and depends on how noisy the dataset is.}
Octrees are effective to reduce the computation time for datasets that contain large homogeneous regions (Isabel and BraTS), but not ideal for very noisy data as the hierarichical construction takes longer than a per-voxel implementation.
Note that our C++ implementation of the per-voxel method took from a few minutes (BraTS) to half an hour (Isabel) for these cases.
We plan to implement the octree method in C++ and consider further acceleration with parallel computing on the GPU.
}
\begin{table}[htb]
  \caption{%
  	Computation time of preprocessing (in seconds) in Matlab. \revcvmmj{For all examples, $t_e=0.01$.}
  }
  \label{tab:preprocessTime}
  \vspace*{1ex}
  \small%
  \centering%
  \begin{tabu}{cccc}
  	\toprule
  	Dataset & Dataset size & Octree  &  Per-voxel   \\
  	\midrule
  	Isabel & $500^2 \times 100 \times5$ & 4605.2 ($t_s$=0.03) & 16844.9 \\
  	& & 10596.5 ($t_s$=0.02) &\\
        BraTS & $240^2 \times155\times4$ & 3583.2 ($t_s$=0.03) & 6876.3 \\

  	\bottomrule
    \end{tabu}
\end{table}

Preprocessed data items computed in Matlab are loaded into our visualization tool for interactive visualization and exploration. 
The performance of the interactive visualization tool is summarized in Table~\ref{tab:interactivePerf}.
\begin{table}[htb]
  \caption{%
  	Performance of the interactive visualization tool (rendering speed in fps). DO: directional occlusion, EO: extinction optimization, t(TF): update time of transfer functions (in seconds).
  }
  \label{tab:interactivePerf}
    \vspace*{1ex}
  \small%
  \centering%
  \begin{tabu}{ccccccc}
  	\toprule
      &  &  \multicolumn{2}{c}{Avg fps}\\
      \cmidrule(lr){3-4}
  	Dataset  & Sampling rate & DO & DO+EO &  Avg t(TF)  \\
  	\midrule
  	Isabel  & 0.3 & 12 & 4  & 5 \\
  	 & 0.5 & 6 & 2  &\\
   	 & 1.0 & 2 & 1  &\\
        BraTS & 0.3 & 14 & 8  & 4 \\
        & 0.5 & 11 & 4  &\\
   	& 1.0 & 4 & 1  &\\
  	\bottomrule
    \end{tabu}
\end{table}

\newcommand{\synthHeight}{4.5cm}
\begin{figure*}[htb]
    \centering
 
    \subfloat[Synthetic data with three different Gaussian distributions]{\includegraphics[width = 0.6\linewidth]{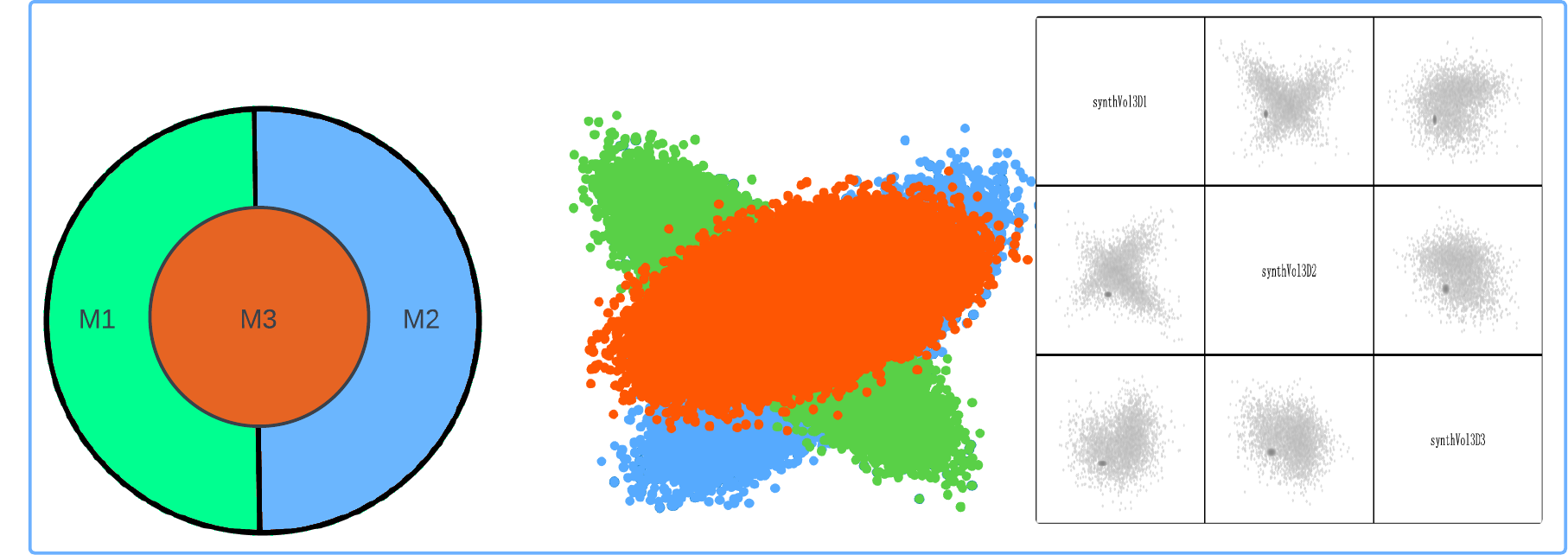}}\hspace{1em}
    \subfloat[Volume rendering using 2D transfer functions]{\includegraphics[width = 0.35\linewidth]{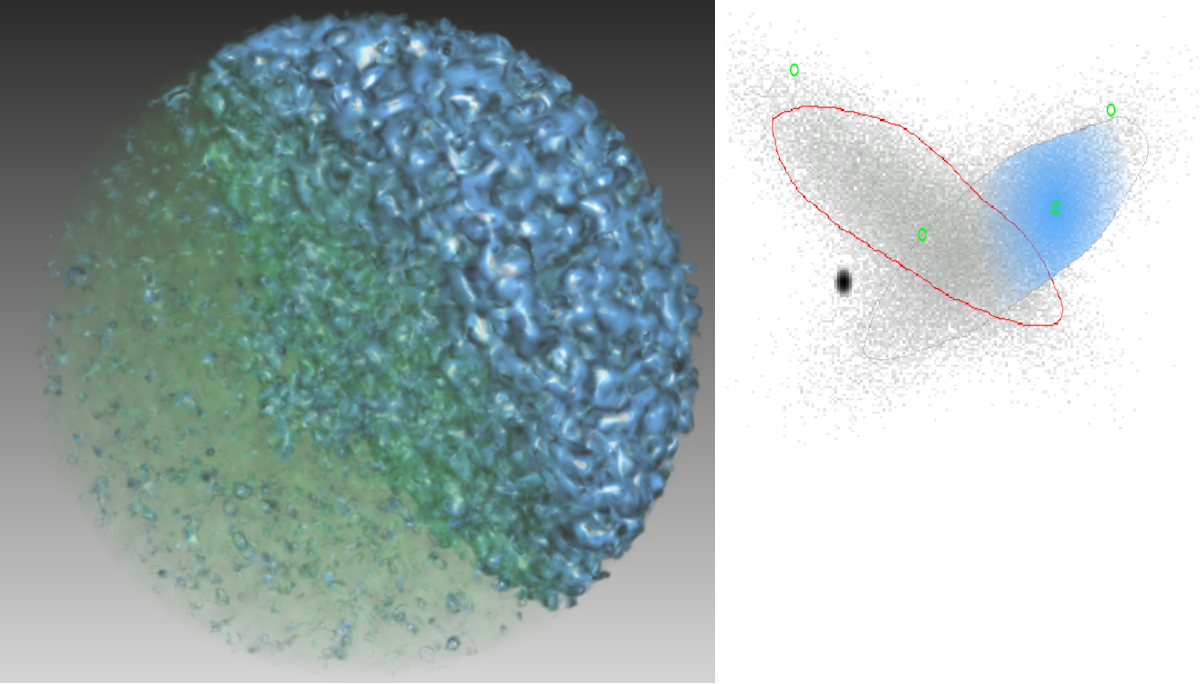}}
    \\ 
    \subfloat[1-flat continuous indexed points]{\includegraphics[height = \synthHeight]{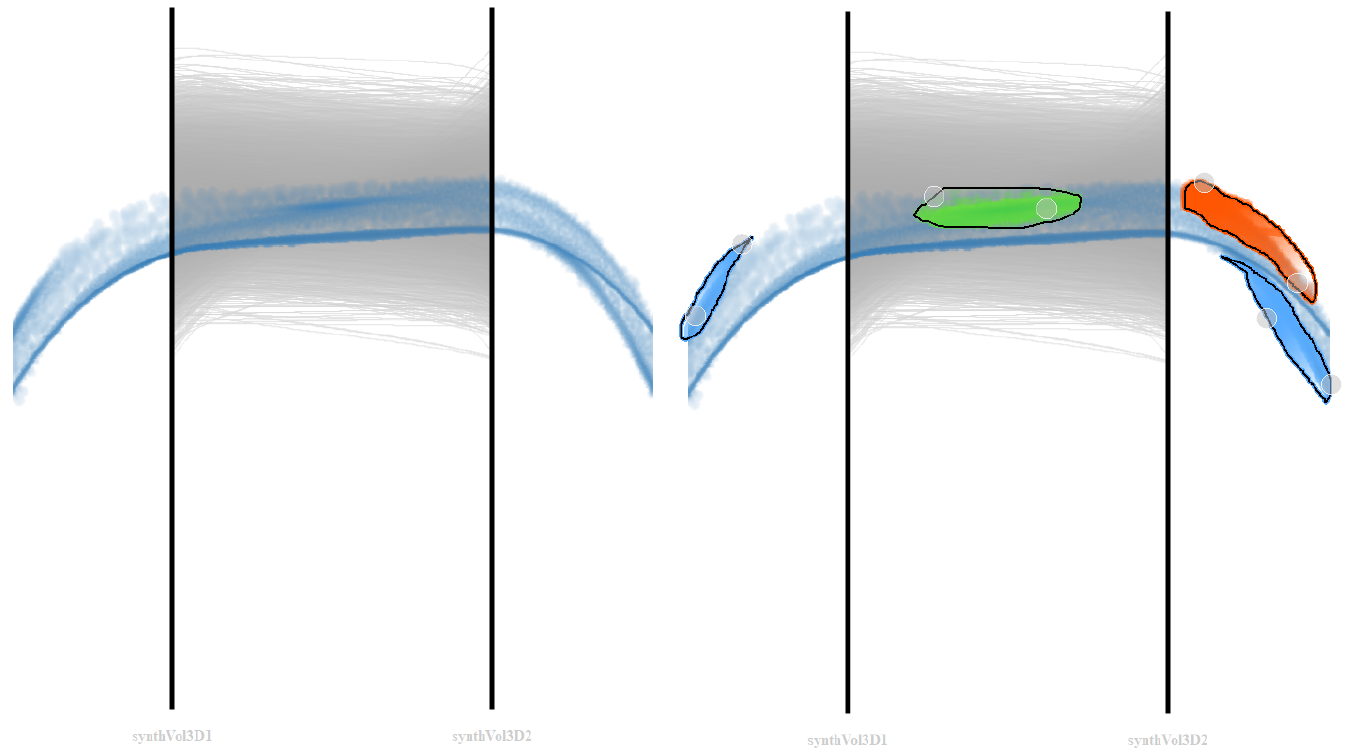}}\hspace{1em}
    \subfloat[Volume rendering]{\includegraphics[height = \synthHeight]{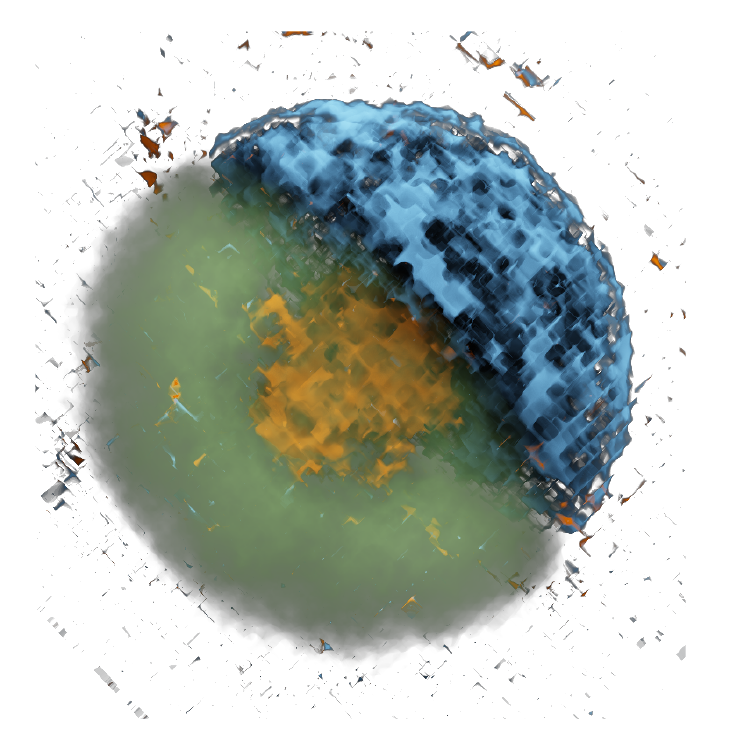}}\hspace{1em}
    \subfloat[Linked scatterplot of the first two attributes]{\includegraphics[height = \synthHeight]{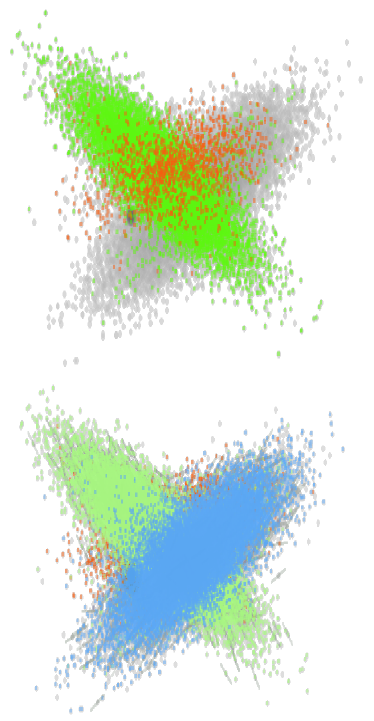}}
    
    \caption{Visualizations of (a) a synthetic dataset containing three structures of different Gaussian distributions. \revcvmmj{Traditional 2D transfer functions (b) cannot classify the orange feature for the first two attributes as the samples overlap in Cartesian coordinates.}
    For the first two attributes, our new method (b--d) can classify all structures by the local fitting information with 1-flat indexed points.
    }
    \label{fig:synth1flatsCompare}
\end{figure*}

\begin{figure*}[htb]
    \centering
 
    \subfloat[Spatial-domain]{\includegraphics[width = 0.3\linewidth]{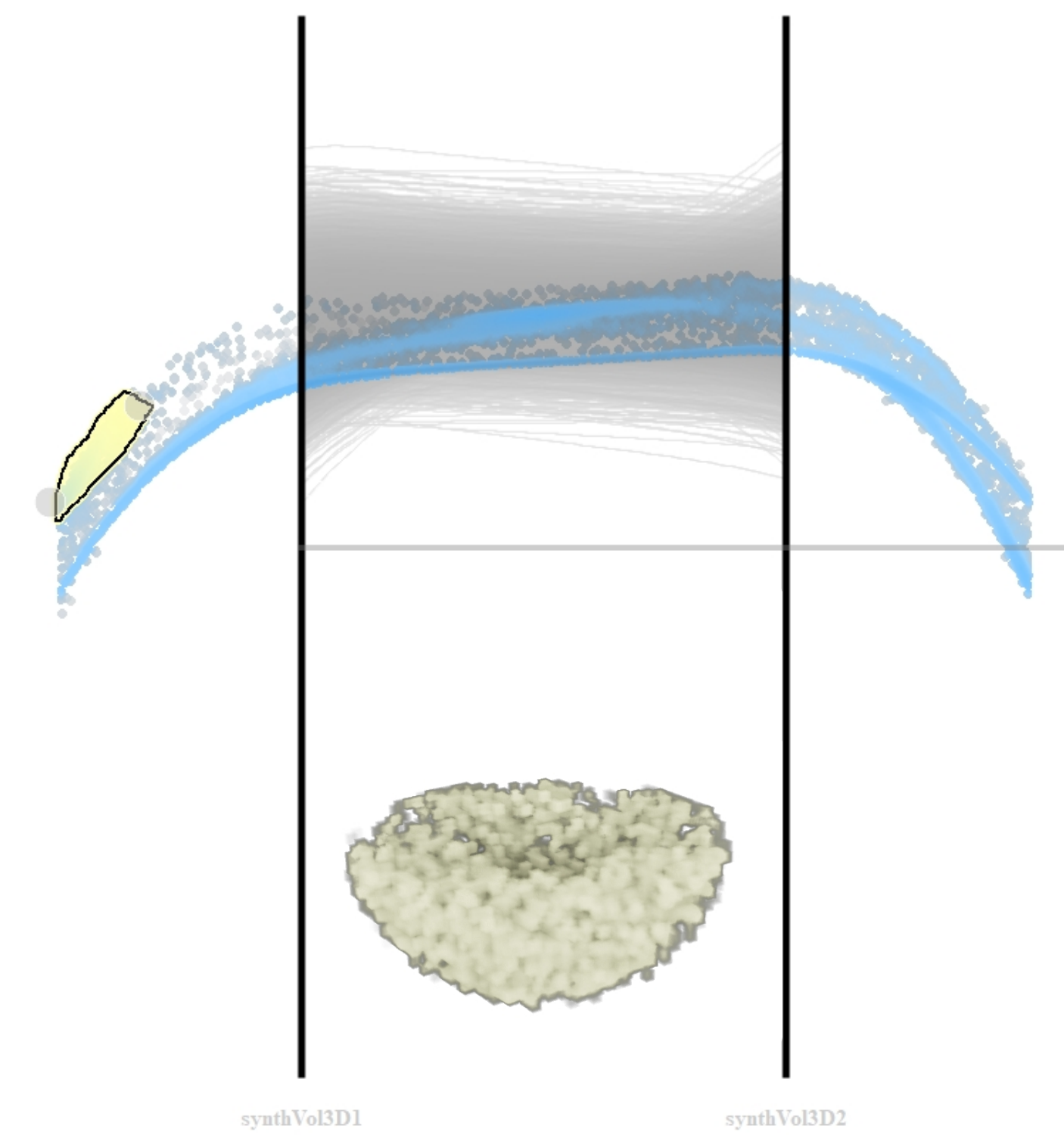}}\hfill
    \subfloat[Data-domain]{\includegraphics[width = 0.3\linewidth]{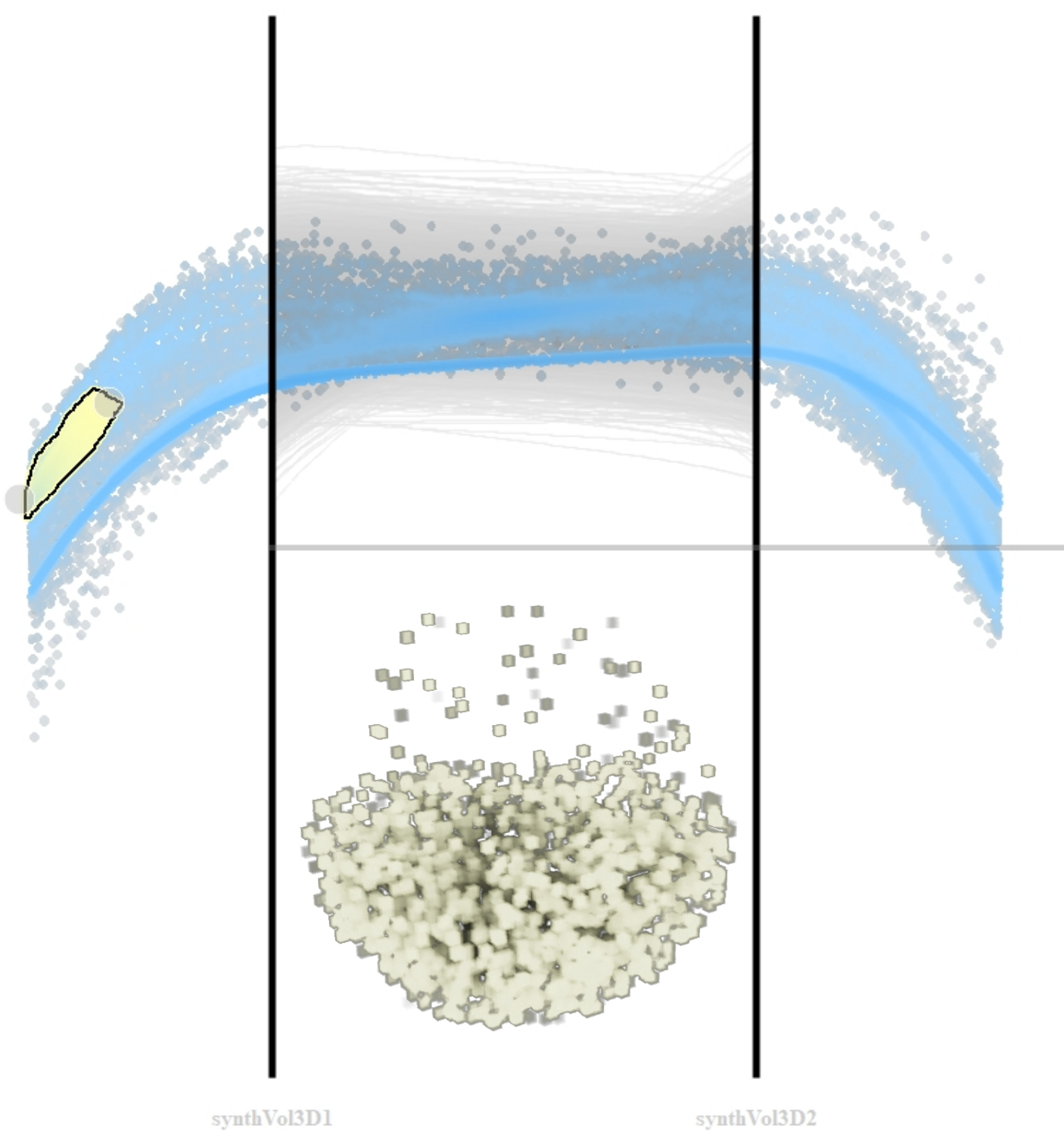}}\hfill
    \subfloat[Data-domain (with a wider vertical lasso)]{\includegraphics[width = 0.3\linewidth]{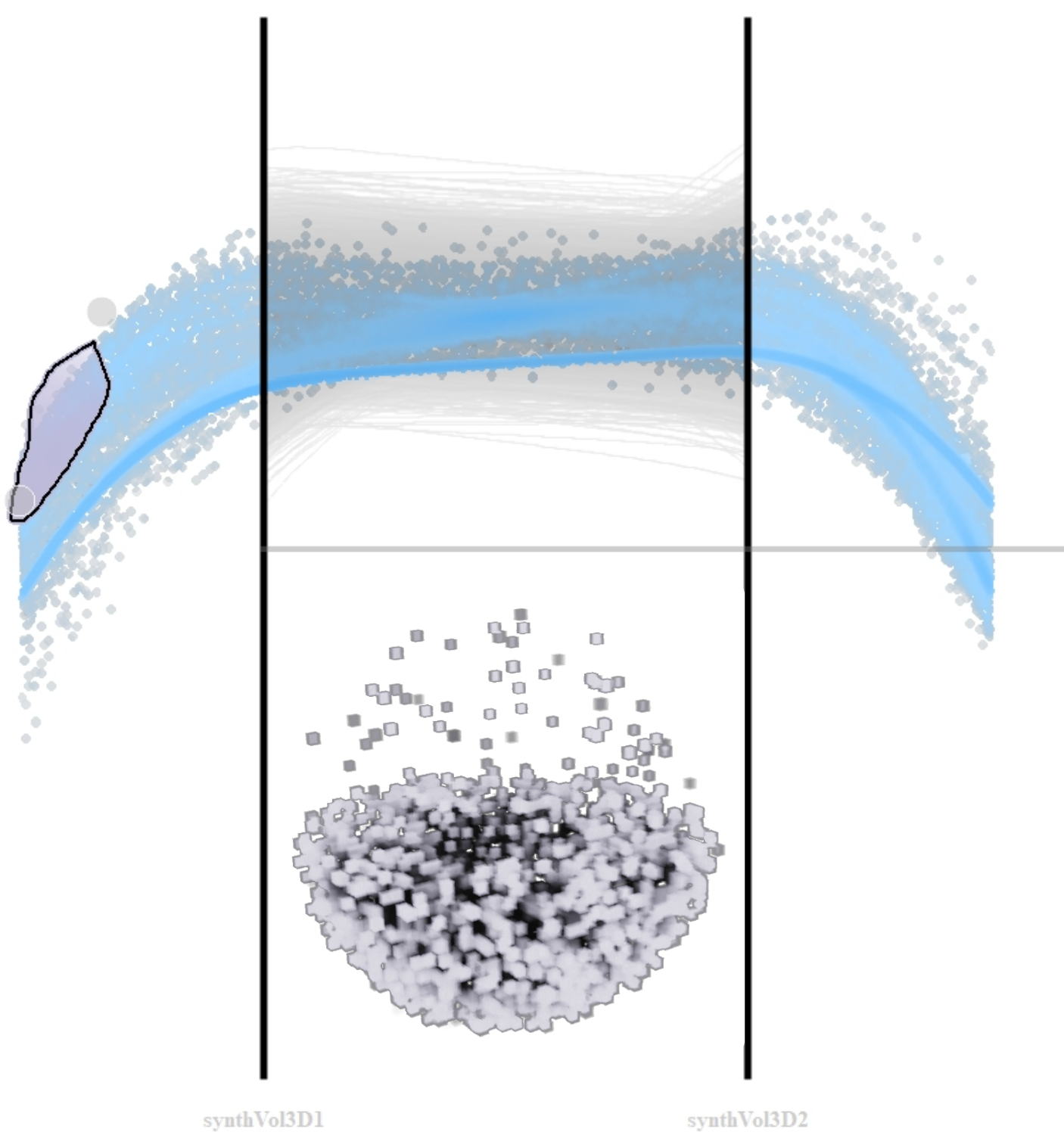}}
    \caption{    
    A comparison of volumetric features selected with a lasso brush on 1-flat indexed points computed from (a) the spatial domain and (b) the data domain computation. In (c), a lasso with a wider vertical span is used to select the smeared feature in the data domain version.     
    }
    \label{fig:synth1flatSpatial}
\end{figure*}

\subsection*{A.3 Interpretation of Indexed Points Patterns}

With a synthetic dataset (Figure~\ref{fig:synth1flatsCompare}), we show continuous indexed point patterns of data features with various correlations.
As illustrated in Figure~\ref{fig:synth1flatsCompare}~(a), the synthetic data is a sphere filled with data values from three 2D Gaussian distributions: the inner sphere contains values from a positively correlated Gaussian with an orientation of $\pi/6$ (orange), and two domes are with values from Gaussian distributions with an orientation of $\pi/4$ (blue) and $-\pi/4$ (green), respectively.
\secRevCvm{
The 2D histogram reveals the overall structure of the data domain, as shown in Figure~\ref{fig:synth1flatsCompare}~(b)-right, where two Gaussian blobs and a dense region (the small black region) of the background can be seen.
However, the third Gaussian blob is not visible as it is occluded by the other blobs.
Therefore, 2D transfer functions cannot classify the inner sphere correctly, as shown in Figure~\ref{fig:synth1flatsCompare}~(b)-left.
}


In contrast, patterns of medium and high densities are shown by continuous indexed points, as in Figure~\ref{fig:synth1flatsCompare}~(c).
A high-density arch that spans the entire horizontal axis of the two attribute axes is shown at the bottom.
A strong negatively correlated pattern is shown between the axes.
A high-density positively correlated pattern can be seen around the left boundary, and two arches of medium density are visible on the region right to the second data axis.
Using lasso widgets, we are able to classify each feature in the volume, as seen in the volume rendering in Figure~\ref{fig:synth1flatsCompare}~(d).
Note that the inner sphere is classified using the orange lasso over the arch on the upper-right. Also noticeable are the two blue lassos that jointly extract the blue dome as the pattern is split into two at the boundaries (with a corresponding angle of 45 degrees of the correlation) of the image plane of the angle-uniform parallel coordinates. 
The classification result can also been seen in the scatterplots in Figure~\ref{fig:synth1flatsCompare}~(e) through brushing-and-linking. Here, the overlapping colored regions indicate that our method successfully classifies features of similar data values using directional information of local~fittings.   


\begin{figure*}[htb]
    \centering

    \subfloat[2-flat continuous indexed points]{\includegraphics[height = 4.5cm]{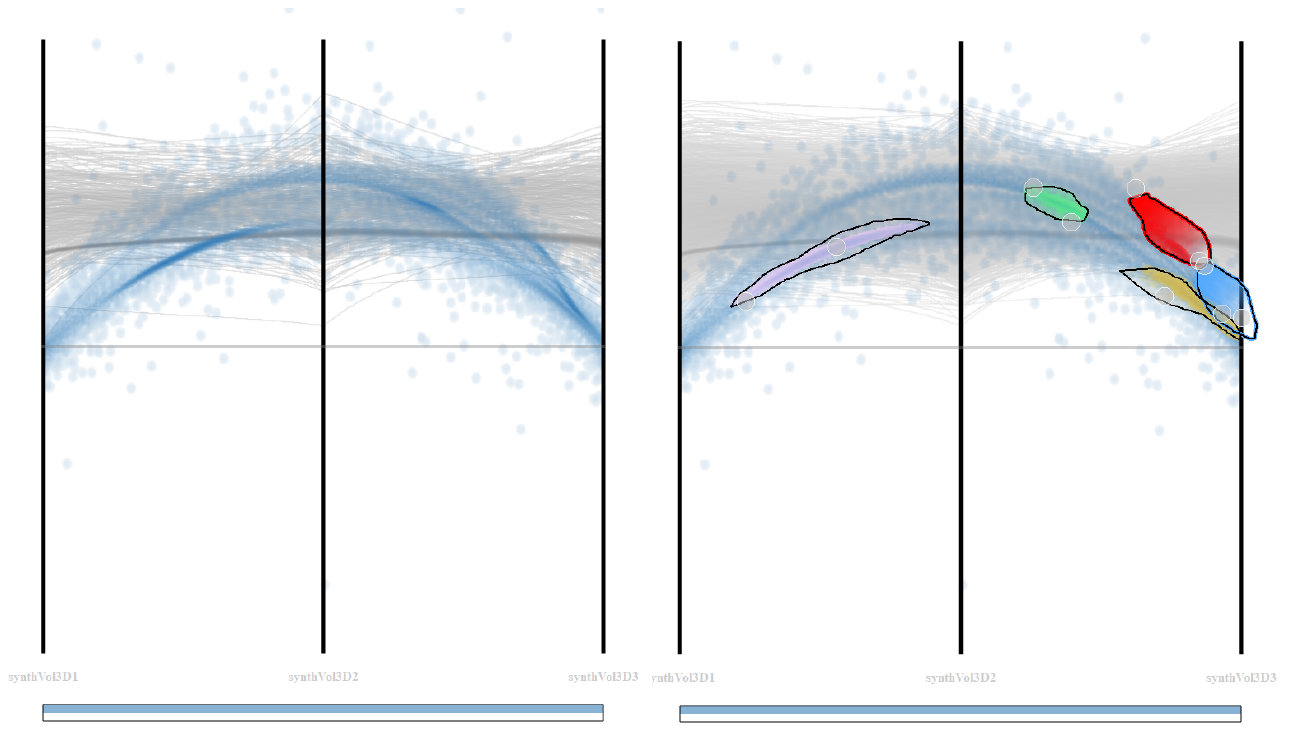}}
    \subfloat[Volume rendering]{\includegraphics[height = 4.5cm]{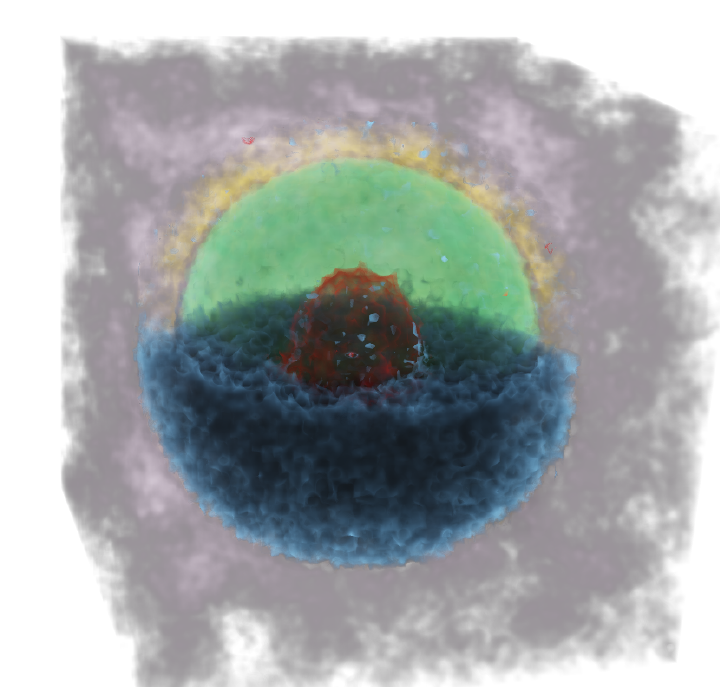}}\hspace{1em}
    \subfloat[Linked SPLOM]{\includegraphics[height = 4cm]{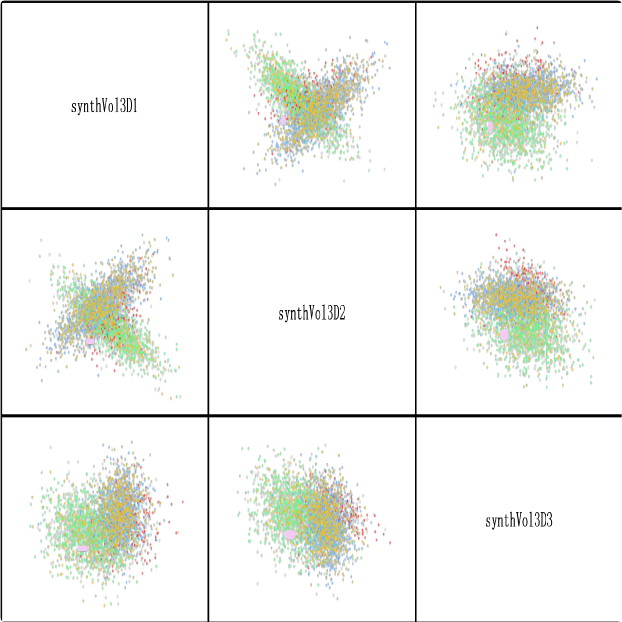}}
    \caption{All three attributes of (b) the synthetic dataset are visualized using (a) 2-flat indexed points. The linked SPLOM with brushed data is shown in (c).}
    \label{fig:synth2flatsCompare}
\end{figure*}
Figure~\ref{fig:synth1flatSpatial} shows the differences of visual signatures of indexed points of the two variants.
\revcvm{The data domain computation (Figure~\ref{fig:synth1flatSpatial}~(b)) creates signatures that are more spread-out than the spatial domain computation (Figure~\ref{fig:synth1flatsCompare}~(a)) with the same lasso brush.
More noisy volume samples from the data domain computation are selected (Figure~\ref{fig:synth1flatSpatial}~(b)), and they are found to be false positives (according to Figure~\ref{fig:synth1flatsCompare}~(a)).
Therefore, the data domain variant is not robust for data with spatial embedding, whereas the spatial computation is more reliable.
A brush with the same horizontal span but a wider vertical span selects the more scattered visual signature in the data domain variant that introduces more noisy volume samples (Figure~\ref{fig:synth1flatsCompare}~(c)). 
}

The case of 2-flat indexed points can be seen in Figure~\ref{fig:synth2flatsCompare}. High-density patterns can be identified in Figure~\ref{fig:synth2flatsCompare}~(a). The classification result is volume rendered with brushing-and-linking (Figure~\ref{fig:synth2flatsCompare}~(b)). The brushed scatterplot matrix is shown in Figure~\ref{fig:synth2flatsCompare}~(c).

\subsection*{A.4 Comparison to Traditional Multidimensional and Multivariate Transfer Functions}
\begin{figure*}[htb]
    \centering
     \subfloat[Volume rendering\\ (1-flat TF)]{\includegraphics[width = 0.2\linewidth]{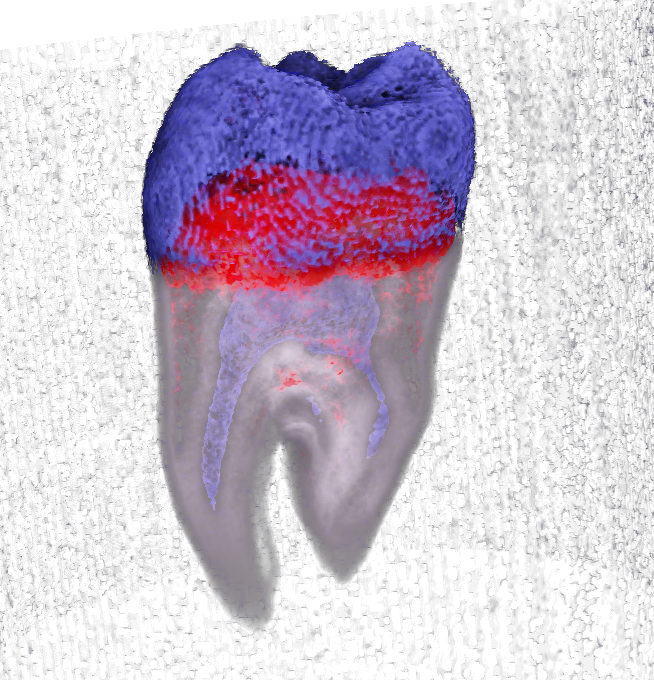}}\hfill
    \subfloat[1-flat indexed points\\ transfer functions]{\includegraphics[width = 0.2\linewidth]{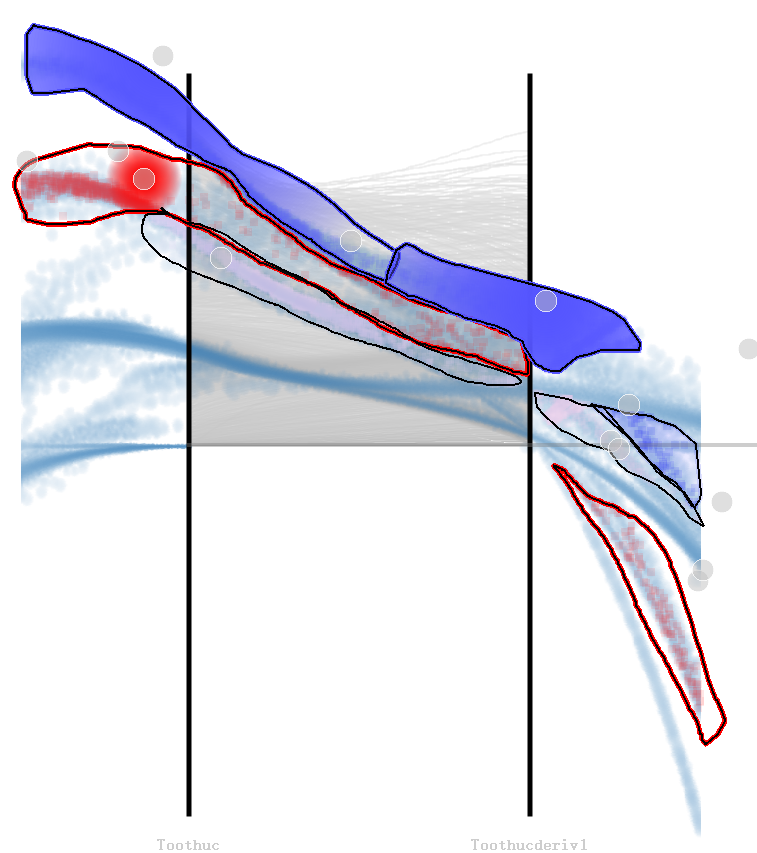}}\hfill
        \subfloat[Linked scatterplot view with brushed data]{\includegraphics[width = 0.2\linewidth]{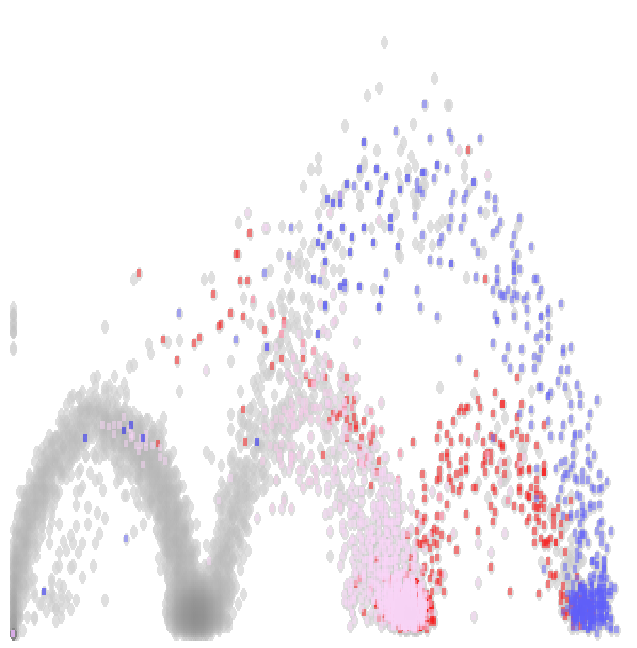}}\hfill
    \subfloat[Volume rendering (2D TF)]{\includegraphics[width = 0.2\linewidth]{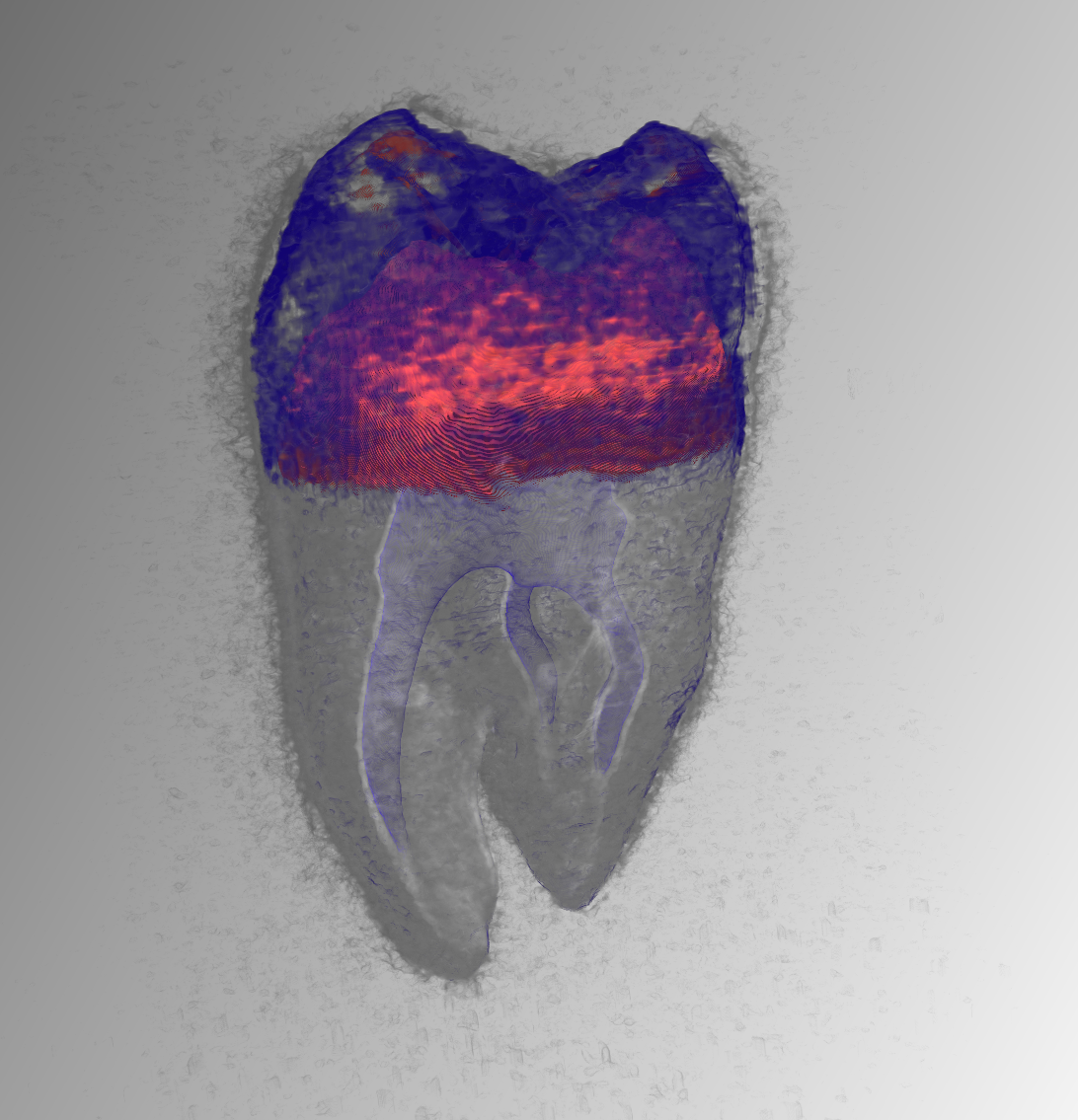}}
     \subfloat[2D transfer functions]{\includegraphics[width = 0.2\linewidth]{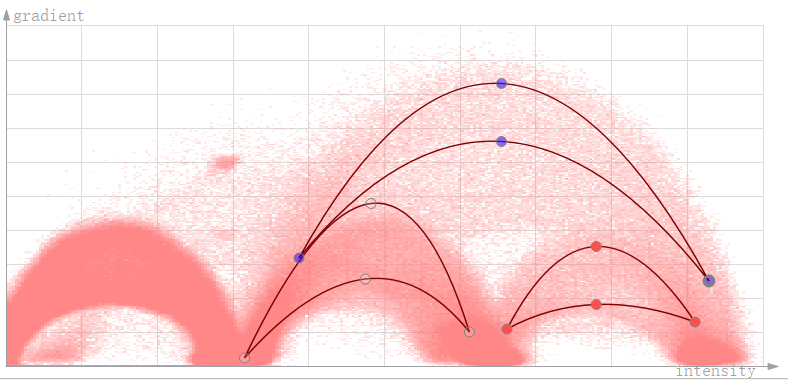}}
    \caption{Visualizations of the CT Tooth dataset. The continuous indexed points method (a--c) effectively replicates the result of the multidimensional transfer function method~\cite{Kniss:2002:MTF:614287.614529} (d, e) using scalar and gradient magnitude. Figures (d, e) are produced using Voreen (\url{https://voreen.uni-muenster.de}). \secRevCvm{Underlying CT Tooth dataset by GE Aircraft Engines~\cite{Pfister2001}.}}
    \label{fig:toothCompare}
\end{figure*}

\begin{figure*}[htb]
    \centering
     \subfloat[Volume rendering\\ (1-flat TF)]{\includegraphics[width = 0.2\linewidth]{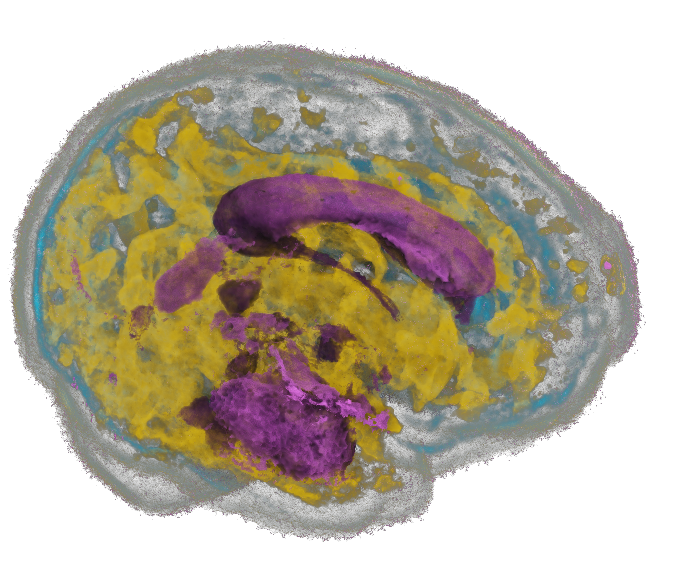}}\hfill
    \subfloat[1-flat indexed points]{\includegraphics[width = 0.2\linewidth]{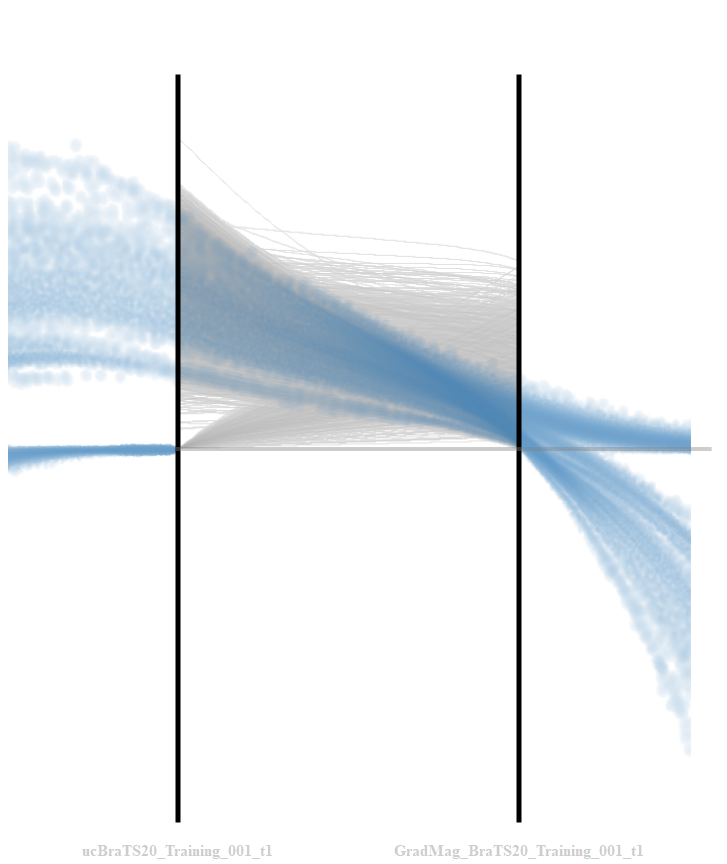}}\hfill
    \subfloat[1-flat indexed points\\ transfer functions]{\includegraphics[width = 0.2\linewidth]{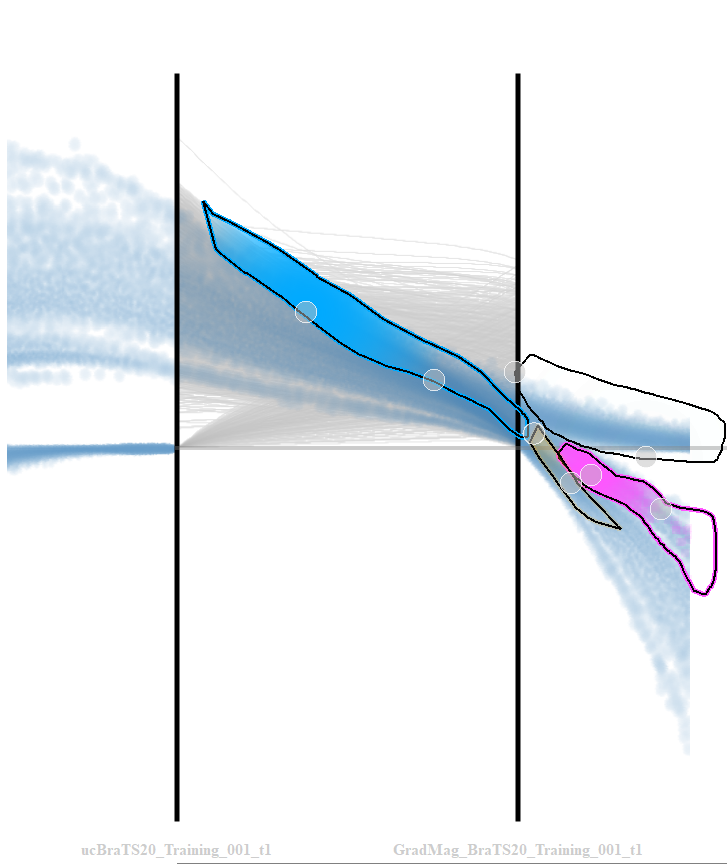}}\hfill
    \subfloat[Volume rendering (2D TF)]{\includegraphics[width = 0.2\linewidth]{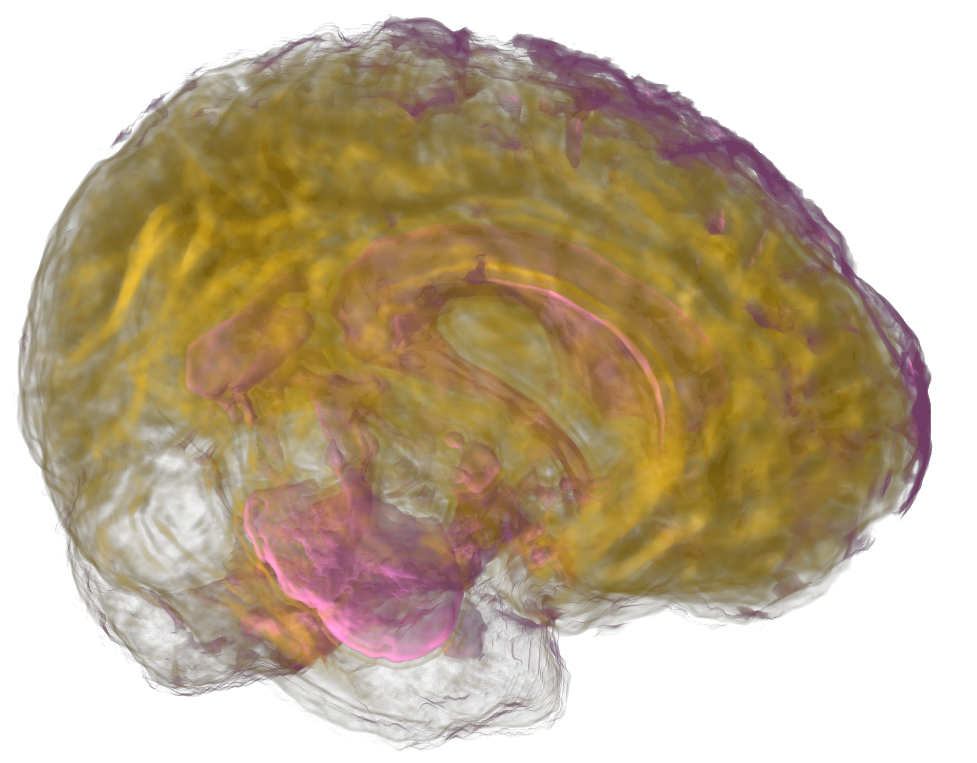}}
     \subfloat[2D transfer functions]{\includegraphics[width = 0.2\linewidth]{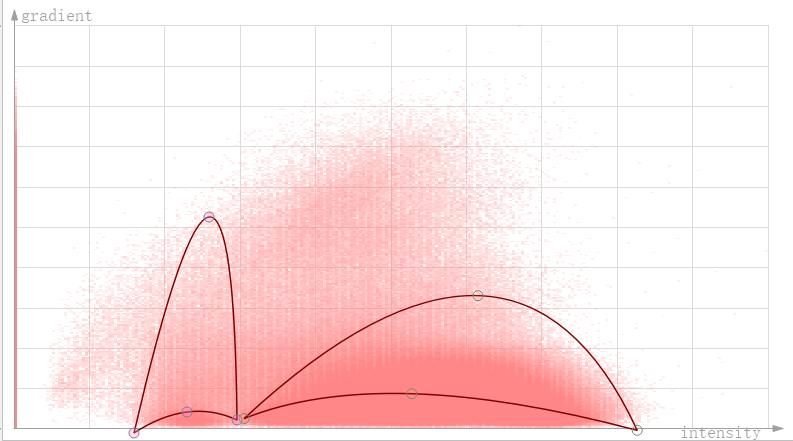}}
    \caption{Visualizations of the Brain MRI T1 dataset of the scalar value and gradient magnitude. Continuous indexed points (a--c) allow us to classify more features than the multidimensional transfer function method (d, e) using scalar and gradient magnitude with Voreen (\url{https://voreen.uni-muenster.de}). \secRevCvm{Underlying MRI dataset by Menze et al.~\cite{Menze2015}.}}
    \label{fig:t1compare}
\end{figure*}

The CT Tooth dataset is a standard benchmark for volume visualization, see Figure~\ref{fig:toothCompare}.
The typical multidimensional transfer functions~\cite{Kniss:2002:MTF:614287.614529} using the scalar value and gradient magnitude can classify boundaries of features by selecting arches in the data domain, as shown in Figure~\ref{fig:toothCompare}~(c).
Here, the top-right arch is brushed with a blue widget, the middle arch at the bottom is pulled out with a gray brush, and the bottom-right arch with a red brush.
The classification can be replicated using our continuous indexed points computed from the same scalar value and gradient magnitude attributes. 
An arch in the scatterplot in Cartesian coordinates becomes a stripe spanning the horizontal locations in the continuous indexed points (Figure~\ref{fig:toothCompare}~(b)): the parts within parallel axes correspond to negatively correlated regions of the arch (downward going), whereas parts outside of parallel axes are associated with positively correlated regions of the arch (upward going).
Therefore, features of the same arches can be classified with lassos as shown in Figure~\ref{fig:toothCompare}~(b), and the corresponding scatterplot is shown in Figure~\ref{fig:toothCompare}~(c). 
A comparable volume rendering is achieved with our method (Figure~\ref{fig:toothCompare}~(a)) even though the original multidimensional volume attributes are not used but their local fitting information. 

\secRevCvm{
Another comparison of 1-flat indexed points and the scalar value and gradient magnitude is shown in Figure~\ref{fig:t1compare} using the Brain T1 MRI data.
Here, no arches are visible in the 2D Cartesian histogram of the scalar value and gradient magnitude (Figure~\ref{fig:t1compare}~(e)) due to the noisy nature of MRI. 
Therefore, although the tumor (purple) can be separated from brain tissues (yellow), no more clear boundaries can be classified with 2D transfer functions (Figure~\ref{fig:t1compare}~(d)).
With 1-flat indexed points (Figure~\ref{fig:t1compare}~(b)), multiple branches of patterns can be seen.
Using lassos over these patterns (Figure~\ref{fig:t1compare}~(c)), the tumor and corpus callosum (purple), different parts of the brain tissues (yellow and blue), and the outer tissue (gray) can be classified, as shown in the volume rendering (Figure~\ref{fig:t1compare}~(a)). 
The correlation information allows us to classify more features compared to the scalar value and gradient magnitude in this case.
}

\newcommand{\dtiheight}{4cm}
\begin{figure*}[htb]
    \centering
    \subfloat[Volume rendering (FA-MD)]{\includegraphics[height = \dtiheight]{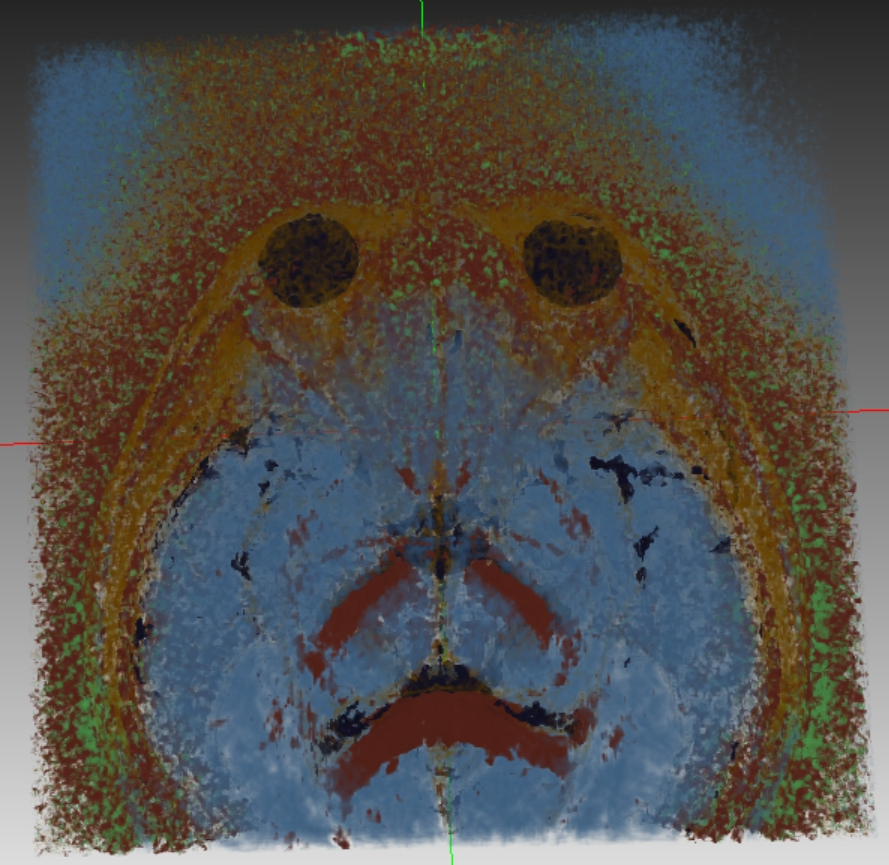}}\hfill
    \subfloat[2D transfer functions (FA-MD)]{\includegraphics[width = 0.25\linewidth]{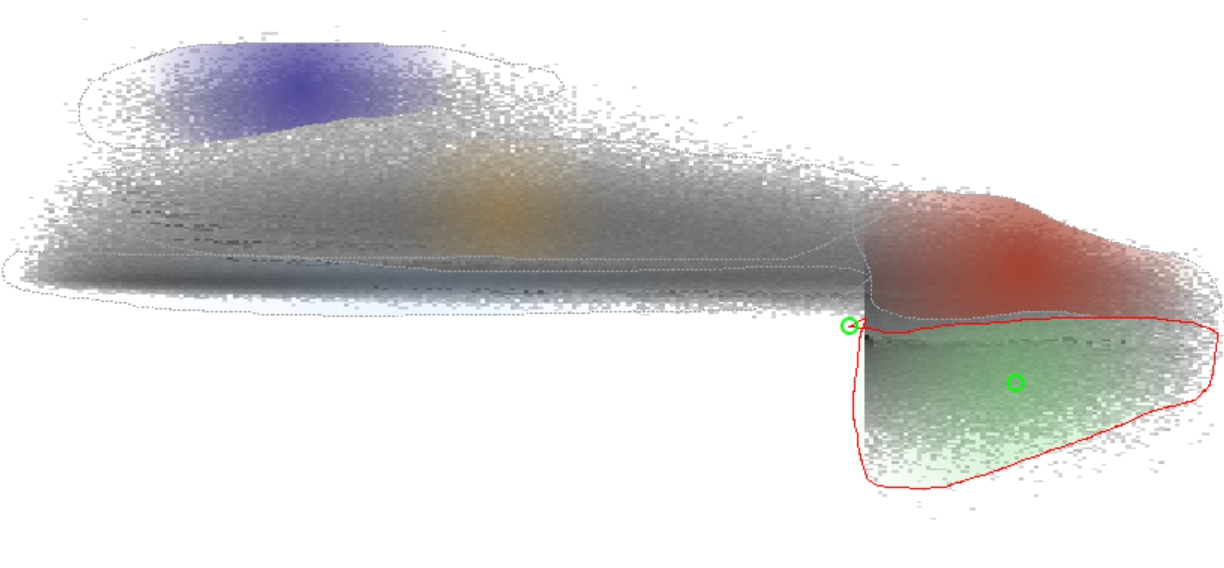}}\hfill
    \subfloat[Volume rendering (MD-RD)]{\includegraphics[height = \dtiheight]{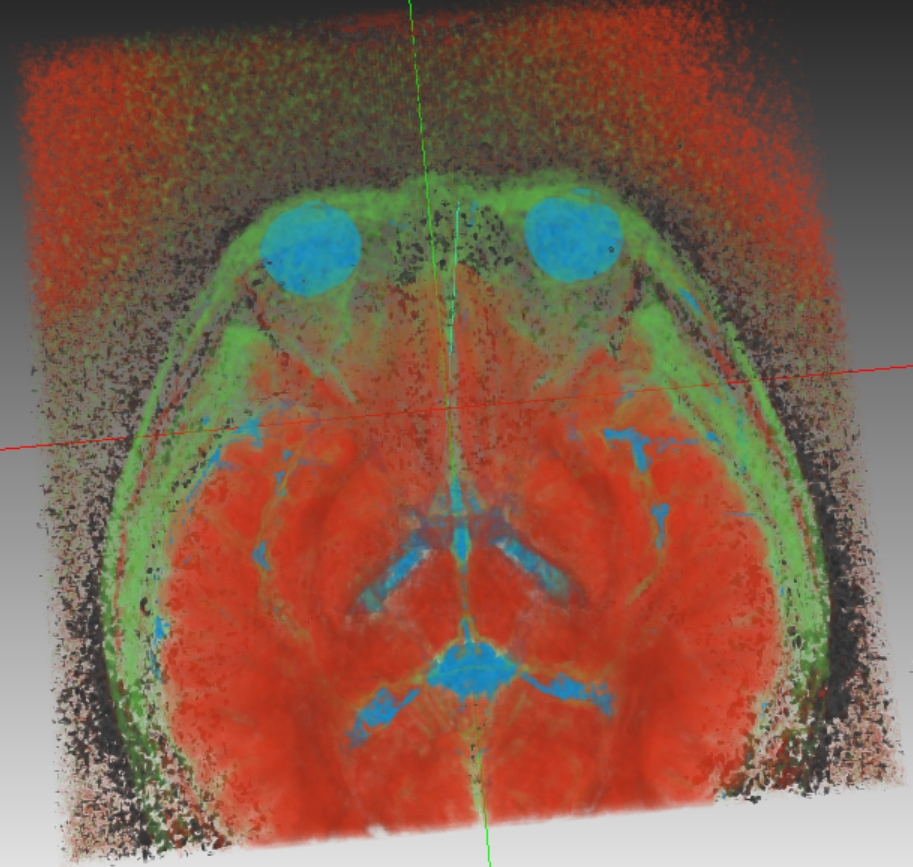}}\hfill
    \subfloat[2D transfer functions (MD-RD)]{\includegraphics[width = 0.25\linewidth]{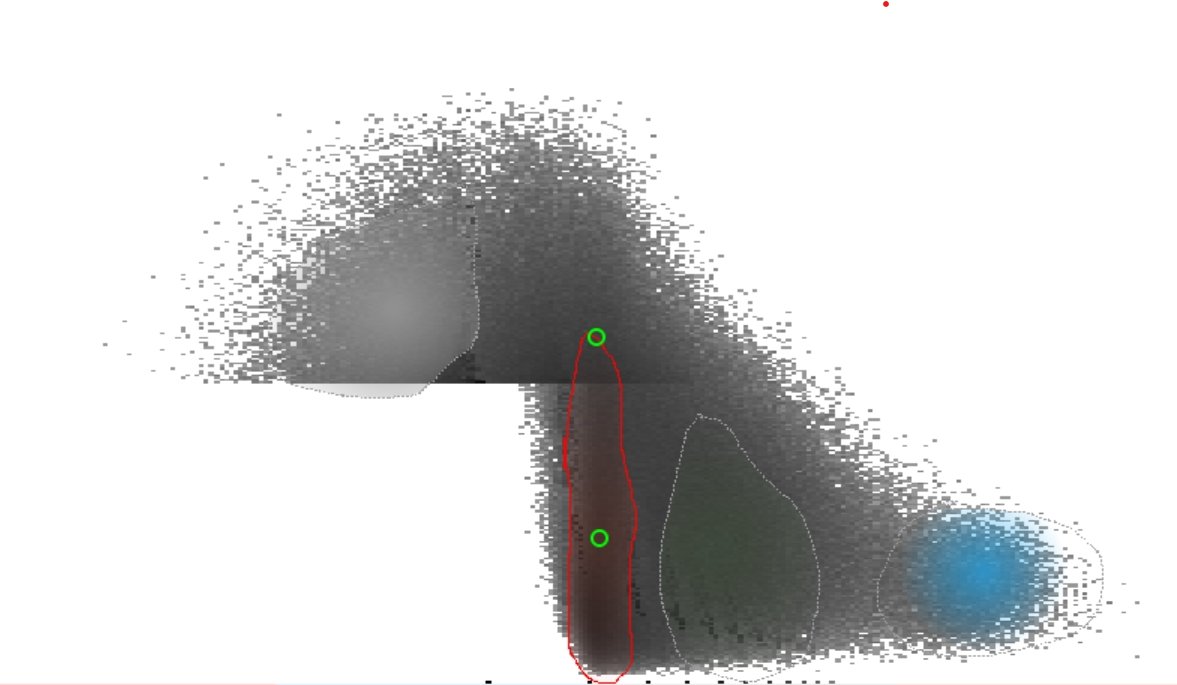}}\\
    \subfloat[Volume rendering (FA-MD-RD)]{\includegraphics[height = \dtiheight]{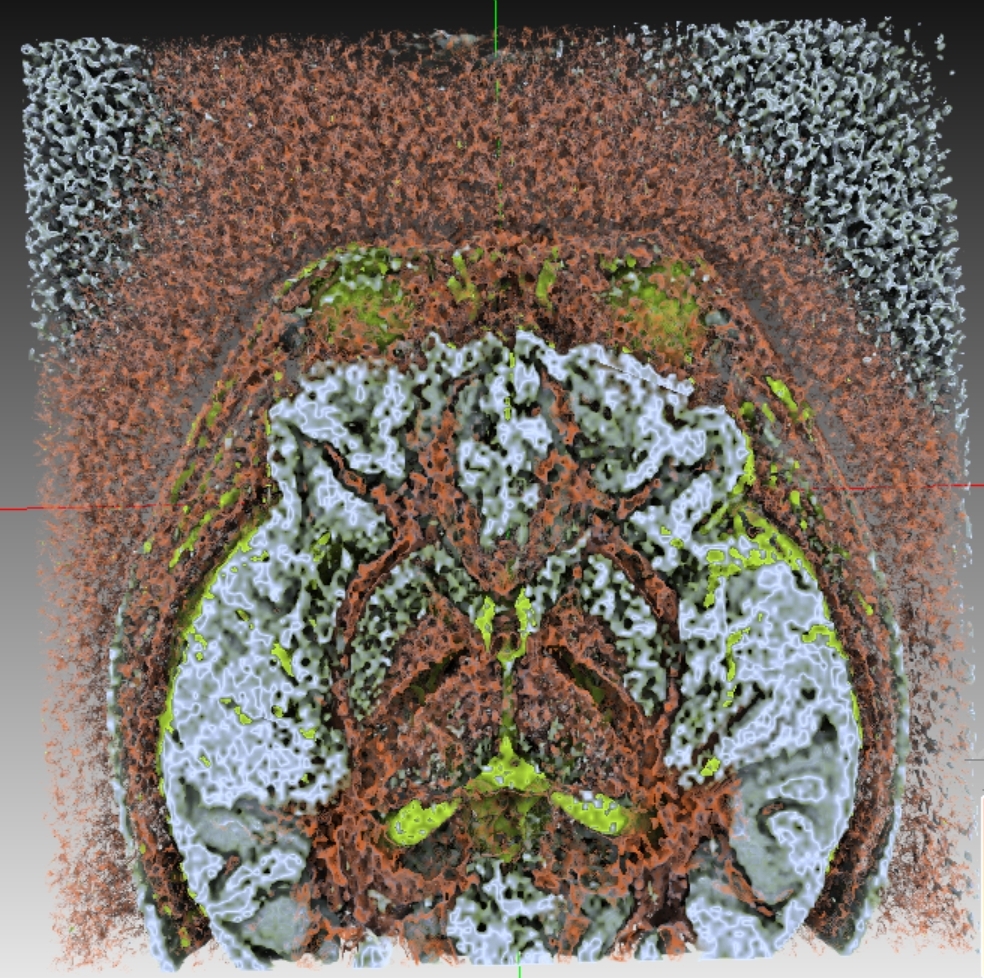}}\;
    \subfloat[3D transfer functions (FA-MD-RD)]{\includegraphics[height = \dtiheight]{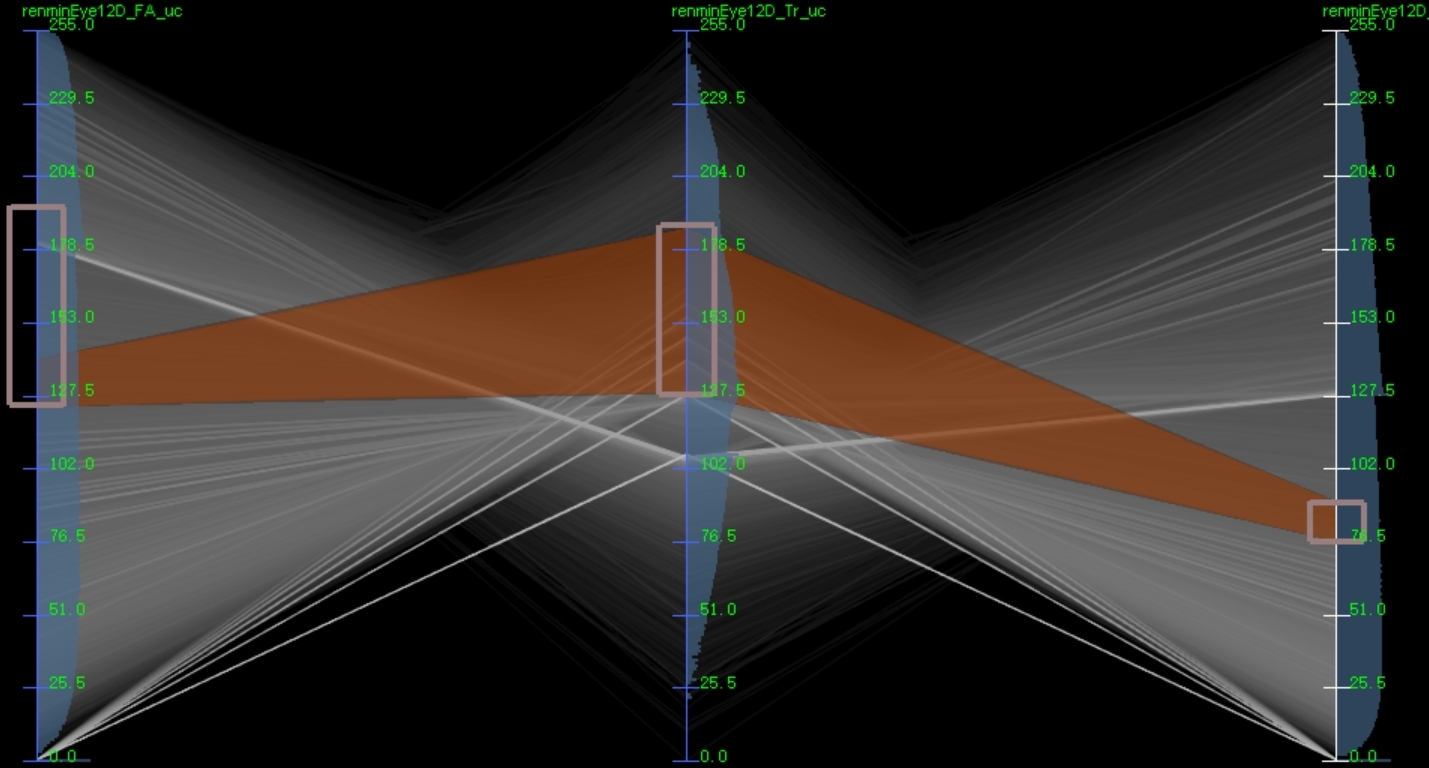}}\hspace{1em}
    \subfloat[1-flat indexed points (FA-MD-RD)]{\includegraphics[height = \dtiheight]{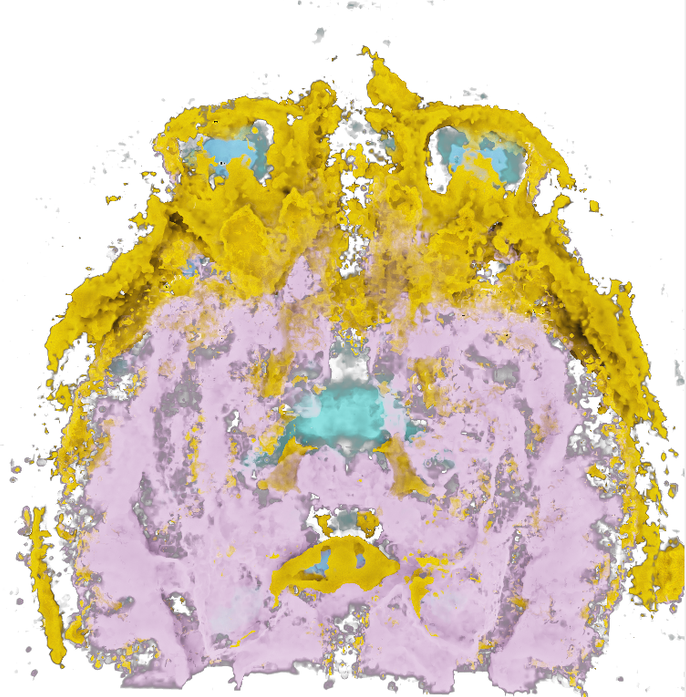}}
    \\
    \caption{Visualizations of the DTI dataset with multivariate transfer functions. The volume renderings (a, c) are generated with (b, d) 2D transfer functions. Another rendering (e) is classified with (f) transfer functions of all three variables. 
    \secRevCvm{For comparison, the volume rendering classified with the 1-flat indexed points is shown in (g). See Figure~\ref{fig:teaser} for a description of the 1-flat indexed points example. Underlying DTI dataset courtesy of Peking University People’s Hospital.}
    }
    \label{fig:DTImdTF}
\end{figure*}

Furthermore, we explore the DTI dataset with multivariate transfer functions, as shown in Figure~\ref{fig:DTImdTF}, to compare with our new method (Figure~\ref{fig:teaser} and Figure~\ref{fig:DTImdTF}~(g)).
Here, the transfer functions are designed on data domains spanned by FA versus MD (Figure~\ref{fig:DTImdTF}~(a, b)), and MD versus RD (Figure~\ref{fig:DTImdTF}~(c, d)), respectively, to be consistent with the case study (Section~\ref{sec:caseStudy}).
It can be seen that visual patterns are scattered and not easily identifiable with the scatterplots (Figure~\ref{fig:DTImdTF}~(b, d)).
We can barely select some high intensity regions and disconnected clusters with lassos---purple and blue regions of Figure~\ref{fig:DTImdTF}~(b, d), respectively, for eye balls and cerebral fluid.
Compared to Figure~\ref{fig:teaser}, multivariate transfer functions create more noisy volume rendered results, cannot separate the brain tissue from the outside environment (blue in Figure~\ref{fig:DTImdTF}~(a), and red in Figure~\ref{fig:DTImdTF}~(c)), and cannot correctly classify muscles that are of the interest of the radiologist.  
Even with all three variables FA, MD, and RD, multivariate transfer functions (Figure~\ref{fig:DTImdTF}~(f)) still cannot classify identifiable muscles (Figure~\ref{fig:DTImdTF}~(e)) after trial-and-error. 

\section*{Declarations}

\subsection*{Availability of Data and Materials}
\secRevCvm{The DTI dataset visualized in Figures~\ref{fig:teaser},~\ref{fig:dti2flats}, and~\ref{fig:DTImdTF} is by courtesy of Peking University People's Hospital.
The MRI dataset shown in Figures~\ref{fig:pipeline},~\ref{fig:renderingCompare},~\ref{fig:ui}, and~\ref{fig:t1compare} is
by Menze et al.~\cite{Menze2015}, and is publicly available at~\url{https://www.kaggle.com/datasets/dschettler8845/brats-2021-task1?select=BraTS2021_Training_Data.tar}.
The Hurricane Isabel dataset visualized in Figures~\ref{fig:idxDomains} and~\ref{fig:isabel} is by NCAR~\cite{HurricaneIsabel:2004:VisContest}, and is publicly available at~\url{https://www.earthsystemgrid.org/dataset/isabeldata.html}.
The CT Tooth dataset shown in Figures~\ref{fig:kderesults} and~\ref{fig:toothCompare} is by GE Aircraft Engines~\cite{Pfister2001}, and is publicly available at~\url{http://volume.open-terrain.org/}.
Other datasets were generated by the authors and will be made publicly available.
The source code for indexed point computation and visualization, and the binary of the interactive visualization tool will be provided too.}

\subsection*{Competing Interests}

The authors have no competing interests to declare that are relevant to the
content of this article.

\subsection*{Funding}
This work was supported in part by the NSFC grant (No.~62372012), and the Deutsche Forschungsgemeinschaft (DFG, German Research Foundation) -- Project-ID 251654672 -- TRR 161.

\subsection*{Ethical Approval}
The user study was approved by the ethics commission of the institution of the last author.

\subsection*{Authors' Contributions}
L. Zhou conceptualized and implemented the method, structured, drafted, provided the supervision of the work, and edited the manuscript.
X. Gou provided medical support as the radiologist and contributed to the relevant parts to the manuscript.
D. Weiskopf conceptualized the method, and contributed to the structuring and editing of the manuscript. 

\subsection*{Acknowledgments}

\bibliographystyle{CVMbib}
\bibliography{idxPtVolumes}

\subsection*{Author biography}
\note{(at least the first author's and the corresponding author's)}

\begin{biography}[zhou]{Liang Zhou} is an Assistant Professor at the National Institute of Health Data Science, Peking University. He received his Ph.D. degree in Computing from the University of Utah, USA, in 2014. His research interests include visualization, visual analytics for health sciences, and extended reality for medicine.

\end{biography}

\begin{biography}[gou]{Xinyi Gou} is a PhD candidate at the Peking University People's Hospital, specializing in Imaging and Nuclear Medicine. Her research focuses on imaging genomics of gastric cancer and multimodal imaging analysis of thyroid-associated ophthalmopathy.
\end{biography}

\begin{biography}[weiskopf]{Daniel Weiskopf}is a Professor at the Visualization Research Center (VISUS) of the University of Stuttgart, Germany. He received his Dr.\,rer.\,nat.\ degree (similar to PhD) in Physics from the University of T\"ubingen, Germany, in 2001, and the Habilitation degree in Computer Science from the University of Stuttgart in 2005. His research interests include visualization, visual analytics, eye tracking, human-computer interaction, XR, computer graphics, and special and general relativity. 
\end{biography}

\vspace*{2.6em}
\subsection*{Graphical abstract}

Graphical abstract is optional yet highly encouraged to supply which can
summarize your content vividly in one picture (at least 600 dpi, 5 cm
$\times $ 8 cm, the ratio of height to length should be less than 1 and
larger than 5/8). We will upload it onto SpringerLink and display it on the
webpage of this paper.

\begin{figure*}
    \centering
    \includegraphics[width = \linewidth]{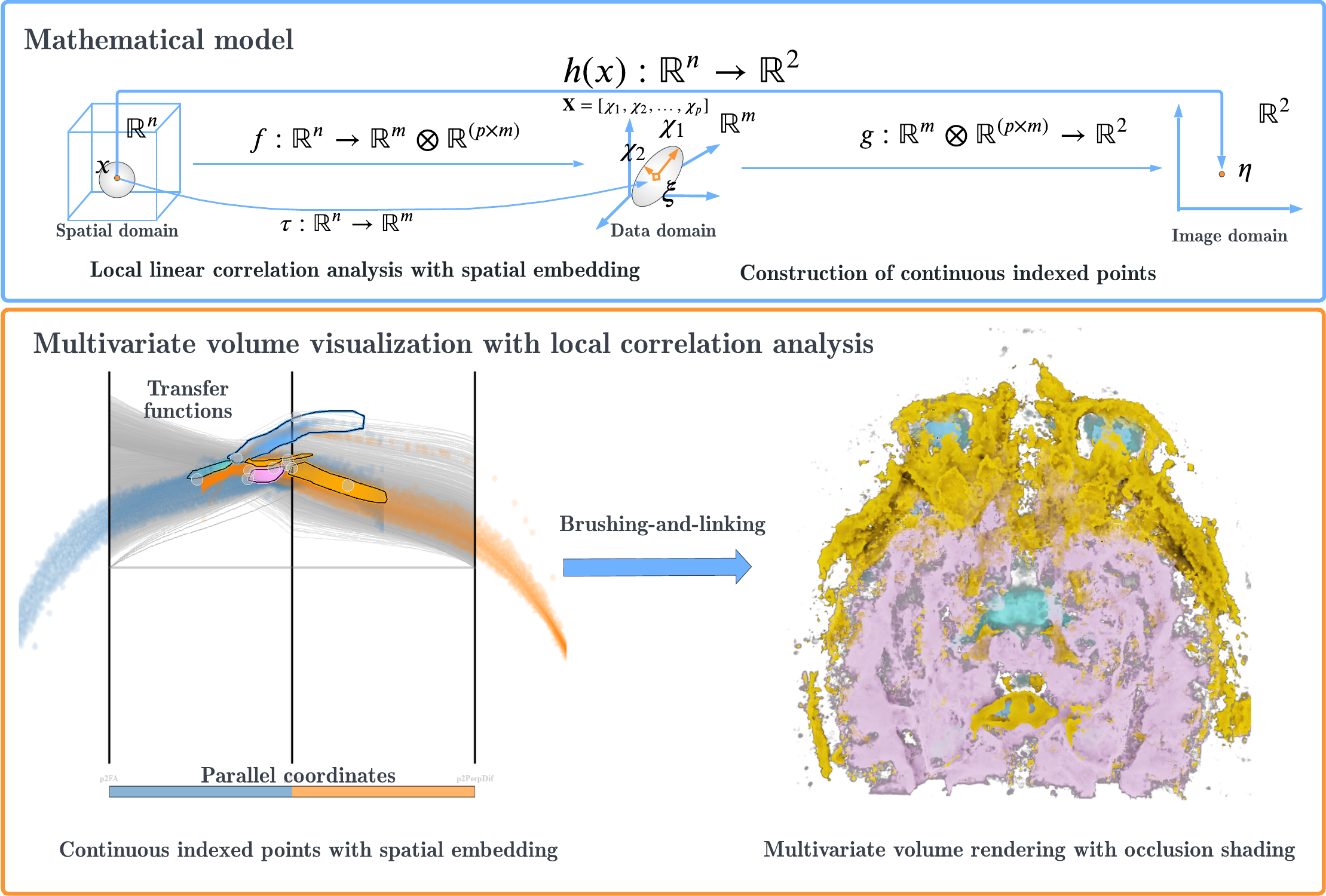}
    \caption{The graphical abstract.}
\end{figure*}

\end{document}